\documentclass[useAMS,usenatbib,10pt]{mn2e}
\usepackage{times}  
\usepackage{listings}
\usepackage{graphics}
\usepackage{subfigure}
\usepackage{graphicx}
\usepackage{amssymb}
\usepackage{hyperref}
\usepackage{aas_macros}

\def\simprop{ \lower .75ex \hbox{$\sim$} \llap{\raise .27ex \hbox{$\propto$}} }

\title[Molecular Hydrogen in EAGLE]{
Molecular hydrogen abundances of galaxies in the EAGLE simulations}
\author[Claudia del P. Lagos et al.]{
\parbox[t]{\textwidth}{
\vspace{-1.0cm}
Claudia del P. Lagos$^{1,2}$\thanks{E-mail: claudia.lagos@icrar.org} Robert A. Crain$^{3,4}$, Joop
Schaye$^{4}$, Michelle Furlong$^{5}$, Carlos S. Frenk$^{5}$, Richard G. Bower$^{5}$, Matthieu Schaller$^{5}$, Tom
Theuns$^{5}$, James W. Trayford$^{5}$, Yannick M. Bah\'e$^{6}$, Claudio Dalla Vecchia$^{7,8}$}
\vspace*{6pt}\\
$^{1}$International Centre for Radio Astronomy Research (ICRAR), M468, University of Western Australia, 35 Stirling Hwy, Crawley, WA 6009, Australia.\\
$^{2}$European Southern Observatory, Karl-Schwarzschild-Strasse 2, 85748,
Garching, Germany.\\
$^{3}$Astrophysics Research Institute, Liverpool John Moores University, 146
Brownlow Hill, Liverpool, L3 5RF, UK.\\
$^{4}$Leiden Observatory, Leiden University, PO Box 9513, 2300 RA Leiden, The
Netherlands.\\
$^{5}$Institute for Computational Cosmology, Department of Physics,
University of Durham, South Road, Durham, DH1 3LE, UK.\\
$^{6}$Max-Planck-Institut fur Astrophysik, Karl-Schwarzschild-Str. 1, D-85748 Garching, Germany.\\
$^{7}$Instituto de Astrof\'isica de Canarias, C/ V\'ia L\'actea s/n, E-38205 La Laguna, Tenerife, Spain.\\ 
$^{8}$Departamento de Astrof\'isica, Universidad de La Laguna, Av. del Astrof\'isico Franciso S\'anchez s/n, E-38206 La Laguna, Tenerife, Spain.
\vspace*{-0.5cm}}

\begin{document}


\pagerange{\pageref{firstpage}--\pageref{lastpage}} \pubyear{2012}

\maketitle

\label{firstpage}

\begin{abstract}
  We investigate the abundance of galactic molecular hydrogen (H$_2$)
  in the ``Evolution and Assembly of GaLaxies and their Environments''
  ({EAGLE}) cosmological hydrodynamic simulations. We assign H$_2$
  masses to gas particles in the simulations in post-processing using
  two different prescriptions that depend on the local dust-to-gas ratio and the 
interstellar radiation field. Both result in H$_2$ galaxy mass
  functions that agree well with observations in the local and
  high redshift Universe. The simulations  reproduce the
  observed scaling relations between the mass of H$_2$ and the stellar
  mass, star formation rate and stellar surface density. Towards high
  redshifts, galaxies in the simulations display larger H$_2$ mass
  fractions and lower H$_2$ depletion timescales,
  also in good agreement with observations. The comoving mass density of H$_2$ in
  units of the critical density, $\Omega_{\rm H_2}$, 
   peaks at $z\approx 1.2-1.5$, later than the
  predicted peak of the cosmic star formation rate activity, at 
  $z\approx 2$. This difference stems from the decrease in gas
  metallicity and increase in interstellar radiation field with
  redshift, both of which hamper H$_2$ formation.  We find that the
  cosmic H$_2$ budget is dominated by galaxies with $M_{\rm
    H_2}>10^9\,\rm M_{\odot}$, star formation rates $>10\,\rm
  M_{\odot}\,\rm yr^{-1}$ and stellar masses $M_{\rm
    stellar}>10^{10}\,\rm M_{\odot}$, which are readily observable in
  the optical and near-IR. The match between the H$_2$
  properties of galaxies that emerge in the simulations and
  observations is remarkable, {particularly since H$_2$ observations were not used to 
  adjust parameters in EAGLE.}
\end{abstract}


\section{Introduction}

The past few years have brought impressive developments in surveys of
molecular gas in resolved and unresolved galaxies, locally and at high
redshift (e.g. \citealt{Leroy08}; \citealt{Saintonge11};
\citealt{Genzel10}; \citealt{Daddi10}; \citealt{Young11};
\citealt{Davis11}; \citealt{Genzel13}; \citealt{Boselli14b}).  Molecular hydrogen (H$_2$)
is an important component of the interstellar medium (ISM) and, in
galaxies like our own, it amounts to a few percent of the stellar
mass \citep{Putman12}.  The exquisite resolution and sensitivity of studies of local
galaxies have enabled important conclusions to be reached.  For
example, on kiloparsec scales, observations show a nearly linear
correlation between the star formation rate (SFR) surface density and
the surface density of H$_2$ (\citealt{Bigiel08};
\citealt{Bigiel10}; \citealt{Leroy08}; \citealt{Schruba11}).  At a
global level (i.e. integrated over the galaxy), the H$_2$ mass
correlates very well with the total SFR (\citealt{Saintonge11};
\citealt{Boselli14b}). These observations place new constraints on
galaxy formation models.

Molecular hydrogen is very difficult to observe in the ISM of galaxies
because it lacks a dipole moment, making its emission extremely weak
 at the typical temperature of the cold ISM.  A
widely-used tracer of H$_2$ is the carbon monoxide (hereafter `CO')
molecule, which is the second most abundant molecule after H$_2$, and
is easily excited (see \citealt{Carilli13} for a recent review on how
CO has been used to trace H$_2$ at $z\lesssim 6$).  Direct CO
detections are available for relatively large samples in the local
Universe ($z<0.1$), from which it has been possible to derive the
$z\approx 0$ $\rm CO(1-0)$ luminosity function \citep{Keres03}, where
$\rm (1-0)$ is the lowest energy rotational transition.  From this
luminosity function, and adopting a Milky-Way like $\rm CO(1-0)$-H$_2$
conversion factor \citep{Bolatto13}, it has been possible to derive
the H$_2$ mass function and the cosmic density of H$_2$, $\rho_{\rm
  H_2}=4.3\pm 1.1\times 10^{7} \, h\,\rm M_{\odot}\,\rm Mpc^{-3}$, at
$z\approx 0.05$ (\citealt{Keres03}; \citealt{Obreschkow09b}).

Campaigns to obtain constraints on $\rho_{\rm H_2}$ at redshifts
higher than $z\approx 0$ have used blind CO surveys tuned to find CO
emission at $z>1$.  For example, \citet{Aravena12} performed a Jansky
Very Large Array\footnote{{\tt
    https://science.nrao.edu/facilities/vla}} blind survey towards a
candidate cluster at $z\sim 1.5$, from which they obtained two
detections and placed constraints on the number density of bright
CO($1-0$) galaxies.  \citet{Walter14} carried out a blind CO survey
with the Plateau de Bure Interferometer and placed constraints on both
the CO($1-0$) luminosity function and $\rho_{\rm H_2}$ in the redshift
range $z\approx 1-3$. Both surveys indicate that $\rho_{\rm H_2}$
increases by an order of magnitude from $z=0$ to $z\approx 2$, but
systematic uncertainties still dominate the measurements.

There is also a wealth of literature using dust mass as a proxy for
H$_2$ mass through the dust-to-gas mass ratio dependence on
metallicity (e.g. \citealt{Boselli02}; \citealt{Santini13};
\citealt{Swinbank14}; \citealt{Bethermin14}). Using this technique,
and with the availability of IR photometry for large samples of
galaxies from Herschel, one can obtain more statistically meaningful
samples than is possible with direct CO imaging.  So far results from
both direct CO detections and dust-derived H$_2$ masses are in good
agreement. For both techniques, scaling relations between the H$_2$
mass and other galaxy properties have been explored locally and at
high redshift.  It has been established that the H$_2$-to-stellar mass
ratio anti-correlates with stellar mass and that such a relation is
present from $z=0$ to at least $z\approx 2.5$
(e.g. \citealt{Saintonge11}; \citealt{Tacconi13}; \citealt{Santini13};
\citealt{Saintonge13}; \citealt{Bothwell14};
\citealt{Dessauges-Zavadsky14}). These studies have also conclusively
shown that there is an overall tendency for an increasing
H$_2$-to-stellar mass ratio with increasing redshift at fixed stellar
mass (e.g. \citealt{Geach11}; \citealt{Saintonge13}).  This evidence
suggests that gas in galaxies at high redshift is denser.

The availability of high-fidelity observations of molecular gas in the
local and distant Universe has opened up new means of testing galaxy
formation models and simulations. This opportunity has been exploited
using semi-analytic models which, when coupled with a star formation
law that relates the SFR directly to the content of H$_2$, reproduce
many of the above correlations (e.g. \citealt{Fu10};
\citealt{Lagos10}; \citealt{Lagos11}; \citealt{Lagos14};
\citealt{Popping14}). However, all these models assume a constant
H$_2$ to SFR conversion efficiency, even though there is observational
evidence contradicting this (e.g. \citealt{Saintonge13};
\citealt{Dessauges-Zavadsky14}; \citealt{Huang14}).

Hydrodynamic simulations are well suited to investigate the relation
between SFR, H$_2$ mass, stellar mass and other galaxy properties
because the gas dynamics is followed self-consistently and no
relations between the global (galaxy-wide) and local properties of the
gas are necessarily imposed (which is the case in semi-analytic
models, where galaxies are unresolved). Thus, one can use observations
of H$_2$ to test whether the simulations reproduce the scaling
relations of the gas component of galaxies and isolate the physical
drivers of the observed relations.

Many of the scaling relations discussed above, for example those
between the surface density of SFR, H$_2$, and other properties of the
ISM of galaxies (such as hydrostatic pressure), have been explored in
simulations of small portions of the ISM and in individual galaxy
simulations in which H$_2$ formation has been implemented
(e.g. \citealt{Pelupessy06}; \citealt{Pelupessy09};
\citealt{Robertson08}; \citealt{Gnedin09}; \citealt{Glover12};
\citealt{Christensen12}; \citealt{Walch14}). For example, the
physical drivers of the relation between the surface density of SFR
and H$_2$ have been explored in \citet{Feldmann11}, \citet{Gnedin11},
and \citet{Glover12}. {\citet{Feldmann11} and \citet{Gnedin11} show that the 
 latter relation is not fundamental in the sense that large variations are expected 
 for varying physical conditions (e.g. gas metallicity and interstellar radiation field). 
In addition, \citet{Glover12} argued against the importance of molecular cooling (e.g. from H$_2$ and CO) 
 as the low temperatures and high-densities needed
to form stars are reached even in situations where the gas is forced to remain
atomic (see also \citet{Schaye04}).} 
 We can implement some of the relations derived from the
high-resolution simulations above in lower-resolution, cosmological
hydrodynamic simulations to explore the scaling relation between the
gas properties and other galaxy properties.

Recently, significant advances have been achieved in cosmological
hydrodynamical simulations, which can now produce a galaxy population
with properties that are broadly consistent with many observations
(see \citealt{Vogelsberger14b} for the {Illustris} simulation and
\citealt{Schaye14} for the Evolution and Assembly of GaLaxies and
their Environments, { EAGLE}, simulations).  For example, {EAGLE}
reproduces the stellar mass function of galaxies at $z\approx 0.1$,
the SFR history of the Universe, and the properties of star-forming
galaxies, in large cosmological volumes, ($\sim (100\,\rm Mpc)^3)$,
with enough particles to resolve the Jeans scale in the warm ISM.

\citet{Genel14} presented molecular gas scaling relations for the {
  Illustris} simulation, and showed that it reproduces the
observed correlation between H$_2$ and stellar mass reasonably well if
the star-forming gas (which has high enough density to allow star
formation to take place) is used as a proxy for the H$_2$ mass.
However, {estimating the H$_2$ fraction from the neutral gas
  introduces a large systematic uncertainty because the range of
  temperatures, densities and metallicities of the star-forming gas
  indicates that, contrary to the assumption of Genel et al., it is
  not always H$_2$-dominated.}

The {OWLS} simulation suite described in \citet{Schaye10}, a
forerunner of EAGLE, explored lower-resolution and smaller
cosmological volumes than EAGLE.  \citet{Duffy12b} presented a
comparison between OWLS and the H$_2$ mass function and scaling
relations for galaxies with $M_{\rm H_2}>10^{10}\,\rm M_{\odot}$,
which was the mass range probed by their simulation.  The authors
found that the H$_2$ content of galaxies decreases with decreasing
redshift, but their galaxies tended to be more H$_2$-rich than
observed and the mass range probed was very limited.  We extend this
work using more sophisticated, theoretically motivated ways of
calculating the H$_2$ fraction, and apply them to EAGLE, which has a
much larger dynamical range than OWLS and, unlike OWLS, reproduces key
properties of the observed galaxy population.

Here, we present a comprehensive comparison between EAGLE and
observations of H$_2$ on galaxy scales. Compared with, for example,
the work of \citet{Duffy12} and \citet{Genel14}, we extract a more
realistic H$_2$ fraction for individual gas particles, based on their
gas metallicity, density, pressure, SFR and Jeans length.  The methods
to compute the H$_2$ fraction of gas particles that we use here employ
simple, analytic prescriptions that are based on more detailed
studies.  This kind of parametrisation avoids the use of expensive
radiative transfer and high-resolution simulations that are required
to account for direct H$_2$ formation within a multi-phase ISM. However,
we note that in parallel to our simple approach, the development of
detailed chemistry networks and their application to individual
galaxies is under way (e.g. \citealt{Glover07}; \citealt{Richings14a};
\citealt{Richings14b}). In future, the application of such networks to
entire galaxies will enable us to test how well the phenomenological
approach that we adopt here describes the ISM in galaxies, and to
identify regimes within which our approach is inaccurate. An
analysis of the atomic hydrogen content of galaxies in {EAGLE} will be
presented in Crain et al. (in prep.), Bah\'e et al. (submitted) and
\citet{Rahmati15}.

This paper is organised as follows. In $\S$~\ref{EagleSec} we give a
brief overview of the simulation and the subgrid physics included in
EAGLE. In $\S$~\ref{PartitioningSec} we describe how we partition the
gas into ionised, atomic and molecular components, the methods we
use to do this and their main differences, with detailed
equations provided in Appendix~\ref{H2Prescriptions}.
Comparisons with local Universe and high-redshift surveys are
presented in $\S$~\ref{localU} and $\S$~\ref{RedEvoSec} respectively.
We discuss the results and present our conclusions in
$\S$~\ref{ConcluSec}.  In Appendix~\ref{ConvTests} we present `weak'
and `strong' convergence tests (terms introduced by \citealt{Schaye14}),
focusing on the H$_2$ properties of galaxies.

\section{The { EAGLE} simulation}\label{EagleSec}

The {EAGLE} simulation suite\footnote{See {\tt
    http://eagle.strw.leidenuniv.nl} and {\tt
    http://icc.dur.ac.uk/Eagle/} for images, movies and data
  products.}  (described in detail by \citealt{Schaye14}, hereafter
S15, and \citealt{Crain15}) consists of a large number of cosmological
simulations with different resolution, box sizes and physical models,
adopting the cosmological parameters of the \citet{Planck14}.  S15
introduced a reference model, within which the parameters of the
subgrid models governing energy feedback were calibrated to ensure a
good match to the $z=0.1$ galaxy stellar mass function, whilst also
reproducing the observed sizes of present-day disk galaxies. \citet{Furlong14} 
presented the evolution of the galaxy stellar mass function for the EAGLE simulations 
and found that the agreement extends to $z\approx 7$.

The
EAGLE suite includes simulations adopting the reference model, run
within cosmological volumes of $(25\,{\rm cMpc})^3$, $(50\,{\rm
  cMpc})^3$ and $(100\,{\rm cMpc})^3$, where cMpc denotes comoving
megaparsecs.  In Table~\ref{TableSimus} we summarise technical details
of the simulations used in this work, including the number of
particles, volume, particle masses, and spatial resolution.  The label
of each simulation indicates box size and particle number.  For
example, L0100N1504 is a simulation of length $100$~cMpc on a side,
using $1504^3$ particles of dark matter and an equal number of
baryonic particles. Note that the spatial resolution of the
intermediate-resolution simulations  (such as L0100N1504) was chosen to
(marginally) resolve the Jeans length in the warm ISM (with a
temperature of $\approx 10^4$~K; see S15 for details). In
Table~\ref{TableSimus}, pkpc denotes proper kiloparsecs.

\begin{table*}
\begin{center}
  \caption{{EAGLE} simulations used in this paper.  The columns list:
    (1) the name of the simulation, (2) comoving box size, (3) number
    of particles, (4) initial particle masses of gas and (5) dark
    matter, (6) comoving gravitational
    softening length, and (7) maximum proper comoving Plummer-equivalent
    gravitational softening length. Units are indicated below the name of
    column. The top two simulations are of intermediate resolution,
    while the bottom two are high-resolution simulations. { In EAGLE 
    we adopt (6) as the softening length at $z\ge 2.8$, and (7) at $z<2.8$. 
    At $z=2.8$, $2.66\,{\rm ckpc}\equiv 0.7\,{\rm pkpc}$ and $1.33\,{\rm ckpc}\equiv 0.35\,{\rm pkpc}$.}}\label{TableSimus}
\begin{tabular}{l c c c c c c}
\\[3pt]
\hline
(1) & (2), & (3) & (4) & (5) & (6) & (7) \\
\hline
Name & $L$ & \# particles & gas particle mass & DM particle mass & Softening length & max. gravitational softening \\
Units & $[\rm cMpc]$   &                &  $[\rm M_{\odot}]$ &  $[\rm M_{\odot}]$ & $[\rm ckpc]$ & $[\rm pkpc]$\\
\hline
Ref-L025N0376 & ~$25$ &   ~$2\times 376^3$  &$1.81\times 10^6$   &   ~~$9.7\times 10^6$ & $2.66$ & ~$0.7$  \\
Ref-L050N0752 & ~$50$  &  ~$2\times 752^3$    &$1.81\times 10^6$ & ~~$9.7\times 10^6$   & $2.66$ & ~$0.7$  \\
Ref-L100N1504 & $100$ &  $2\times 1504^3$   &$1.81\times 10^6$   &   ~~$9.7\times 10^6$ & $2.66$ & ~$0.7$  \\
Ref-L025N0752 & ~$25$ &   ~$2\times 752^3$  &$2.26\times 10^5$   &   $1.21\times 10^6$  & $1.33$ &  $0.35$ \\
Recal-L025N0752 & ~$25$ & ~$2\times 752^3$  &$2.26\times 10^5$   &   $1.21\times 10^6$  & $1.33$ &  $0.35$ \\
\hline
\end{tabular}
\end{center}
\end{table*}

The EAGLE simulations were performed using an extensively modified
version of the parallel $N$-body smoothed particle hydrodynamics (SPH)
code {\tt gadget-3} (\citealt{Springel08}; see also
\citealt{Springel05b} for the publicly available code {\tt gadget-2}).
{\tt Gadget-3} is a Lagrangian code, where a fluid is represented by a
discrete set of particles, from which the gravitational and
hydrodynamic forces are calculated. SPH properties, such as the
density and pressure gradients, are computed by interpolating across
neighbouring particles.  The main modifications to the standard {\tt
  Gadget-3} code for the EAGLE project include updates to the
hydrodynamics, as described in Dalla Vecchia (in prep.; see also
Appendix~A in S15), which are collectively referred to as
`Anarchy'. The impact of these changes on cosmological simulations are
discussed in Schaller et al. (in prep.).  Among the main features of
Anarchy are (i) the pressure-entropy formulation of SPH described in
\citet{Hopkins13b}, (ii) the artificial viscosity switch of
\citet{Cullen10}, (iii) an artificial conduction switch described in
\citet{Price08}, (iv) a C2 \citet{Wendland95} kernel with $58$
neighbours to interpolate SPH properties across neighbouring
particles, and (v) the timestep limiter from \citet{Durier12}, which
is required to model feedback events accurately.

Another major aspect of the EAGLE project is the use of
state-of-the-art subgrid models that capture the unresolved physics.
We briefly discuss the subgrid physics modules adopted by EAGLE in
$\S$~\ref{subgridsec}.  In order to distinguish models with different
parameter sets, a prefix is used. For example, Ref-L100N1504
corresponds to the reference model adopted in a simulation with the
same box size and particle number of L100N1504.  A complete
description of the model can be found in S15 and an analysis of the
impact of variations in the parameters of the subgrid physics on
galaxy properties is given in \citet{Crain15}. Crain et al. also presented 
the motivation for the parameters adopted in each physics module in EAGLE.

S15 introduced the concept of `strong' and `weak' convergence
tests. Strong convergence refers to the case where a simulation is
re-run with higher resolution (i.e. better mass and spatial resolution)
adopting exactly the same subgrid physics and parameters. Weak
convergence refers to the case when a simulation is re-run with higher
resolution but the subgrid parameters are recalibrated to recover, as
far as possible, similar agreement with the adopted calibration
diagnostic (in the case of EAGLE, the $z=0.1$ galaxy stellar mass
function and disk sizes of galaxies). S15 argue that simulations that do not resolve the cold
ISM, nor the detailed effects of feedback mechanisms, require
calibration of the subgrid model for feedback, and that such
simulations should be recalibrated if the resolution is changed,
because in practice changing the resolution leads to changes in the
physics of the simulation as better spatial resolution allows higher
gas densities to be reached. In this spirit, S15 introduced two
higher-resolution versions of {EAGLE}, both in a box of
($25$~cMpc)$^{3}$ and with $2\times 752^3$ particles, Ref-L025N0752
and Recal-L025N0752. These simulations have better spatial and mass
resolution than the intermediate resolution simulations by factors of
$2$ and $8$, respectively.

Comparisons between the simulations Ref-L025N0376 and Ref-L025N0752
represent strong convergence tests, whilst comparisons between the
simulations Ref-L025N0376 and Recal-L025N0752 test weak convergence.
In the simulation Recal-L025N0752, the values of $4$ parameters have
been slightly modified with respect to the reference simulation. These
are related to the efficiency of feedback associated with star
formation and active galactic nuclei (AGN), which are adjusted to
reach a similar level of agreement with the galaxy stellar mass
function at $z=0.1$ as obtained for the Ref-L025N0376 simulation (see
S15 for details). In Appendix~\ref{ConvTests} we present strong and
weak convergence tests explicitly involving the H$_2$ abundance of
galaxies.

\subsection{Subgrid physics included in { EAGLE}}\label{subgridsec}

\begin{itemize}
\item {\it Radiative cooling and photoheating.}  Radiative cooling and
  heating rates are computed on an element-by-element basis for gas in
  ionisation equilibrium exposed to the Cosmic Microwave Background
  Radiation (CMB), and the model for the UV and X-ray background
  radiation from \citet{Haardt01}.  Cooling tables are produced with
  {\tt CLOUDY} (version 07.02; \citealt{Ferland98}) for the $11$
  elements that dominate the cooling rate
  (i.e. H,~He,~C,~N,~O,~Ne,~Mg,~S,~Fe,~Ca,~Si), which are followed
  individually. For more details see \citet{Wiersma09b} and S15.
 
\item {\it Star formation.} Gas particles that have cooled to reach
  densities greater than $n^{\ast}_{\rm H}$ are eligible for
  conversion to star particles, where $n^{\ast}_{\rm H}$ is defined
  as:
\begin{equation}
n^{\ast}_{\rm H}=10^{-1}\, {\rm cm^{-3}} \left(\frac{Z}{0.002}\right)^{-0.64},
\label{critn}
\end{equation}

\noindent which has a dependence on gas metallicity, $Z$, as described
in \citet{Schaye04} and S15.  Gas particles with densities greater
than $n^{\ast}_{\rm H}$ are assigned a SFR, $\dot{m}_{\star}$
\citep{Schaye08}:
\begin{equation}
\dot{m}_{\star}=m_{\rm g}\,A\,(1\,{\rm M}_{\odot}\,{\rm pc}^{-2})^{-n}\, \left(\frac{\gamma}{G}\,f_{\rm g}\,P\right)^{(n-1)/2},
\label{SFlaw}
\end{equation}

\noindent where $m_{\rm g}$ is the mass of the gas particle,
$\gamma=5/3$ is the ratio of specific heats, $G$ is the gravitational
constant, $f_{\rm g}$ is the mass fraction in gas (which is unity for
gas particles), $P$ is the total pressure. $A=1.515\times 10^{-4}\,\rm
M_{\odot}\,\rm yr^{-1}\,kpc^{-2}$ and $n=1.4$ are chosen to reproduce
the observed Kennicutt-Schmidt relation \citep{Kennicutt98}, scaled to
a Chabrier initial mass function (IMF).  A global temperature floor,
$T_{\rm eos}(\rho)$, is imposed, corresponding to a polytropic
equation of state,

\begin{equation}
P\propto \rho^{\gamma_{\rm eos}}_{\rm g}, 
\label{EoS}
\end{equation} 

\noindent where $\gamma_{\rm eos}=4/3$. Eq~\ref{EoS} is normalised 
to give a temperature $T_{\rm eos}=8\times 10^3$~K at $n_{\rm H}=10^{-1}\,\rm cm^{-3}$, 
which is typical of the warm ISM (e.g. \citealt{Richings14a}).

\item {\it Stellar evolution and enrichment.}
The contribution from mass and metal loss from 
Asymptotic Giant Branch (AGB) stars, massive stars and 
supernovae (both core collapse and type Ia) are tracked using 
the yield tables of \citet{Portinari98}, \citet{Marigo01}, 
and \citet{Thielemann03}. The mass and metals that are lost from stars are 
added to the gas particles that are within the SPH kernel of the given star particle 
(see \citealt{Wiersma09} and S15 for details). The adopted stellar IMF 
is that of \citet{Chabrier03}, with minimum and maximum masses of $0.1\,\rm M_{\odot}$ and $100\,\rm M_{\odot}$.

\item {\it Stellar Feedback.} 
  The method used in {EAGLE} to represent energetic feedback
  associated with star formation (which we refer to as `stellar
  feedback') was introduced by \citet{DallaVecchia12}, and consists of
  a stochastic selection of neighbouring gas particles that are heated
  by a temperature of $10^{7.5}$~K. A fraction of the energy from
  core-collapse supernovae is injected into the ISM $30$~Myr after the
  star particle forms. {This fraction, $f_{\rm th}$, depends on the local
  gas metallicity and density as,}

  \begin{equation}
     f_{\rm th}=f_{\rm th,min} + \frac{f_{\rm th,max}-f_{\rm th,min}}{1+\left(\frac{Z}{0.1\,Z_{\odot}}\right)^{n_{\rm Z}}\left(\frac{n_{\rm H,birth}}{n_{\rm H,0}}\right)^{-n_{\rm n}}},
  \label{SNfeedback}
  \end{equation}

 \noindent {where $Z$ is the star particle metallicity, $n_{\rm H,birth}$ is the gas density of the star particle's parent 
gas particle, when the star particle was formed. The parameters $f_{\rm th,min}$ and $f_{\rm th,max}$ are fixed at 
$0.3$ and $3$, respectively, for the simulations analysed here. In the reference model,  
$n_{\rm Z}=n_{\rm n}=2/{\rm ln}(10)$ and $n_{\rm  H,0}=0.67\,\rm cm^{-3}$, while in the recalibrated model $n_{\rm n}$ is reduced to 
$1/{\rm ln}(10)$ and $n_{\rm  H,0}$ to $0.25\,\rm cm^{-3}$. In principle, $f_{\rm th,max}>1$ may seem unphysical. 
However, the mechanism of ‘stellar feedback’
in EAGLE includes several forms of feedback in the
ISM: supernovae, stellar winds, radiation pressure, etc. Moreover, a
subgrid efficiency greater than unity can be justified if there are
numerical radiative losses. S15 quoted the mean and median values of $f_{\rm th}$ at $z=0.1$ which are 
$1.06$ and $0.7$ for the Ref-L100N1504 simulations, and $1.07$ and $0.93$ for the Recal-L025N0752, respectively.}

  The metallicity dependence in Eq.~\ref{SNfeedback} aims to take
  into account the faster cooling that takes place in regions of high
  metallicity, while the density dependence compensates for numerical
  radiative losses that occur at high density as a result of finite
  resolution (see \citealt{Crain15} and S15 for the motivation of this
  model). The need for this phenomenological model comes from the
  fact that stellar feedback is not simulated from first principles.

\item {\it Black hole growth and AGN feedback.} When halos become more
  massive than $10^{10}\,{\rm M}_{\odot} \,h^{-1}$, they are seeded
  with black holes of mass $10^5\,{\rm M}_{\odot}
  \,h^{-1}$. Subsequent gas accretion episodes make black holes grow
  at a rate that is computed following the modified Bondi-Hoyle
  accretion rate of \citet{Rosas-Guevara13}. In this modification, the
  angular momentum of the gas can reduce the accretion rate compared
  to the standard Bondi-Hoyle rate if the tangential velocity of the
  gas is similar to, or larger than, the local sound speed. A
  viscosity parameter, $C_{\rm visc}$, controls the degree of modulation
  of the Bondi rate in high circulation flows, and it is a free
  parameter of the model.  The value of $C_{\rm visc}$ in the
  reference simulations is $C_{\rm visc}=2\pi$, and in the
  recalibrated, higher resolution simulation it is $C_{\rm
    visc}=2\pi\times 10^3$ (see \citealt{Crain15} for analysis of the
  effect of $C_{\rm visc}$ on galaxy properties).  The Eddington limit
  is imposed as an upper limit to the accretion rate onto black holes.
  In addition, black holes can grow by merging with other black holes.

  For AGN feedback, a similar model to the stochastic model of
  \citet{DallaVecchia12} is applied.  Particles surrounding the black
  hole are chosen randomly and heated by a temperature $\Delta T_{\rm
    AGN}=10^{8.5}$~K in the reference simulation and $\Delta T_{\rm
    AGN}=10^{9}$~K in the recalibrated simulation (see
  Table~\ref{TableSimus} for the prefixes). 

\end{itemize}

Throughout the paper we make extensive comparisons between stellar
mass, SFR, H$_2$ mass and metallicity. Following S15, all these
properties are measured in $3$-dimensional apertures of $30$~pkpc. 
The effect of the aperture is minimal, even if we double
its size to $60$~pkpc differences are within $15$\% and much smaller
for most galaxies.

\section{Calculating HI and H$_2$ gas fractions}\label{PartitioningSec}

We describe how we calculate the neutral gas fraction of each gas
particle in $\S$~\ref{NeutralFraction} and how much of that neutral
gas is in molecular form in $\S$~\ref{H2FractionSec}. We discuss two
prescriptions for calculating the H$_2$ fraction and some basic
predictions of each of them. For clarity, we introduce the number
density of hydrogen, $n_{\rm H}\equiv n_{\rm HII} + n_{\rm HI} +
2\,n_{\rm H_2}$, and of neutral hydrogen, $n_{\rm n}\equiv n_{\rm HI}
+ 2\,n_{\rm H_2}$. We also refer to the fraction $f_{\rm H_2}\equiv
\Sigma_{\rm H_2}/\Sigma_{\rm n}$, where $\Sigma_{\rm H_2}$ and
$\Sigma_{\rm n}$ are the mass surface densities of H$_2$ and neutral
hydrogen, respectively.

\subsection{Calculating the neutral fraction}\label{NeutralFraction}

Before we calculate the mass of H$_2$ associated with each gas
particle, we need to determine the transition from H${\rm II}$ to
neutral hydrogen. \citet{Rahmati13} studied the neutral gas fraction
in cosmological simulations by coupling them to a full radiative
transfer calculation with the aim of predicting the neutral gas column
densities.  Rahmati et al. presented fitting functions to their
results, which enable the calculation of the neutral fraction on a
particle-by-particle basis from the gas temperature and density, and
the total ionisation rate (photoionisation plus collisional
ionisation). This is presented in Appendix~A of \citet{Rahmati13}.
Here, we do not repeat their calculations but instead refer the reader
to \citet{Rahmati13}.  Using these fitting functions we calculate
$\eta=n_{\rm n}/n_{\rm H}$.

\subsection{Calculating the H$_2$ gas fraction}\label{H2FractionSec}

In this section we briefly describe the prescriptions of
\citet{Gnedin11} and \citet{Krumholz13} to calculate the H$_2$ gas
fraction for individual gas particles, which we apply to EAGLE.  Both
prescriptions give H$_2$ fractions that depend on the gas metallicity
and the interstellar radiation field.

We use the SPH kernel-smoothed metallicities rather than the particle metallicity 
with the aim of alleviating the consequences of the lack of metal mixing that arises from 
the fact that metals are fixed to particles \citep{Wiersma09}.
We assume the interstellar radiation field to be proportional to the SFR surface density, $\Sigma_{\rm SFR}$ (see 
Appendix~\ref{H2Prescriptions} for details).

\subsubsection{The prescription of \citet{Gnedin11} applied to {EAGLE}}\label{GnedinSec}

\citet{Gnedin11} developed a phenomenological model for H$_2$
formation and applied it to a set of zoom-in cosmological simulations
at high resolution that also included gravity, hydro-dynamics,
non-equilibrium chemistry combined with equilibrium cooling rates for
metals, and a $3$-dimensional, on the fly, treatment of radiative
transfer, using an Adaptive Mesh Refinement (AMR) simulation. The
highest spatial resolution achieved in these simulations is $260$ comoving parsecs (cpc),
and the gas mass of individual resolution elements ranges from
$10^3\,\rm M_{\odot}$ to $10^6\,\rm M_{\odot}$.  From the outcome of
the simulations, the authors parametrised in analytic formulae the
fraction of H$_2$-to-total neutral gas, $f_{\rm H_2}$, as dependent on
the dust-to-gas ratio and the interstellar radiation field.
We assume the dust-to-gas mass ratio scales with the local
metallicity, and the radiation field scales with the local surface
density of star formation, which we estimate from the properties of
gas particles assuming local hydrostatic equilibrium
(\citealt{Schaye01}; \citealt{Schaye08}).  The equations used to
implement this prescription into EAGLE are given in
Appendix~\ref{H2Prescriptions}. We refer to this prescription as `GK11'.
{\citet{Gnedin14} (GD14) presented an updated version of the GK11 model taking into account the 
line blending in the Lyman-Werner bands. We find that this model gives results that different from the  
GK11 prescription by $\approx 50\%$, in a way that the resulting H$_2$ masses after applying the GD14 prescription 
are lower than when applying the GK11 prescription. This difference decreases with redshift, and even reverses at $z\gtrsim 3.5$, with 
the H$_2$ masses in the GD14 model being higher than in the GK11 model. The differences between the GD14 and GK11 prescriptions 
are of a similar magnitude as the differences between the GK11 and \citet{Krumholz13} prescriptions ($\S$~\ref{DiffPrescriptions}).
However, applying the GD14 or GK11 prescriptions leads to the same conclusions we present in this work, and 
therefore we will hereafter only use the GK11 prescription. In Appendix~\ref{GnedinSec} we show comparisons between 
the GK11 and GD14 prescriptions.}

\begin{figure}
\begin{center}
\includegraphics[width=0.23\textwidth]{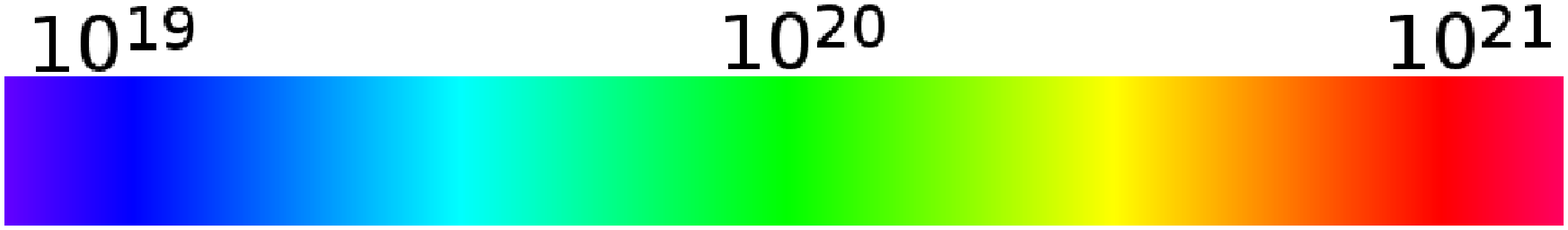}
\includegraphics[width=0.23\textwidth]{Figs/TableColoursrgb.eps}\\
\includegraphics[width=0.23\textwidth]{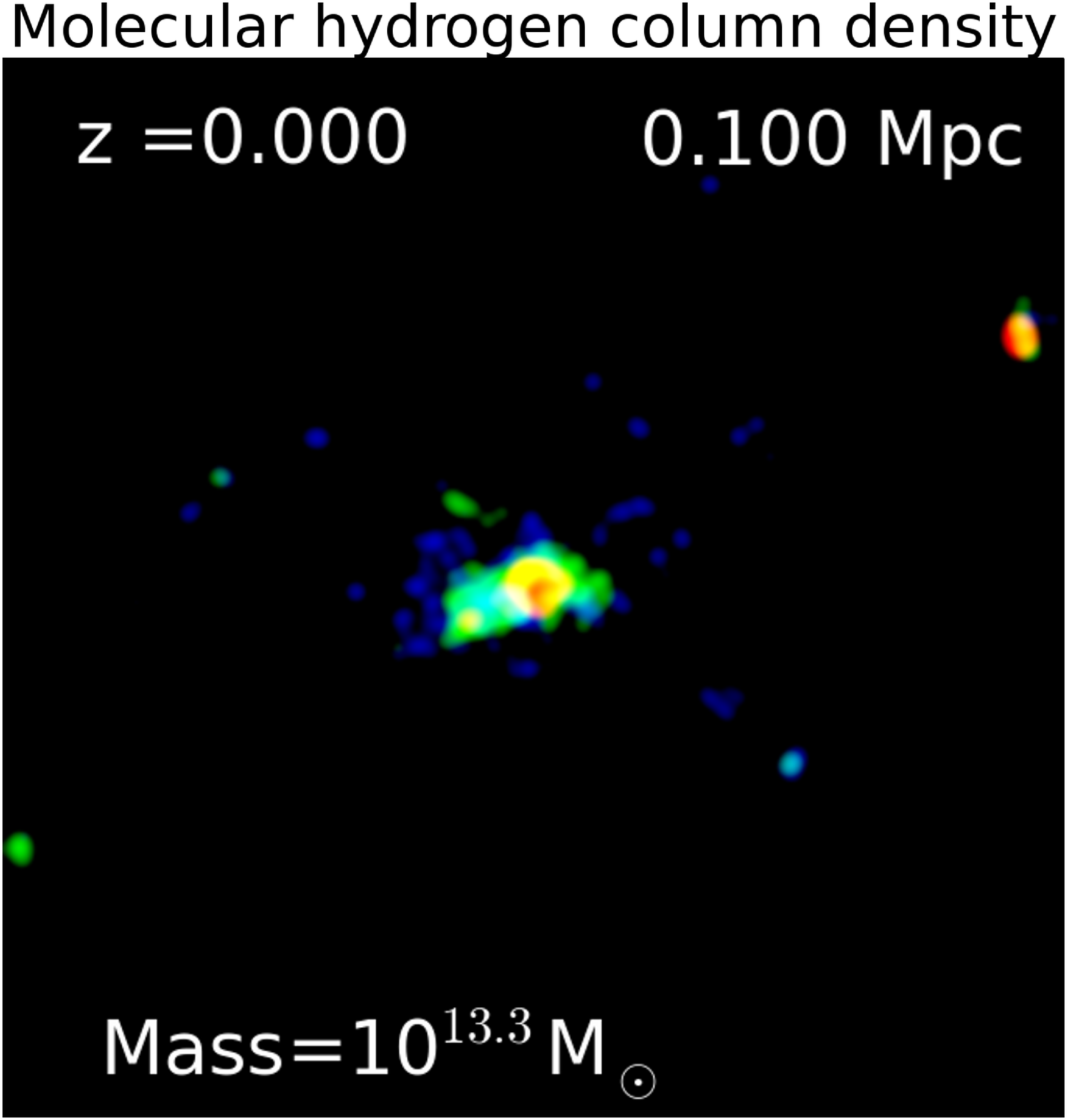}
\includegraphics[width=0.23\textwidth]{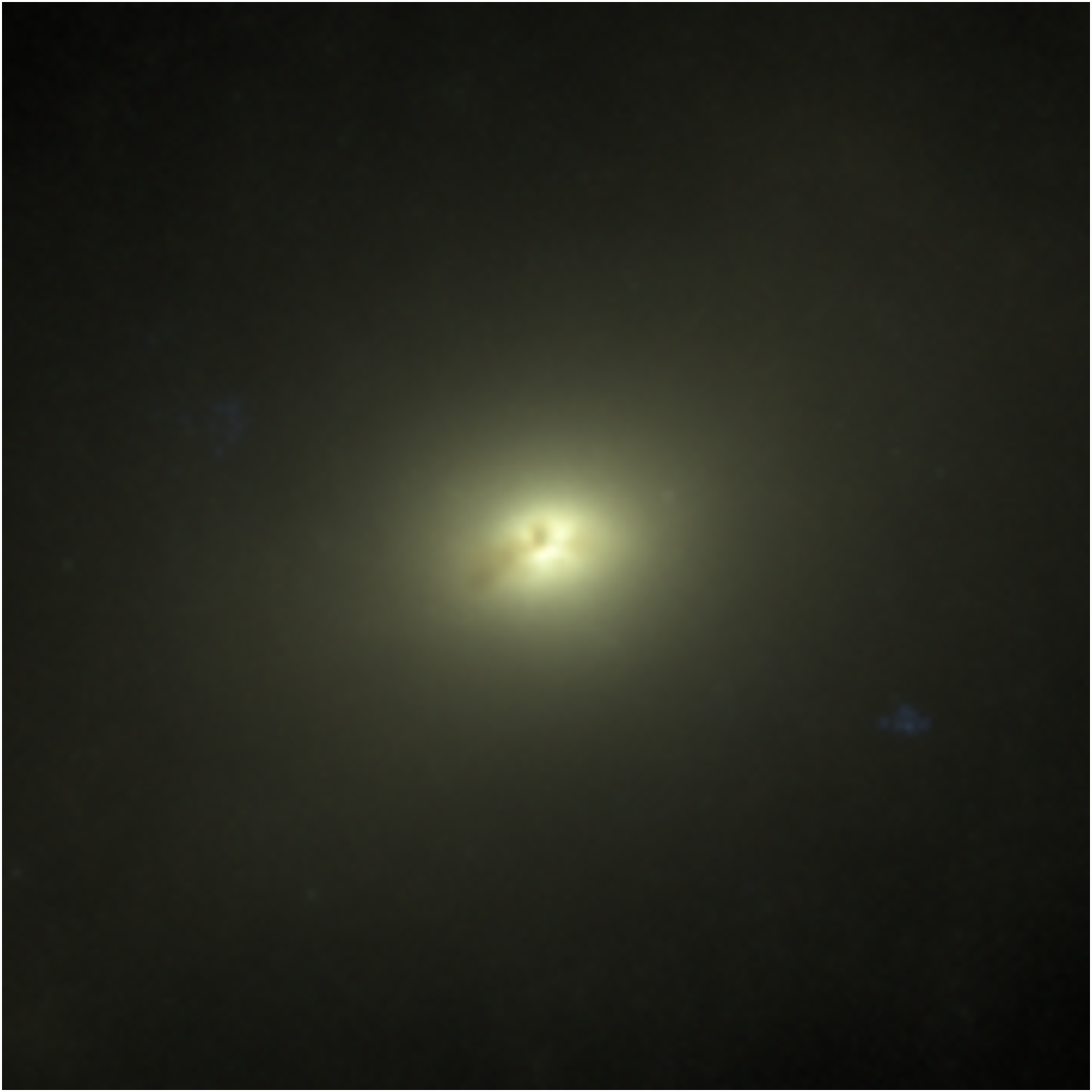}\\
\includegraphics[width=0.23\textwidth]{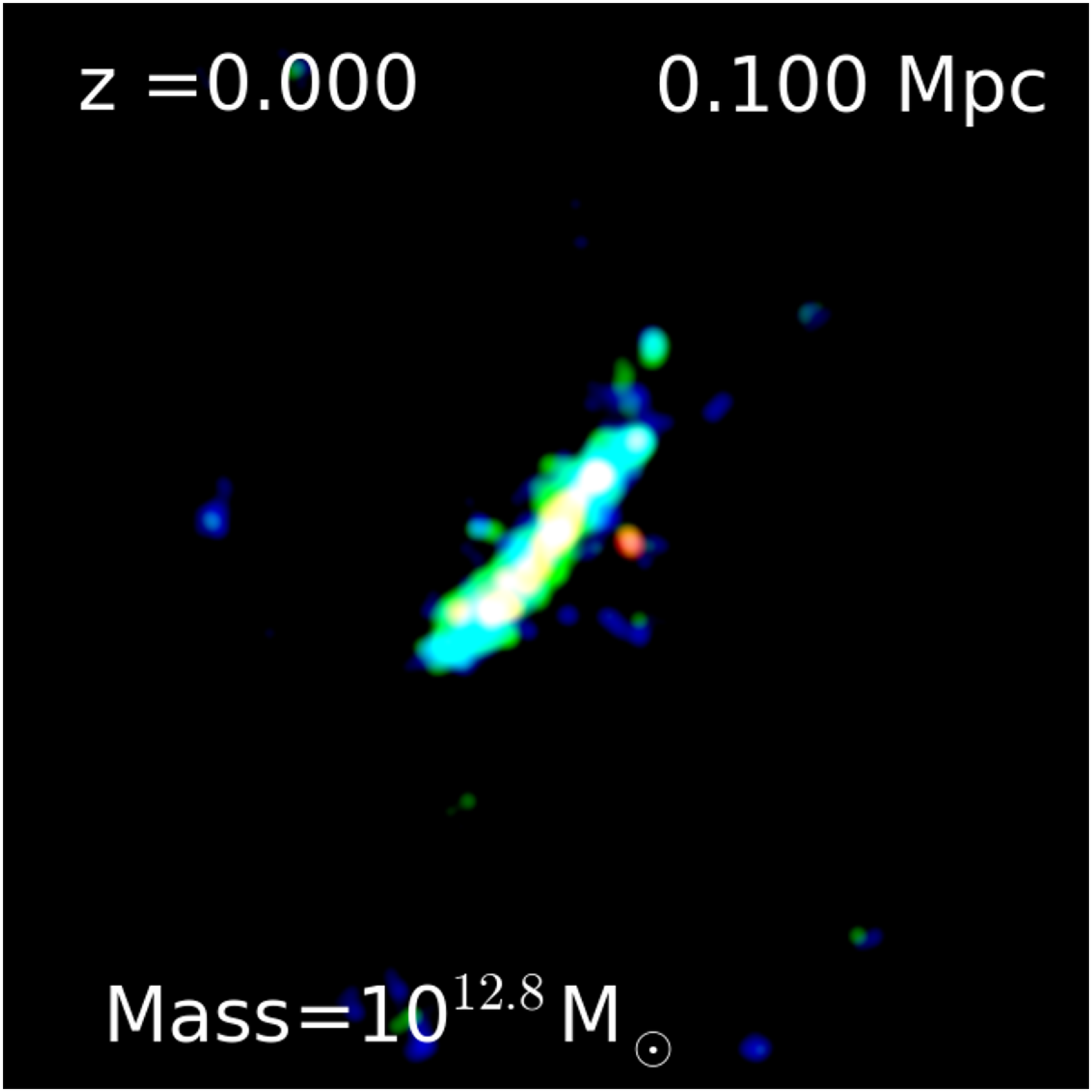}
\includegraphics[width=0.23\textwidth]{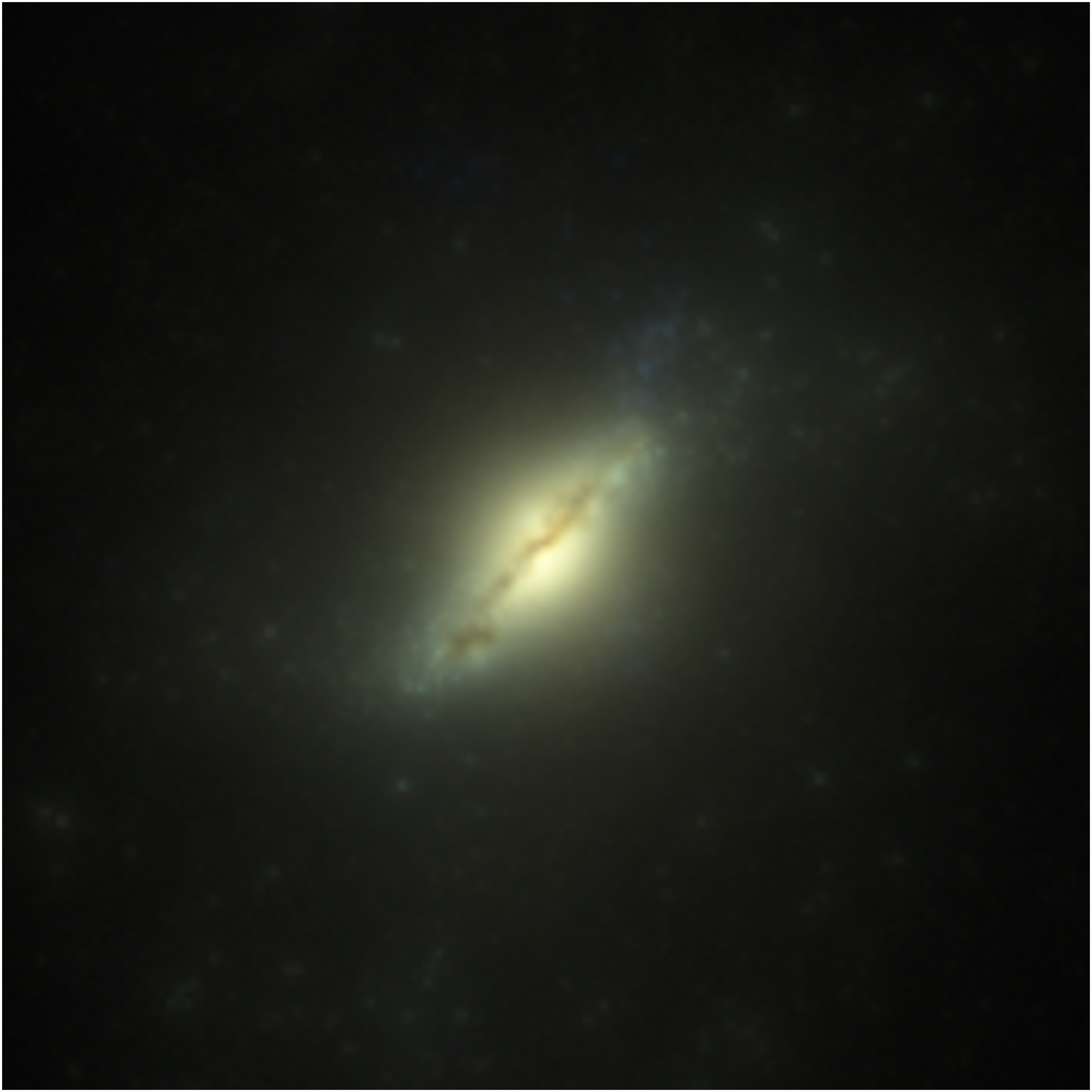}\\
\includegraphics[width=0.23\textwidth]{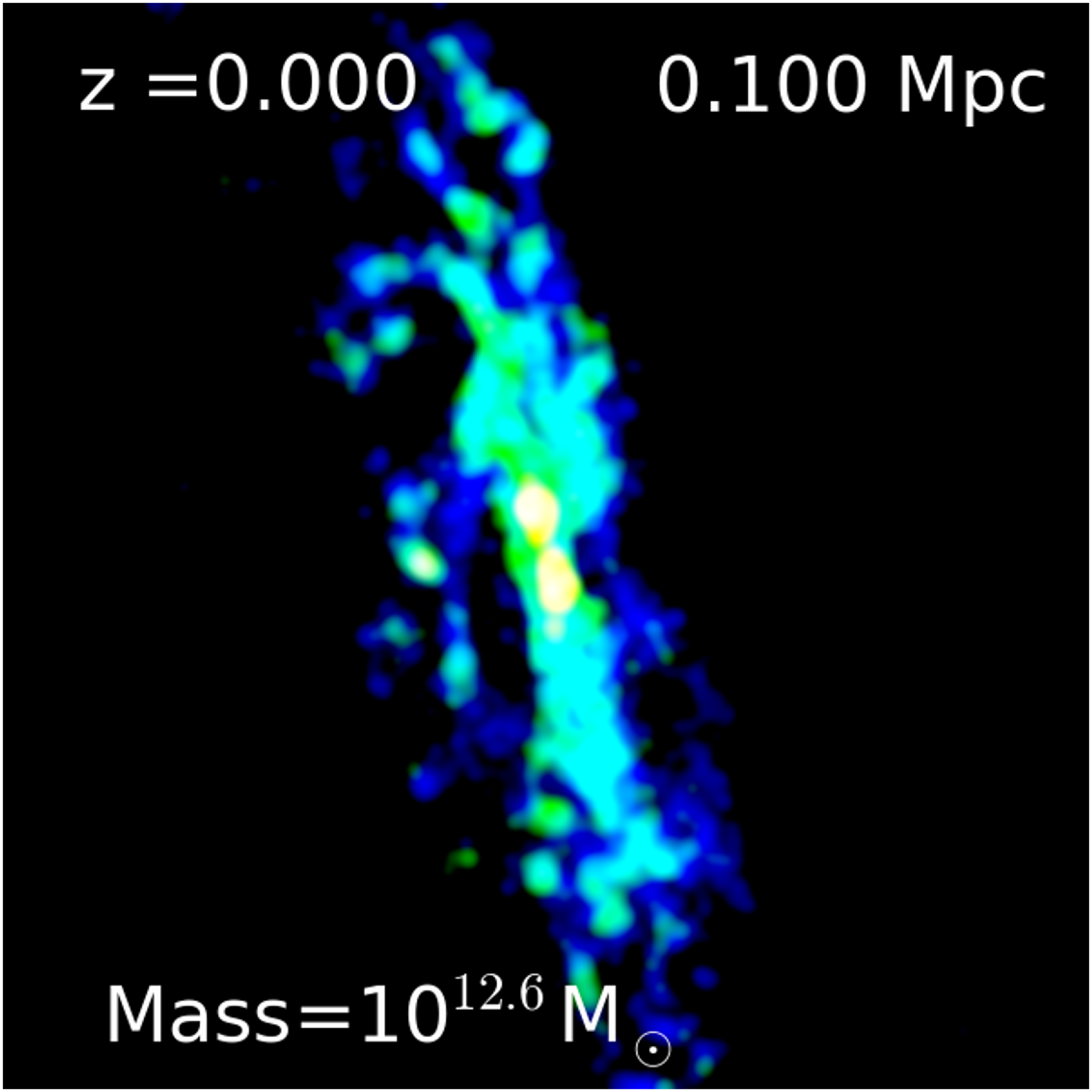}
\includegraphics[width=0.23\textwidth]{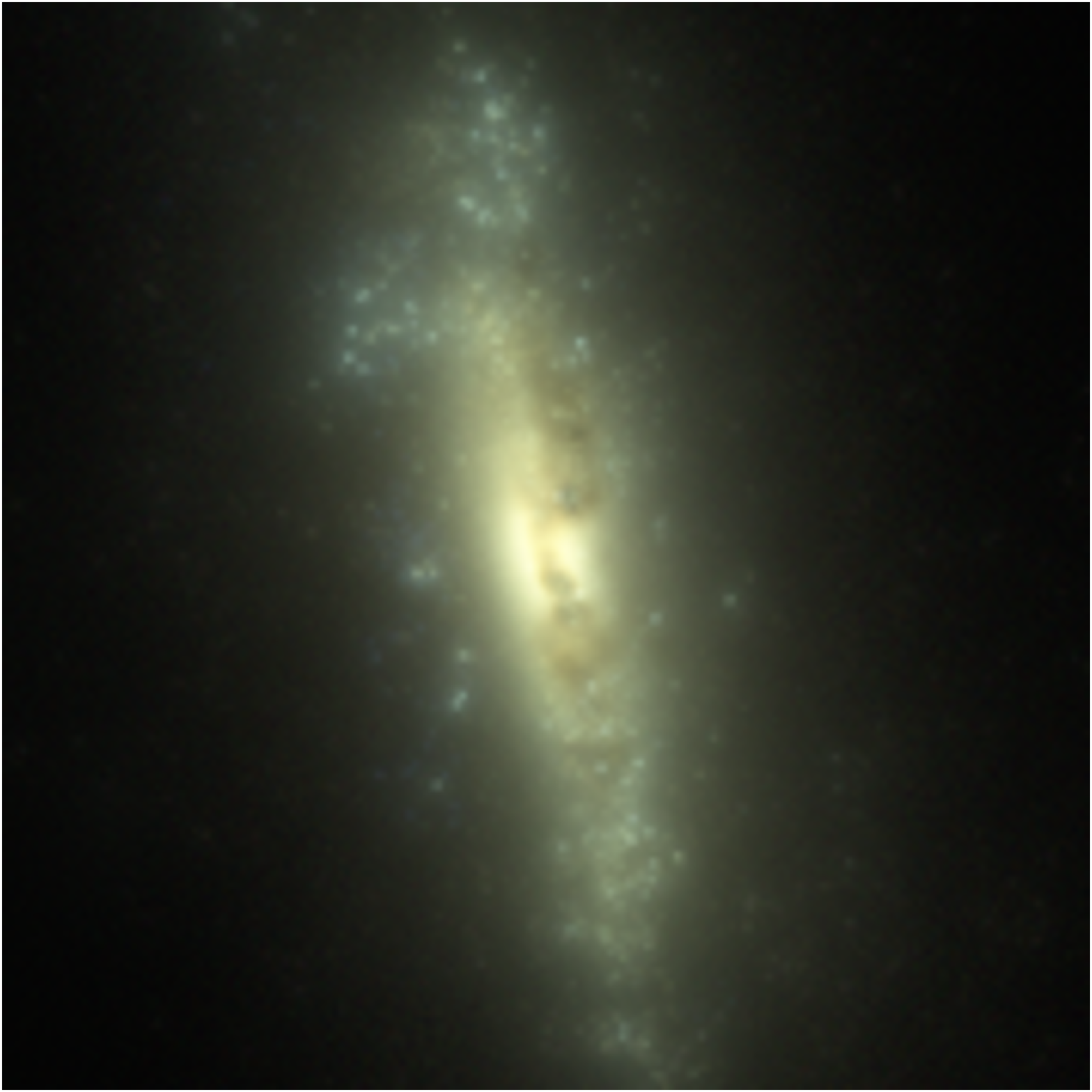}\\
\includegraphics[width=0.23\textwidth]{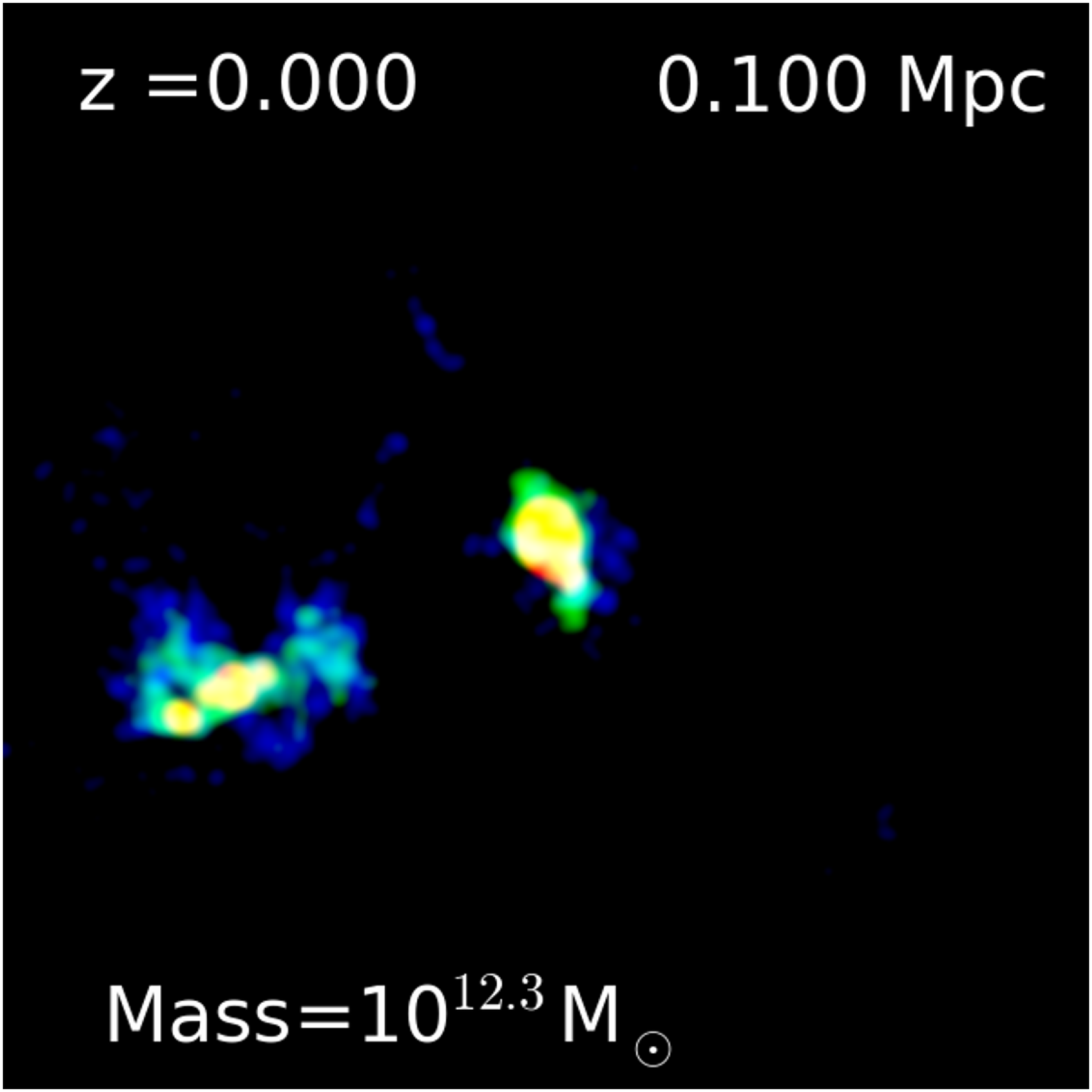}
\includegraphics[width=0.23\textwidth]{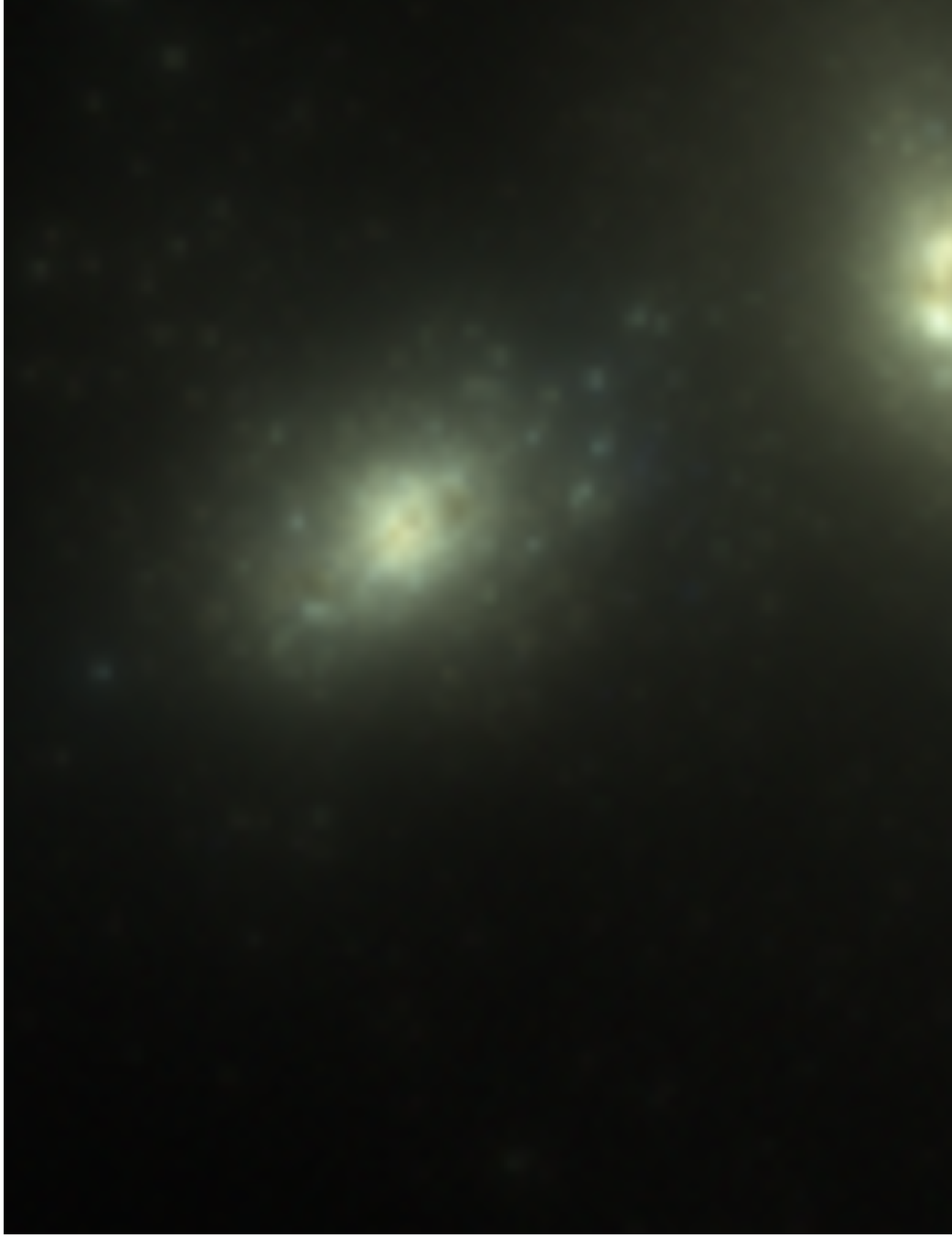}\\
\includegraphics[width=0.23\textwidth]{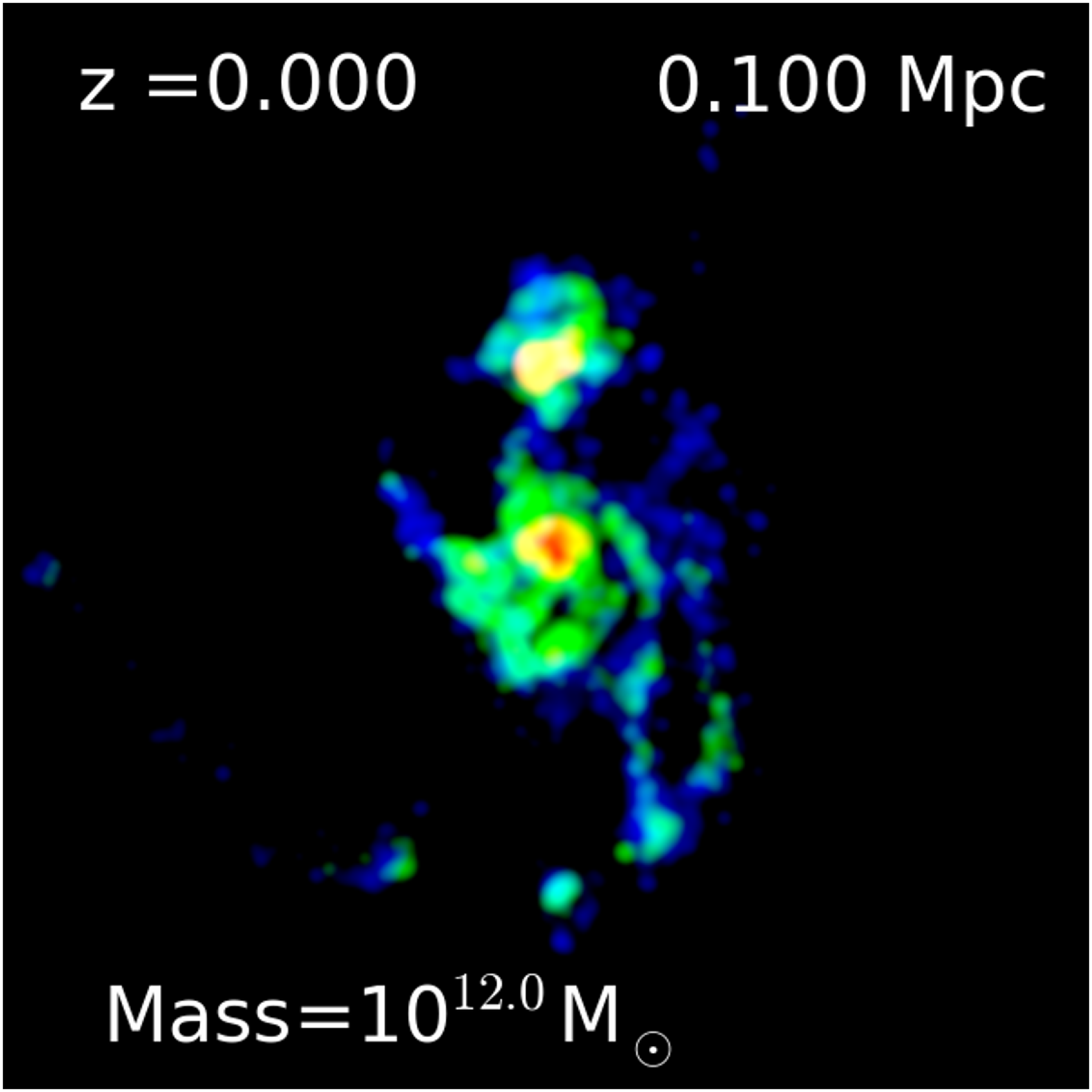}
\includegraphics[width=0.23\textwidth]{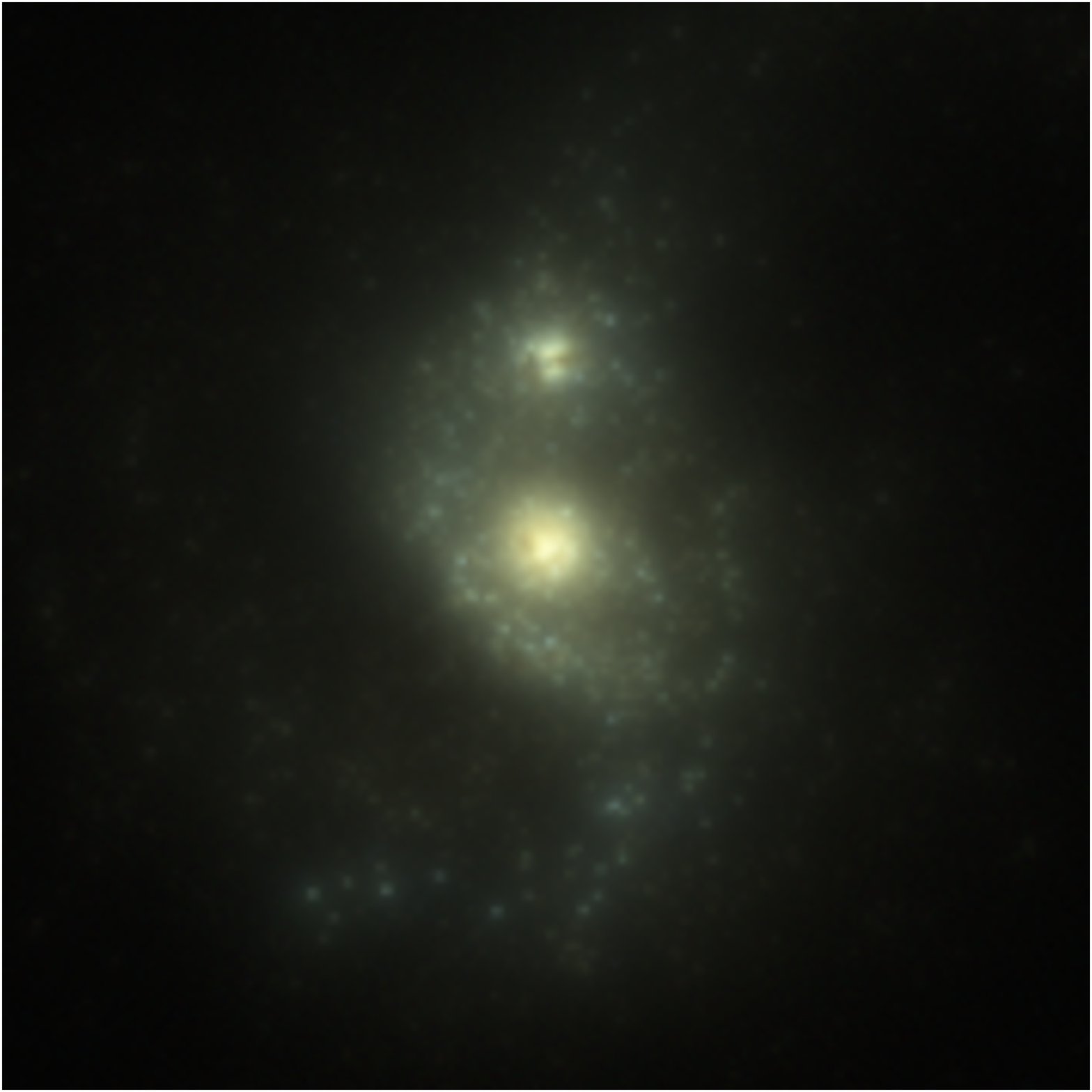}
\caption{Visualisation of $5$ galaxies at $z=0$ in Ref-L025N0376, 
which have been randomly chosen (and oriented) from different ranges
of subhalo masses (indicated at the bottom of the left panels). Maps are coloured by 
 H$_2$ column density (calculated using the GK11 prescription; left
 panels), and the right panels show  the stellar light based on
 monochromatic $u$, $g$ and $r$-band SDSS filter means and accounting
 for dust extinction (see text for details). 
Colours in the H$_2$ column density maps are as in the colour bars at the top of the top, where 
column densities are in units of $\rm cm^{-2}$.
Particles are smoothed by $1$ckpc 
in the $N_{\rm H_2}$ maps.  Map sizes are $0.1\times0.1$~pMpc$^2$.}
\label{Images}
\end{center}
\end{figure}

In order to visualise how the H$_2$ column density is distributed
with respect to the stars, we show in Fig.~\ref{Images} five randomly
chosen galaxies embedded in halos of different masses. The optical
images were created using three monochromatic radiative transfer
simulations with the code {\tt skirt} \citep{Baes11} at the effective
wavelengths of the Sloan Digital Sky Survey (SDSS) $u$, $g$ and $r$
filters. Dust extinction was implemented using the metal distribution
directly from the simulations, and assuming that $30$\% of the metal
mass is locked up in dust grains. Only material within a spherical
aperture of radius of $30$~pkpc was included in the radiative transfer
calculation. The method will be presented in detail in Trayford et
al. (in prep.).

The variety of morphologies in the EAGLE simulations is well captured
in Fig.~\ref{Images} (see also Fig.~$2$ in S15), where we see early-type galaxies (top row),
S0-like galaxies (second row) and galaxy mergers (fourth row). In the
case of the early-type galaxy example in the top row of
Fig.~\ref{Images}, one can see that the H$_2$ is relatively
concentrated compared to the stars, and is also connected to a dust
lane.  This is in qualitative agreement with spatially resolved
observations of early-type galaxies \citep{Young11}, in which H$_2$ is
typically concentrated in the centre and in a relatively thin disk.

\subsubsection{The prescription of \citet{Krumholz13} applied to EAGLE}

\citet{Krumholz13} developed a theoretical model for the transition
from HI to H$_{\rm 2}$, that depends on the total column density of
neutral hydrogen, the gas metallicity and the interstellar radiation
field. The fraction $f_{\rm H_2}$ is determined by the balance between
photodissociation of H$_2$ molecules by the interstellar radiation
field and the formation of molecules on the surfaces of dust grains.

A key property in the \citet{Krumholz13} model is the density of the
cold neutral medium (CNM).  At densities $n_{\rm H}\gtrsim 0.5\,\rm
cm^{-3}$, the transition from HI to H$_2$ is mainly determined by the
minimum density that the CNM must have to ensure pressure balance with
the warm neutral medium (WNM, which is HI dominated). The assumption
is that the CNM is supported by turbulence, while the WNM is thermally
supported (see also \citealt{Wolfire03}). At $n_{\rm H}\lesssim
0.5\,\rm cm^{-3}$ the transition from HI to H$_2$ is mainly determined
by the hydrostatic pressure, which has three components: the
self-gravity of the WNM ($\propto \Sigma^2_{\rm HI}$), the gravity
between the CNM and WNM ($\propto \Sigma_{\rm HI}\Sigma_{\rm H_2}$),
and the gravity between the WNM and the stellar plus dark matter
component ($\propto \Sigma_{\rm HI}\Sigma_{\rm sd}$, where
$\Sigma_{\rm sd}$ is the surface density of stars plus dark
matter). Note that the exact value of $n_{\rm H}$ at which the
transition between these two regimes takes place is a strong function
of gas metallicity. The equations used to implement this prescription
into EAGLE are given in Appendix~\ref{H2Prescriptions}.  We will refer
to the \citet{Krumholz13} prescription as `K13'.

\subsubsection{Resolution limit on the H$_2$ galaxy masses}

We define the resolution limit of the intermediate-resolution
simulations in terms of the subhalo masses rather than H$_2$ masses
and present the resolution analysis in Appendix~\ref{ConvTests}. We
find that halos with $M_{\rm Tot}>10^{10}\,\rm M_{\odot}$, where
$M_{\rm Tot}$ is the dark plus baryonic subhalo mass, are converged in
their H$_2$ abundances in the intermediate-resolution simulations.

\subsubsection{Differences between the GK11 and K13 prescriptions}\label{DiffPrescriptions}

\begin{figure}
\begin{center}
\includegraphics[width=0.49\textwidth]{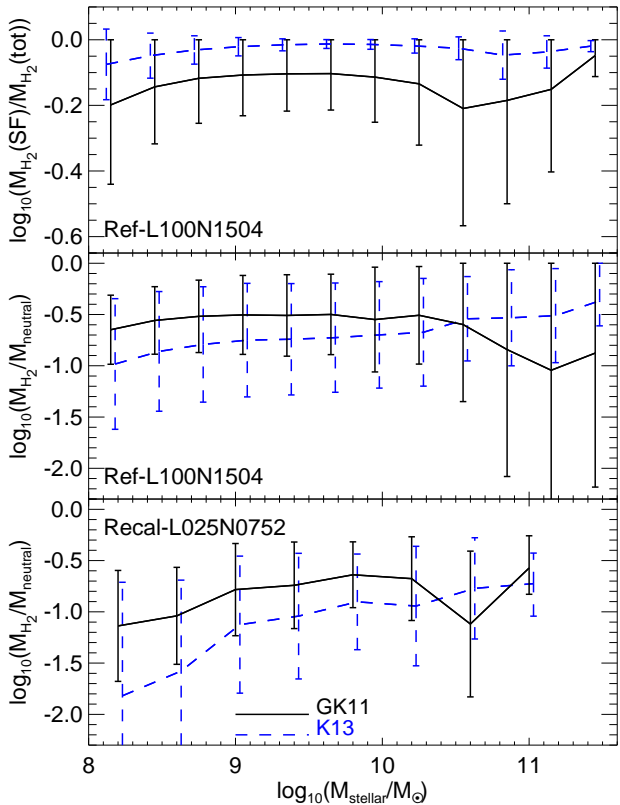}
\caption{{\it Top panel:} The ratio of the H$_2$ mass that 
is locked in star-forming particles to total H$_2$ mass, as a function of stellar mass at $z=0$ in
the simulation Ref-L100N1504 after applying the GK11 (solid line) and
K13 (dashed line) prescriptions. 
Lines and error bars indicate the median and $16^{\rm th}$ and $84^{\rm th}$ percentiles.
{\it Middle panel}: The ratio of the H$_2$ mass to total neutral hydrogen mass, as a function of stellar mass at $z=0$ in
the simulation Ref-L100N1504 after applying the GK11 (solid line) and
K13 (dashed line) prescriptions. Lines and error bars are as in the top panel.
{\it Bottom panel:} As in the middle panel but for the simulation Recal-L025N0752.}
\label{H2FracVsMs}
\end{center}
\end{figure}

The lower two panels of Fig.~\ref{H2FracVsMs} show the predicted
ratios of the total H$_2$ mass to total neutral hydrogen gas mass
(atomic plus molecular) as a function of stellar mass when the GK11
(top panel) and K13 (bottom panel) prescriptions are applied. There
is a weak positive correlation between the H$_2$ to total neutral
hydrogen mass ratio and stellar mass in both the Ref-L100N1504 and
Recal-L025N752 simulations when the K13 prescription applied, but such
a trend is not seen in the case of the GK11 prescription. The width
of the distributions (as shown by the $16^{\rm th}$ and $84^{\rm th}$
percentiles) seen in both the Ref-L100N1504 and Recal-L025N752
simulations when the GK11 prescription is applied is much larger at
$M_{\rm stellar}>10^{10}\,\rm M_{\odot}$ than when the K13
prescription is applied. Most of the dispersion seen in
Fig.~\ref{H2FracVsMs} is due to gas metallicity. For example, for the
simulation Ref-L100N1504 with the GK11 prescription applied and at
$M_{\rm stellar} \approx 10^{8}\,\rm M_{\odot}$, the metallicity
increases from $0.1\,\rm  Z_{\odot}$ at $M_{\rm H_2}/M_{\rm neutral}\approx
0.05$ to $1\,\rm Z_{\odot}$ at $M_{\rm H_2}/M_{\rm neutral}\approx 0.8$
(see also Fig.~\ref{CompBR06} for the effect of metallicity on $f_{\rm
  H_2}$). {Observations favour a positive correlation between 
$M_{\rm H_2}/M_{\rm neutral}$ and stellar mass \citep{Leroy08}, similar to the 
predictions of the Recal-L025N752 simulation. The very weak dependency predicted by the Ref-L100N1504 simulation 
contradicts observational results. We find that this is due to the gas metallicities in the Ref-L100N1504 simulation 
being higher than in the Recal-L025N752 simulation and observations \citep{Schaye14}. We discuss this 
in $\S$~\ref{SecHighMetallicity}.}

The top panel of Fig.~\ref{H2FracVsMs} shows the ratio between the
H$_2$ that is locked up in star-forming particles to the total H$_2$,
which also includes the contribution from non-star-forming particles.
The GK11 prescription applied to {EAGLE} results in a relative
contribution to the total H$_2$ mass from the H$_2$ locked up in
non-star-forming particles that is on average $20-40$ per cent and
does not show a strong dependence on stellar mass. However, when the
K13 prescription is applied, this contribution decreases to $5-20$ per
cent (see the discussion in Appendix~\ref{H2Prescriptions} for the
contribution to the total H$_2$ mass of gas particles that are on the
equation of state). {This difference is significant, given that the errors on the median assessed via boostrapping  
are very small ($\lesssim 0.05$~dex). The source of the difference between the GK11 and K13 prescriptions, is that the K13 prescription 
leads to $f_{\rm H_2}\equiv 0$ at gas surface densities $\Sigma_{\rm H}\lesssim 1\,\rm M_{\odot}\,pc^{-2}$ in 
ISM conditions that are MW-like, while leading to $f_{\rm H_2}\equiv 0$ at higher $\Sigma_{\rm H}$ when the ISRF 
increases. On the contrary, 
the GK11 prescription leads to small, but nonzero H2 fractions at these
low gas surface densities.}

It has been observed in nearby galaxies that
there is a component of the molecular gas that is not associated with
star formation and that is characterised by a high velocity dispersion
consistent with the HI disk (\citealt{Pety13};
\citealt{Caldu-Primo13}). This component contributes $\approx
30-50$\% of the total H$_2$ mass inferred for those galaxies.  This
observed diffuse component could be associated with the H$_2$ locked
up in non-star-forming particles in EAGLE, as their contribution to
the total H$_2$ in galaxies (at least when the GK11 prescription is
applied) is of a similar magnitude as the observed contribution from
the diffuse H$_2$ to the total inferred H$_2$ mass.  This suggests
that EAGLE can shed some light onto the nature of such a component of
the ISM of nearby galaxies.  We investigate this in more detail in a
separate paper (Lagos et al. in preparation).  Additional differences
between the GK11 and K13 prescriptions, as well as comparisons with
previous work can be found in Appendix~\ref{H2Prescriptions}.

Fig.~\ref{H2FracVsMs} shows that the simulations Ref-L100N1504 and
Recal-L025N0752 display significant differences both for the GK11 and
K13 prescriptions, particularly in galaxies with stellar masses
$M_{\rm stellar}<10^{10}\,\rm M_{\odot}$. This is mainly driven by the
gas metallicity in the Recal-L025N0752 simulation being lower on
average than in the Ref-L100N1504 simulation (see Fig.~$13$ in S15 for
the mass-metallicity relations predicted by the simulations  compared
here). For example, at a stellar mass of $10^9\,\rm M_{\odot}$,
Recal-L025N0752 simulation predicts a gas metallicity $0.4$~dex lower
than the metallicity of galaxies of the same mass in the Ref-L100N1504
simulation, on average. However, for $M_{\rm stellar}>10^{10}\,\rm
M_{\odot}$ differences are small. In $\S$~\ref{SecHighMetallicity} we
discuss the effect the gas metallicities have on the H$_2$ masses of
galaxies in {EAGLE}.

\section{Abundance and scaling relations of H$_2$ in the local Universe}\label{localU}

In this Section we compare the results of EAGLE with observations of
H$_2$ in the local Universe. We focus on the H$_2$ mass function in
$\S$~\ref{secH2MF} and on H$_2$ scaling relations in
$\S$~\ref{SecScalingrelations}.

Note that in previous papers using EAGLE, 
simulation results and observations are compared at $z=0.1$ in order to best correspond 
to the median redshift of optical surveys. However, 
here we compare the EAGLE predictions with observations at $z=0$ because 
observations of CO, which here we convert to H$_2$, in the local Universe 
 correspond to galaxies at $z<0.1$.

\subsection{The H$_2$ mass function}\label{secH2MF}

\begin{figure}
\begin{center}
\includegraphics[width=0.49\textwidth]{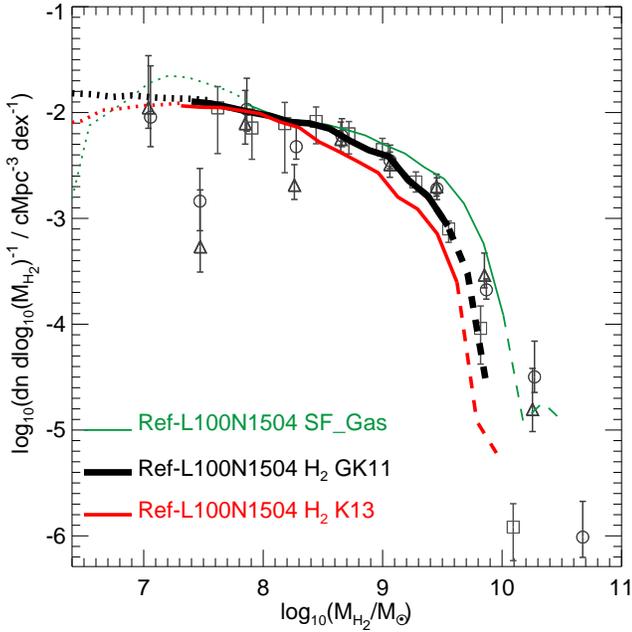}
\caption{The mass functions of H$_2$ (thick lines) and of the hydrogen
  component of the star-forming gas (thin line) in the simulation
  Ref-L100N1504 at $z=0$.  For H$_2$ we show the mass functions after
  applying the GK11 (black line) and K13 (red line) prescriptions.
  The part of the mass functions that is affected by low number
  statistics (less than $10$ objects per dex bin) is shown as dashed
  lines, while the part that is affected by resolution is shown as
  dotted lines (see Appendix~\ref{ConvTests} for details).
  Observational estimates correspond to the \citet{Keres03} $B$-band
  (triangles) and  $60\,\mu$m (circles) selected samples of galaxies,
  adopting the Milky-Way like $\rm H_2$-to-CO(1-0) conversion factor
  $\rm X=2$ as universal (see Eq.~\ref{XCO} for the definition of
  $X$).  We also show the observational estimates of
  \citet{Obreschkow09b} in square symbols, in which a
  luminosity-dependent $X$ was applied to the observations of
  \citet{Keres03}.  The star-forming gas mass function differs
  significantly from the H$_2$ mass functions obtained using either
  the GK11 or the K13 prescriptions at $M_{\rm H_2}>5\times 10^8\,\rm
  M_{\odot}$, and can therefore not be used as a reliable tracer of
  the H$_2$ mass function. The H$_2$ mass function predicted by EAGLE
  is in good agreement with the observations.}
\label{H2MFEagleSF}
\end{center}
\end{figure}

The curves in Fig.~\ref{H2MFEagleSF} show the $z=0$ H$_2$ mass
function calculated using either the GK11 or K13 prescriptions, and
the mass function of the hydrogen component of the star-forming gas,
for the simulation Ref-L100N1504.  These are compared to observations
at $z\approx 0$, which are shown as symbols with $1\sigma$ errorbars.
We can see that the star-forming gas mass is a poor proxy for the
H$_2$ mass. The median ratio between H$_2$ mass and the hydrogen
component of star-forming gas is $\approx 0.7$, but fluctuations
around this value are large, with the $16^{\rm th}$ and $84^{\rm th}$
percentiles being $0.33$ and $1.25$, respectively. We qualitatively discussed the 
large systematics introduced by the assumption that the star-forming gas 
is a good proxy for H$_2$ mass in $\S$~$1$ in the context of previous work 
(e.g. \citealt{Genel14}), and here we have quantified them.

As we discussed in $\S$~$1$, H$_2$ is not directly observed, but
constraints on the H$_2$ mass function at $z=0$ have been provided
by surveys of CO(1-0) in local galaxies. The CO(1-0)-to-$\rm H_2$
conversion factor is the largest uncertainty in the observations.  The
conversion factor, $X$, is defined as

\begin{equation}
\frac{N_{\rm H_2}}{{\rm cm^{-2}}}=X\times10^{-20}\, \left(\frac{I_{\rm CO(1-0)}}{\rm K\,km\,s^{-1}}\right),
\label{XCO}
\end{equation}

\noindent where $N_{\rm H_2}$ is the column density of H$_2$ and
$I_{\rm CO}$ is the integrated CO$(1-0)$ line intensity per unit
surface area. Systematic variations of $X$ are both theoretically
predicted and observationally inferred. For example, variations are
expected as a function of gas metallicity and the interstellar
radiation field, such that $X$ increases in low metallicity and/or
intense interstellar radiation field environments
(e.g. \citealt{Bell07}; \citealt{Meijerink07}; \citealt{Bayet09}; \citealt{Shetty11}; \citealt{Narayanan12}; \citealt{Feldmann12}; 
\citealt{Bibas15}). In
addition to this, different values of $X$ have been measured using
different methods. For example, in the Milky Way, variations in the
method for measuring $X$ lead to values that differ by a factor of
$\approx 3$ (see \citealt{Bolatto13} for a recent review).  This
systematic error clearly dominates the uncertainty in the H$_2$ mass
over the uncertainties in the measurement of the CO emission line.  We
keep these systematics in mind when comparing the EAGLE predictions
with the observations.

For the entire range of observed H$_2$ masses,
$10^7\rm\,M_{\odot}\lesssim {\it M}_{\rm H_2}\lesssim
10^{10}\rm\,M_{\odot}$, the H$_2$ mass function of {EAGLE}, calculated
either using the GK11 or the K13 prescriptions, is in good agreement
with the observations. We see the agreement is the best with the H$_2$
mass function derived by \citet{Obreschkow09b} from the CO(1-0)
surveys of \citet{Keres03} using a luminosity-dependent $\rm
H_2$-to-CO(1-0) conversion factor. Compared to the H$_2$ mass
functions derived using a fixed $X=2$, we find some tension at $M_{\rm
  H_2}\gtrsim 5\times 10^{9}\,\rm M_{\odot}$. However, the differences
are well within the systematic uncertainties of the observational
measurements, which are well illustrated by the difference between the
observational inferences of \citet{Obreschkow09b} and \citet{Keres03}.

\subsubsection{Variations due to gas metallicity}\label{SecHighMetallicity}
\begin{figure}
\begin{center}
\includegraphics[width=0.42\textwidth]{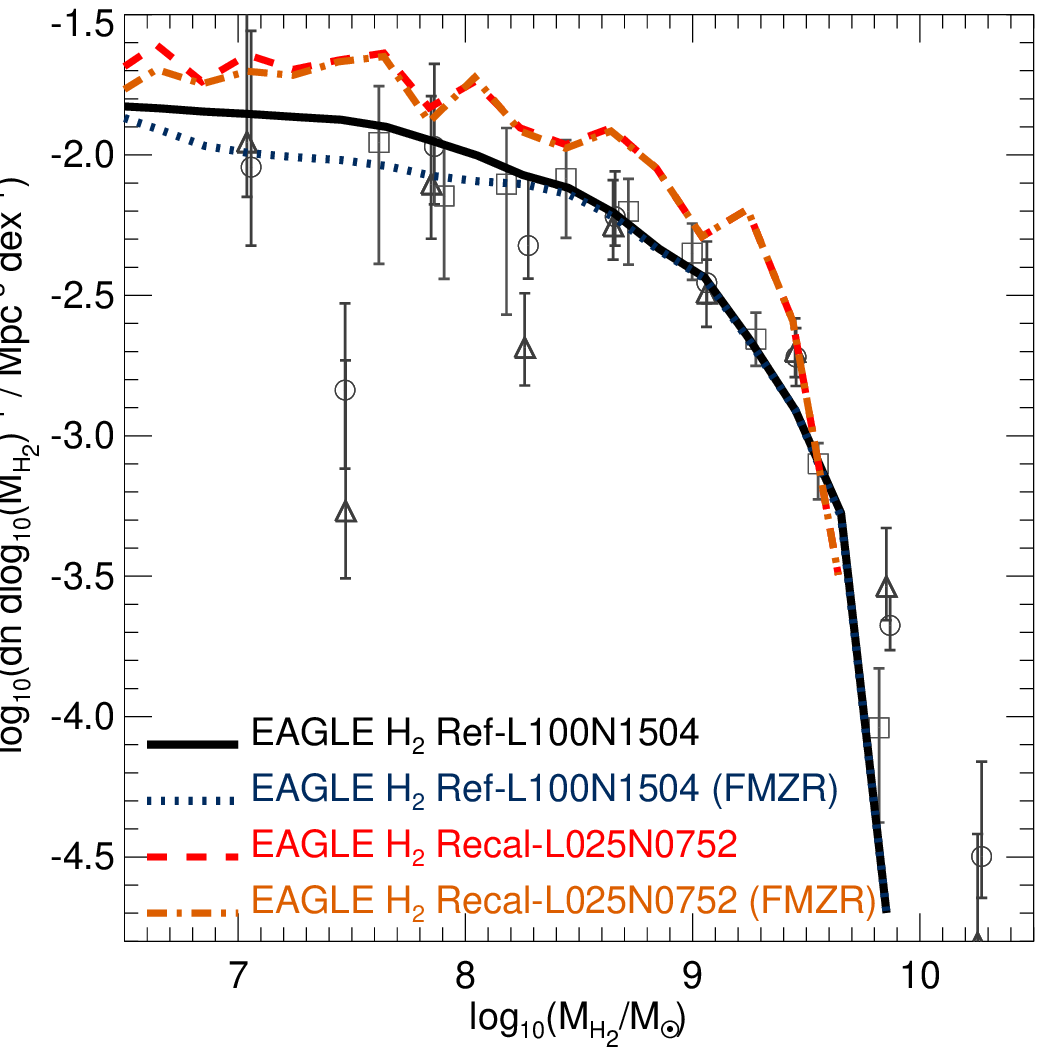}
\includegraphics[width=0.42\textwidth]{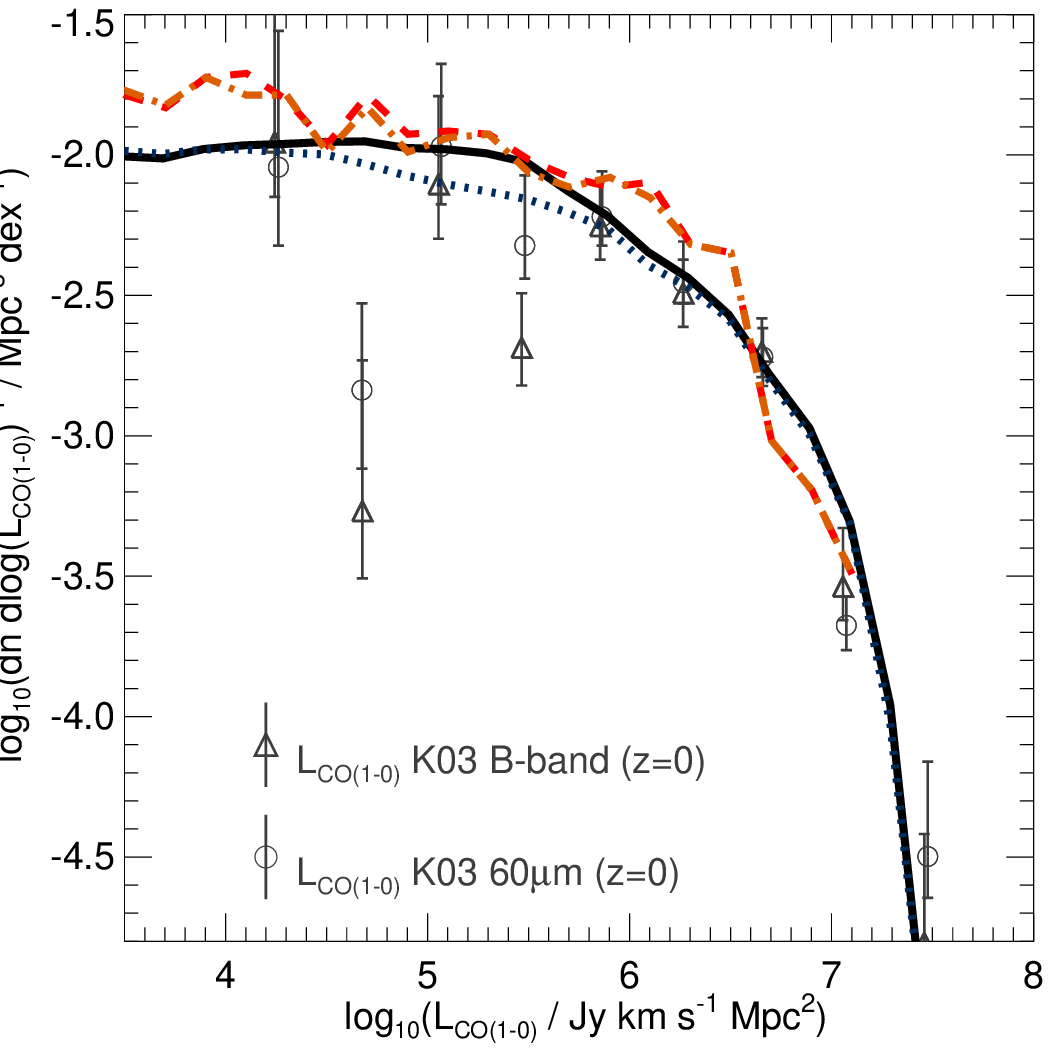}
\caption{{\it Top panel:} The H$_2$ mass function at $z=0$ for the
  simulations Ref-L100N1504 and Recal-L025N0752 using the GK11
  prescription, before and after the gas metallicities of galaxies in
  EAGLE are rescaled to satisfy the observed MZR of \citet{Zahid14}.
  We refer to the simulations after rescaling as `FMZR'. As shown in
  Fig.~\ref{H2MFEagleSF}, the effects of resolution in the
  Ref-L100N1504 start to appear at H$_2$ masses $\lesssim 5\times
  10^{7}\,\rm M_{\odot}$. The effect of sampling, on the other hand,
  appears at number densities of $-2.5$ in the Recal-L025N0752 and
  $-4.3$ in the Ref-L100N1504 (in the units of the
  $y$-axis). Observations are as in Fig.~\ref{H2MFEagle}. The
  rescaling of gas metallicities has a significant effect at $M_{\rm
    H_2}<10^9\,\rm M_{\rm H_2}$. {\it Bottom panel:} CO(1-0)
  luminosity function for the same models of the top panel, and
  applying the conversion factor of Eq.~\ref{XCOZdep} to convert the
  predicted H$_2$ masses into CO(1-0) luminosities. Observations
  correspond to the $B$-band and $60$ microns selected samples of
  \citet{Keres03}, as labelled.}
\label{FMZR}
\end{center}
\end{figure}

S15 compared the mass-metallicity relation (MZR) predicted by {EAGLE}
with observations and showed that the intermediate-resolution
simulation Ref-L100N1504 exhibits ISM metallicities that are higher
than the observations by up to $0.3$~dex at a stellar mass of
$\lesssim 10^9\,\rm M_{\odot}$. The higher resolution simulation,
Recal-L025N0752, on the other hand, displays ISM metallicities that
are lower than the lower-resolution simulations at fixed stellar mass,
and are in good agreement with the observations.

The gas metallicity is important for two reasons. First, the GK11 and
K13 prescriptions have an explicit dependence on gas metallicity, as
dust is the catalyst for H$_2$ formation, and the abundance of dust is
assumed to be proportional to the gas metallicity.  Second, the
formation of CO, and therefore the reliability of CO as a tracer of
H$_2$, depends on the gas metallicity
(e.g. \citealt{Bolatto13}). Observations suggest a strong dependence
of $X$ on gas metallicity. For example, \citet{Boselli02} reported the
following relation between $X$ and $Z$,

\begin{equation}
{\rm log_{\rm 10}}({\it X})=0.5_{-0.2}^{+0.2}-1.02_{-0.05}^{+0.05}{\rm log_{\rm 10}}(\rm Z/Z_{\odot}).
\label{XCOZdep}
\end{equation}

In order to quantify the effect that the predicted higher
metallicities have on the H$_2$ content of galaxies in EAGLE, we first
repeat the calculation presented in $\S$~\ref{H2FractionSec} but
forcing galaxies to follow the observed MZR. In practice, we compare
the neutral gas mass-weighted metallicity of individual EAGLE galaxies
with the observed median gas metallicity of \citet{Zahid14} at the
same stellar mass and estimate how much higher the predicted
metallicity is compared to the median in the observations. If the
difference is greater than $50$\% of the observed median, we rescale the
metallicities of all gas particles by the factor needed to bring the
galaxy metallicity into agreement with the observations, and
recalculate the H$_2$ fraction and the $30$~pkpc aperture H$_2$ mass.
The $50$\% factor is included to allow for a dispersion around the
median of the same magnitude as the $1\sigma$ distributions reported
in the observations. We convert the metallicities of Zahid et al. to
our adopted metal abundances: $12+\rm log_{\rm 10}(O/H)_{\odot}=8.69$
and $Z_{\odot}=0.0127$. For stellar masses $M_{\rm stellar}<10^9\,\rm
M_{\odot}$ we extrapolate the MZR of Zahid et al. to get an `observed'
metallicity at that stellar mass.  The H$_2$ mass functions using the
GK11 prescription before and after the rescaling of gas metallicities
are shown for the simulations Ref-L100N1504 and Recal-L025N0752 in the
top panel of Fig.~\ref{FMZR}.

The differences in the H$_2$ mass functions before and after the
metallicity rescaling are mild. The metallicity rescaling has no
visible effect on the H$_2$ mass function of Recal-L025N0752, while
for Ref-L100N1504, the number density of galaxies with $M_{\rm
  H_2}\lesssim 5\times 10^8\,\rm M_{\odot}$ decreases slightly after
rescaling.
When integrating over all H$_2$ masses, the average density of H$_2$
in Ref-L100N1504 decreases by $5$\% after the gas metallicity
rescaling. This difference is well within the error bars of the
observations (see $\S$~\ref{OmegaH2Sec}). {However, larger differences are obtained 
for the relation between the $M_{\rm H_2}/M_{\rm neutral}$ ratio and stellar mass (Fig.~\ref{H2FracVsMs}), in which 
the gas metallicity rescaling leads to $M_{\rm H_2}/M_{\rm neutral}$ ratios lower by up to $\approx 0.5$~dex 
at $10^8\,\rm M_{\odot}\lesssim M_{\rm stellar} \lesssim 5\times 10^8\,\rm M_{\odot}$, for both prescriptions. 
A steeper relation is then obtained, which removes the tension with the observations described in $\S$~\ref{DiffPrescriptions}.} 

As a second experiment, we use Eq.~\ref{XCOZdep} to convert the
predicted H$_2$ masses in EAGLE to CO(1-0) luminosities and construct
the CO(1-0) luminosity function, in order to show the effect of a
metallicity-dependent $X$. This is shown in the bottom panel of
Fig.~\ref{FMZR}, along with the observations of CO(1-0) of
\citet{Keres03}.  The level of agreement between the Ref-L100N1504 and
Recal-L025N0752 simulations and the observations is similarly good to
that for the H$_2$ mass functions, and the effect of rescaling the gas
metallicity is also similar. This shows that the main effect of gas
metallicities is in the H$_2$ formation rather than its later
conversion to CO(1-0). {Note that the CO(1-0) luminosity functions predicted by the 
simulations Ref-L100N1504 and Recal-L025N0752 are closer to each other than the predicted H$_2$ mass functions. 
This is because the simulation Recal-L025N0752 has, on average, lower gas metallicities than the 
 Ref-L100N1504 simulation, which in practice leads to higher $X$ values.} 

The effects of the higher gas metallicities on the H$_2$ masses of
galaxies is mild and therefore we continue to use the predicted
gas metallicities (rather than the rescaled ones) for the rest of the
paper.

\subsection{H$_2$ Scaling relations}\label{SecScalingrelations}

Measurements of the H$_2$ and HI mass content, as well as other galaxy
properties, are available for relatively large samples of local
galaxies (running into the hundreds), enabling the characterisation of
scaling relations between the cold gas and the stellar mass content
(e.g. \citealt{Bothwell09}; \citealt{Catinella10};
\citealt{Saintonge11}; \citealt{Boselli14b}).  Recently, H$_2$ and HI
have been studied in homogeneous samples of relatively massive
galaxies by \citet{Saintonge11} and \citet{Catinella10}, respectively,
with the aim of measuring the fundamental relations between the
stellar content of galaxies and their cold gas mass. The survey of
\citet{Saintonge11}, the CO Legacy Database for the GALEX Arecibo SDSS
Survey (COLD GASS) is the first stellar mass-limited CO survey. The
strategy used was to select all galaxies with $M_{\rm
  stellar}>10^{10}\,\rm M_{\odot}$ at $z<0.05$ from the SDSS and
follow-up a subsample of those at millimetre (mm) wavelengths to detect
CO(1-0). Saintonge et al. integrated sufficiently long to enable H$_2$
gas fractions, $M_{\rm H_2}/M_{\rm stellar}>0.015$, at stellar masses
$M_{\rm stellar}>10^{10.6}\,\rm M_{\odot}$, to be detected, or H$_2$
masses $M_{\rm H_2}>10^{8.8}\,\rm M_{\odot}$ in galaxies with stellar
masses, $10^{10}\,\rm M_{\odot}<M_{\rm stellar}<10^{10.6}\,\rm
M_{\odot}$.  The simple and well-defined selection function of COLD
GASS makes this survey an instructive testbed for { EAGLE}.

\begin{figure}
\begin{center}
\includegraphics[width=0.5\textwidth]{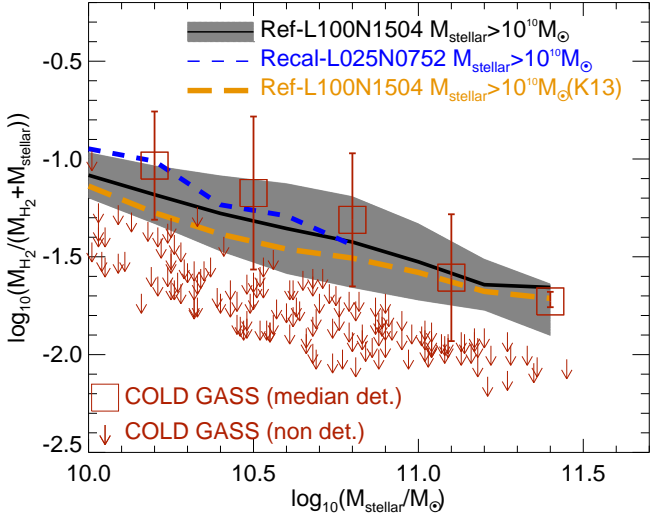}
\caption{Molecular hydrogen fraction, $M_{\rm H_2}/(M_{\rm H_2}+M_{\rm
    stellar})$, as a function of stellar mass for the simulations
  Ref-L100N1504 (solid line and filled region) and Recal-L025N0752
  (dashed line with error bars) for galaxies with $M_{\rm
    stellar}>10^{10}\,\rm M_{\odot}$ and with H$_2$ masses above the
  sensitivity limit of COLD GASS at $z=0$. Here we use the GK11
  prescription to calculate H$_2$ masses. We also show the simulation
  Ref-L100N1504 when the K13 prescription is applied (long-dashed
  line) for the same redshift and stellar mass selection.  Lines show
  the medians of the simulated galaxies for which the H$_2$ mass
  exceeds the COLD GASS detection limit, while the filled region shows
  the $16^{\rm th}$ and $84^{\rm th}$ percentiles for the
  Ref-L100N1504 simulation when the GK11 prescription is applied. The
  dispersions in the Ref-L100N1504 simulation with the K13
  prescription and in the Recal-L025N0752 simulation are of similar
  size.  Squares with error bars show the median and $16^{\rm th}$ and
  $84^{\rm th}$ percentiles of galaxies in the COLD GASS survey that
  have the CO(1-0) emission line detected, while arrows show upper
  limits for the non-detections \citep{Saintonge11}. The predictions
  from EAGLE are in very good agreement with the data.}
\label{COLDGASScomparison}
\end{center}
\end{figure}

Fig.~\ref{COLDGASScomparison} shows the H$_2$ fraction, $M_{\rm
  H_2}/(M_{\rm H_2}+M_{\rm stellar})$, as a function of stellar mass
for galaxies with $M_{\rm stellar}>10^{10}\,\rm M_{\odot}$ and 
H$_2$ gas fractions above the sensitivity limit defined by
\citet{Saintonge11} for COLD GASS at $z=0$ in {EAGLE}. Note that both
observations and simulation results account for molecular hydrogen
only (without Helium correction).
We show the median and $16^{\rm th}$ and $84^{\rm th}$ percentiles for galaxies in COLD GASS, for which the 
 CO(1-0) emission line is detected as squares with error bars, while we show the upper limits for the non-detections as arrows. 
Note that uncertainties in the measurement of the integrated flux in the CO(1-0) emission line are small (approximately $1$\%).
Here the systematic errors are dominant, and 
they are mainly related to how the CO flux is converted into the H$_2$ mass, as we discussed in $\S$~\ref{secH2MF}.

From Fig.~\ref{COLDGASScomparison} we see that {EAGLE} predicts that the
ratio $M_{\rm H_2}/(M_{\rm H_2}+M_{\rm stellar})$ decreases with
$M_{\rm stellar}$. The amplitude of this relation is slightly higher,
by $\lesssim 0.1$~dex, when the GK11 prescription is applied than when
the K13 prescription is applied.
The high-resolution simulation Recal-L025N0752 predicts a trend that
is similar to that in Ref-L100N1504, although with systematically
higher $M_{\rm H_2}/(M_{\rm H_2}+M_{\rm stellar})$ (except for the
largest galaxies, for which we have only a small number in the
Recal-L025N0752 simulation). This is due to the slightly higher SFRs
and H$_2$ masses predicted in the higher resolution simulation at a
fixed halo mass.  
{S15 showed that the Ref-L100N1504 simulation predicts SFRs that are lower 
than the observations by $\approx 0.1-0.15$~dex for galaxies with $M_{\rm stellar}>10^{10}\,M_{\odot}$, 
while the Recal-L025N0752 simulation predicts 
higher SFRs, in better agreement with the observations. Such differences 
also appear in Fig.~\ref{COLDGASScomparison}, where the median 
in the Ref-L100N1504 simulation is $\approx 0.1$~dex lower 
than the median H2 fraction in COLD GASS. This offset is smaller than
the scatter in the observations. Thus, the agreement seen here is fully
consistent with the analysis of the specific SFRs in S15.}
The differences between the Ref-L100N1504 and
Recal-L025N0752 simulations are also similar to those discussed for the
H$_2$-subhalo mass relation in Appendix~\ref{ConvTests}.
 
The predictions from the EAGLE simulations shown in
Fig.~\ref{COLDGASScomparison} agree very well with the observations.
Galaxies in COLD GASS for which the CO(1-0) emission line is detected
lie on the main sequence of star-forming galaxies (in the SFR-stellar
mass plane), which is also the case for galaxies in EAGLE that have
H$_2$ masses above the detection limit of COLD GASS. This is a
  significant outcome, since none of these observations were used to
  constrain the calibrated parameters governing the efficiency of star formation and feedback, nor 
 those governing the partitioning of gas into ionised, atomic and molecular components in {EAGLE}. This is the first time
  that a hydrodynamic simulation performs so well when confronted by 
  observational inferences of H$_2$.

\begin{figure}
\begin{center}
\includegraphics[width=0.5\textwidth]{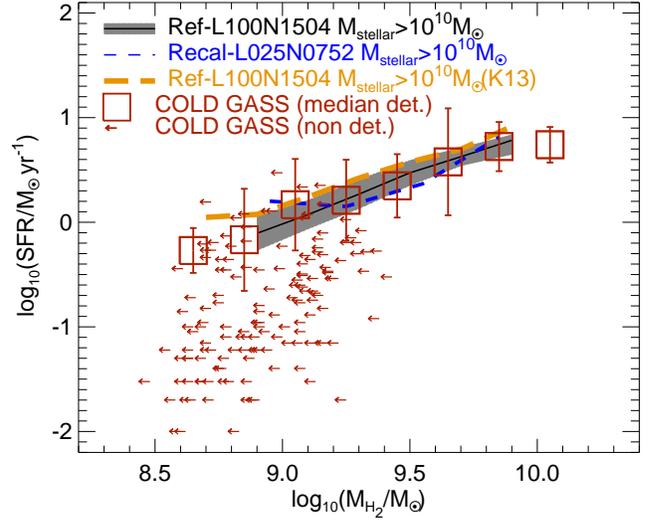}
\caption{The SFR as a function of the molecular hydrogen mass at $z=0$
  for galaxies with $M_{\rm stellar}>10^{10}\,\rm M_{\odot}$ and with
  H$_2$ gas fractions above the sensitivity limit of COLD GASS in the
  simulations Ref-L100N1504 (solid line and filled region) and
  Recal-L025N0752 (dashed line) with the GK11 prescription, and the
  simulation Ref-L100N1504 with the K13 prescription (long dashed
  line), where lines show the medians.  Observations are as in
  Fig.~\ref{COLDGASScomparison}. The simulations and observations are
  in good agreement.}
\label{COLDGASScomparison2}
\end{center}
\end{figure}

\begin{figure}
\begin{center}
\includegraphics[width=0.5\textwidth]{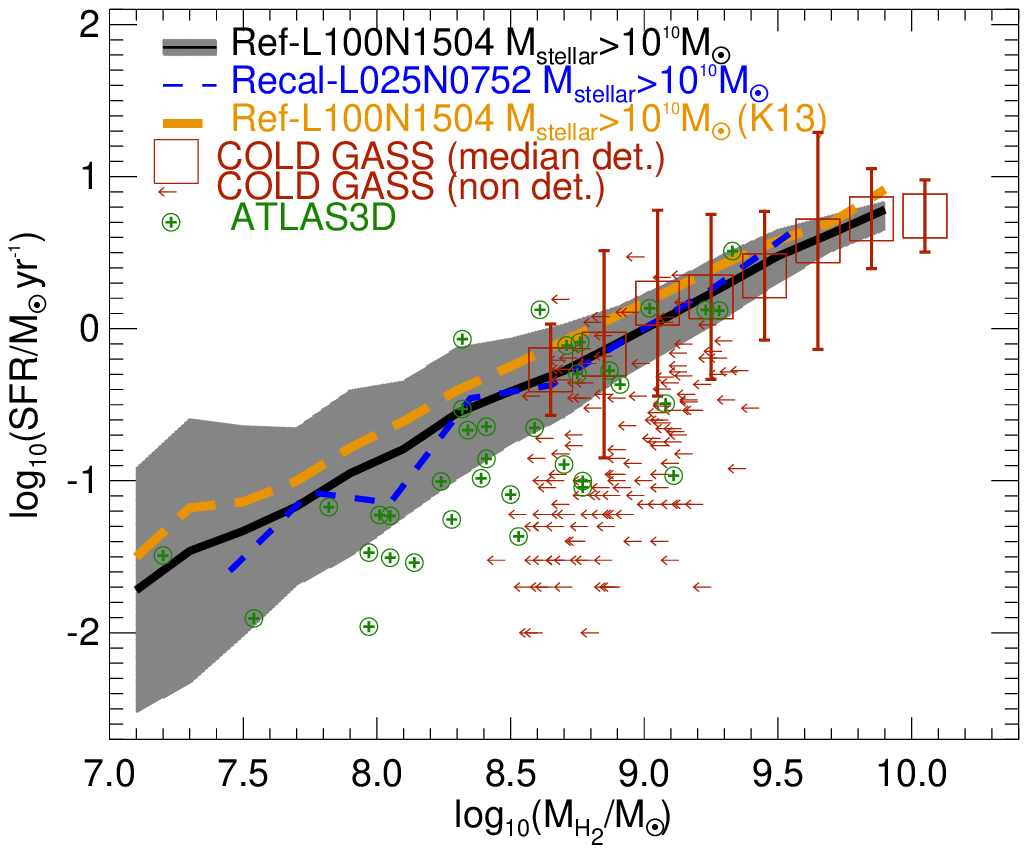}
\caption{As Fig.~\ref{COLDGASScomparison2} but for galaxies with
  $M_{\rm H_2}/M_{\rm stellar}>0$ and $M_{\rm stellar}>10^{10}\,\rm
  M_{\odot}$. {Here the filled region corresponds to the 
  $2.5^{\rm th}$ and $97.5^{\rm th}$ percentiles for the simulation Ref-L100N1504 with the GK11 prescription applied.}
  To the observations shown in
  Fig.~\ref{COLDGASScomparison2}, we add those of the ATLAS$^{\rm 3D}$
  survey (\citealt{Young11}; \citealt{Davis14}), as labelled, which
  sample a similar range of stellar mass as COLD GASS, but go much
  deeper in the CO(1-0) emission line. The main difference is that
  ATLAS$^{\rm 3D}$ only includes early-type galaxies, while COLD GASS
  does not distinguish between different morphological types. {Thus, 
  ATLAS$^{\rm 3D}$ cannot be directly compared to EAGLE (given the selection) but one can conclude 
  that about $17$\% of the early-type galaxies in ATLAS$^{\rm 3D}$ are $>2\sigma$ from the median 
  relation predicted by EAGLE.}}
\label{AllGASScomparison}
\end{center}
\end{figure}

In Fig.~\ref{COLDGASScomparison2} we show the SFR as a function of the
H$_2$ mass for {EAGLE} galaxies with $M_{\rm stellar}>10^{10}\,\rm
M_{\odot}$ and H$_2$ gas fractions above the sensitivity limit of COLD
GASS at $z=0$ and for the COLD GASS observations. In {EAGLE} there is
a tight correlation between these two quantities, because star
formation and most of the H$_2$ coexist in the same gas phase (see
Fig.~\ref{H2FracVsMs}). This is also the reason why the relation
between SFR and $M_{\rm H_2}$ is the tightest of all the scaling
relations discussed in this section. Simulation Recal-L025N0752 displays a
lower normalisation than Ref-L100N1504 (with the same prescription
applied) at H$_2$ masses $M_{\rm H_2}\gtrsim 10^{9.5}\,\rm M_{\odot}$
by a factor of $\approx 1.5$, consistent with the typically greater
H$_2$ mass associated with galaxies at a fixed stellar mass in the
Recal-L025N0752 simulation, as discussed in Appendix~\ref{ConvTests}.
{Note that the tight relation between the SFR and the H$_2$ mass is not surprising, because 
the star formation law adopted in EAGLE relates the SFR of particles with their gas mass, above a 
given density threshold. However, the component of the H$_2$ mass that is 
 non star-forming gas, in addition to the dependence on gas metallicity, result in differences 
between the  different simulations with the K13 or GK11 prescriptions to look different 
in the SFR-$M_{\rm H_2}$ relation (with differences being of  $\approx 0.2$~dex), even though 
the star formation law adopted is the same in every case.}

The relation between SFR and the H$_2$ mass predicted by EAGLE agrees
with the reported detections in COLD GASS. Galaxies for which the
CO(1-0) emission line is not detected have significantly lower SFRs.
EAGLE predicts that the H$_2$ masses of these galaxies should be an
order of magnitude (or more) below the detection limit of COLD GASS.
This is because EAGLE predicts that the tight relation between SFR and
H$_2$ extends down to very low values in both quantities. This can be
seen in Fig.~\ref{AllGASScomparison}, where we lower the threshold in
$M_{\rm H_2}/M_{\rm stellar}$ to $M_{\rm H_2}/M_{\rm stellar}>0$
(while maintaining the threshold in stellar mass).

In Fig.~\ref{AllGASScomparison} we also show the observations of the
ATLAS$^{\rm 3D}$ survey \citep{Cappellari11}, which covers a similar
range in stellar mass as COLD GASS but only includes early-type
galaxies (ellipticals and lenticulars). The complementary aspect of
ATLAS$^{\rm 3D}$ relative to COLD GASS is that it detects much fainter
CO(1-0) fluxes (see \citealt{Young11} and \citealt{Young13}). SFRs in
ATLAS$^{\rm 3D}$ come from a combination of UV and mid-IR photometry
(see \citealt{Davis14} for details). {We find that $\approx 17$\% of the early-type galaxies
in ATLAS$^{\rm 3D}$ lie $>2\sigma$ from the median relation in EAGLE, 
pointing to a possible conflict between EAGLE and the observations. However, this 
cannot be confirmed here because we would need to make a thorough comparison with ATLAS$^{\rm 3D}$ 
by morphologically selecting galaxies to be early-type in EAGLE.
This is beyond the scope of this paper.} 
If the discrepancy was real, it could be due 
to the star
formation law used in EAGLE, which is constructed from the observed
\citet{Kennicutt98} relation between the surface densities of SFR and
gas. Galaxies in ATLAS$^{\rm 3D}$ as well as the upper limits from
COLD GASS (shown in Fig.~\ref{AllGASScomparison}) are offset from the
\citet{Kennicutt98} relation in the sense that their efficiency of star
formation is inferred to be lower (see also conclusions of \citealt{Davis14}).

\begin{figure}
\begin{center}
\includegraphics[width=0.5\textwidth]{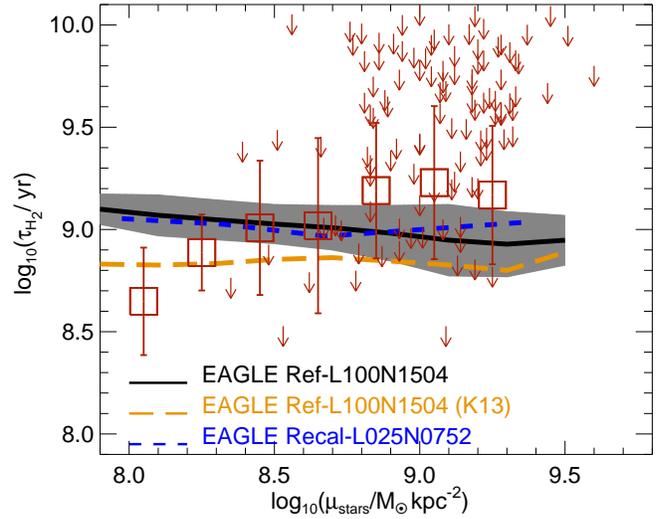}
\caption{The molecular hydrogen depletion timescale, $\tau_{\rm
    H_2}\equiv M_{\rm H_2}/{\rm SFR}$, as a function of the stellar
  mass surface density, $\mu_{\star}=M_{\rm stellar}/2\pi r^2_{\rm
    50}$ at $z=0$ (see text for details on how $\mu_{\star}$ is
  calculated in observations and EAGLE), for galaxies in EAGLE with
  $M_{\rm stellar}>10^{10}\rm \rm M_{\odot}$ and H$_2$ gas fractions
  above the sensitivity limit defined for COLD GASS).  Simulations and
  observations shown are as in Fig.~\ref{COLDGASScomparison2}.}
\label{COLDGASScomparison3}
\end{center}
\end{figure}

\begin{figure}
\begin{center}
\includegraphics[width=0.5\textwidth]{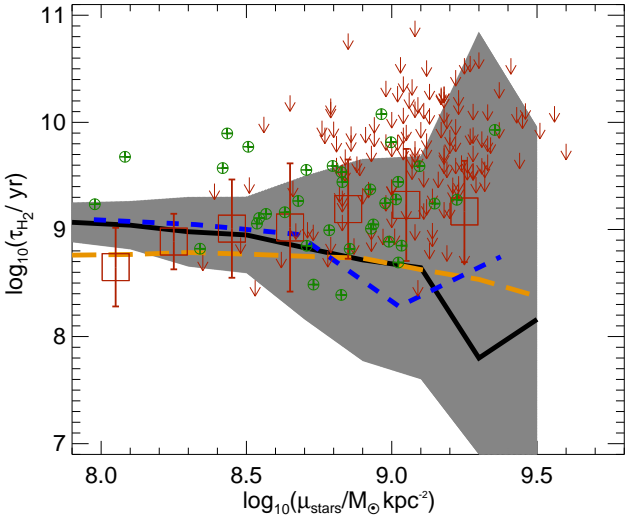}
\caption{As Fig.~\ref{COLDGASScomparison3} but for galaxies with
  $M_{\rm H_2}/M_{\rm stellar}>0$. {Here the filled region corresponds to the
  $2.5^{\rm th}$ and $97.5^{\rm th}$ percentiles of the simulation Ref-L100N1504 with the GK11 prescription applied.} 
  As in Fig.~\ref{AllGASScomparison}
  we add the observations of ATLAS$^{\rm 3D}$, for which $\mu_{\star}$
  is calculated using the $r$-band half-mass radius. { 
  ATLAS$^{\rm 3D}$ cannot be directly compared to EAGLE (given the selection) but one can conclude 
  that about $7$\% of the early-type galaxies in ATLAS$^{\rm 3D}$ are $>2\sigma$ from the median 
  relation predicted by EAGLE.}}
\label{COLDGASScomparison32}
\end{center}
\end{figure}

Fig.~\ref{COLDGASScomparison3} shows the H$_2$ depletion timescale,
$\tau_{\rm H_2}\equiv M_{\rm H_2}/{\rm SFR}$, as a function of the
stellar surface density at $z=0$ for the simulation Ref-L100N1504 with
the GK11 prescription and the median and $16^{\rm th}$ and
$84^{\rm th}$ percentiles of the observations of COLD GASS.  The H$_2$
depletion timescale is defined as the time that would be required to
exhaust the current H$_2$ mass if the SFR were constant.  The stellar
surface density is defined in {EAGLE} and in the observations as
$\mu_{\star}=(M_{\rm stellar}/2)/(\pi r^2_{\rm 50})$, where $M_{\rm
  stellar}$ and $r_{\rm 50}$ in {EAGLE} are the stellar mass in the
subhalo and the projected radius that contains half of that stellar
mass, respectively. In COLD GASS the stellar mass is derived from the
broad-band photometry and $r_{\rm 50}$ is the radius encompassing half
of the Petrosian flux in the $z$-band, which is measured in circular
apertures (see \citealt{Saintonge12} for details).  We show the median
and $16^{\rm th}$ and $84^{\rm th}$ percentiles of galaxies in COLD
GASS for which the CO(1-0) emission line is detected as squares with
error bars, while upper limits are shown as arrows.  

Even though the {EAGLE} predictions lie in a similar region to the
observations in Fig.~\ref{COLDGASScomparison3}, there are some
  interesting discrepancies.  It has been suggested from the
  observations shown in Fig.~\ref{COLDGASScomparison3} that there is a
  tendency for $\tau_{\rm H_2}$ to increase with increasing
  $\mu_{\star}$ (\citealt{Saintonge12}; \citealt{Martig13};
  \citealt{Davis14}). This trend is apparent if upper limits are
  considered, but less clear if only detections in COLD GASS are
  considered.  It has been suggested that such a trend can arise if
  the central stellar surface density increases the surface density threshold for gas
  instability, therefore lowering the SFR (\citealt{Saintonge12};
  \citealt{Martig13}; \citealt{Davis14}), in other words, as
  dynamically driven star formation suppression.  Such a trend is not
  seen in the reference simulation, Ref-L100N1504, when the GK11
  prescription is applied, but instead the predicted relation is
  consistent with a constant $\tau_{\rm H_2}$.

  To gain more insight into the relation between $\tau_{\rm H_2}$ and
  $\mu_{\star}$, we again relax the limit imposed on $M_{\rm
    H_2}/M_{\rm stellar}$ to $M_{\rm H_2}/M_{\rm stellar}>0$, while
  maintaining the cut in stellar mass, $M_{\rm stellar}>10^{10}\,\rm
  M_{\odot}$. The results are shown in Fig.~\ref{COLDGASScomparison32}
  where we also include the observations of the ATLAS$^{\rm 3D}$
  survey, in which $\mu_{\star}$ is measured from the $K$-band derived
  stellar masses and the half-light radius in the $r$-band. Since most
  early-type galaxies in ATLAS$^{\rm 3D}$ do not have significant
  ongoing star formation, measuring $r_{\rm 50}$ in the $r$-band is
  similar to measuring it in the infrared (as was done for COLD GASS).

  The observations of ATLAS$^{\rm 3D}$ confirm the existence of a
  population of galaxies with larger values of $\tau_{\rm H_2}$ than
  the galaxies in COLD GASS with CO(1-0) line detections. The trends
  predicted by EAGLE are as in Fig.~\ref{COLDGASScomparison3},
  although the dispersion around the medians becomes much larger at
  $\mu_{\star}>10^{8.5}\,\rm M_{\odot}\rm kpc^{-2}$. This means that
  some galaxies in {EAGLE} that have large values of $\mu_{\star}$
  also have high values for $\tau_{\rm H_2}$, consistent with the
  observations, but not the majority of them. However, { we find that 
  only $7$\% of early-type galaxies in ATLAS$^{\rm 3D}$ are $>2\sigma$ 
  from the median relation predicted by EAGLE, and again, we cannot confirm this is 
  a real issue in the simulation given that we are not selecting galaxies in EAGLE to be 
  early type.}
  Although we have not
  applied the same definition of $\mu_{\star}$ used in the
  observational data, we find that the trends of
  Fig.~\ref{COLDGASScomparison3} are not strongly sensitive to the
  definition of $\mu_{\star}$. For example, similar trends and values
  for $\mu_{\star}$ and $\tau_{\rm H_2}$ are found if we take the
  half-mass radius measured in apertures of $30$pkpc and $20$pkpc.

  We conclude that EAGLE captures the main scaling relations between
  stellar mass, SFR, stellar mass surface density and H$_2$ mass of
  galaxies that are on the main sequence of star formation.  Some
  discrepancies are present when we consider galaxies with high values of
  $\tau_{\rm H_2}$, which in EAGLE are rare but observationally
  represent a large fraction of the early-type galaxy population in
  the local Universe.  These tests are important since none of these
  observations were used to fix the free parameters in EAGLE. A study
  of the radial profiles of these quantities is ongoing and will be
  presented in a future paper (Lagos et al., in preparation).

\section{The evolution of the H$_2$ abundance of galaxies}\label{RedEvoSec}

In this section we present the analysis of the evolution of the H$_2$
mass function ($\S$~\ref{H2MFevolution}), H$_2$ scaling relations
($\S$~\ref{ScalRelationsHighz}), the stellar masses and SFRs of
galaxies selected by their H$_2$ mass ($\S$~\ref{H2selectedSamps}),
and the global H$_2$ density ($\S$~\ref{OmegaH2Sec}), over the
redshift range $0\le z \le 5$.

\subsection{The evolution of the H$_2$ mass function}\label{H2MFevolution}

\begin{figure*}
\begin{center}
\includegraphics[width=0.99\textwidth]{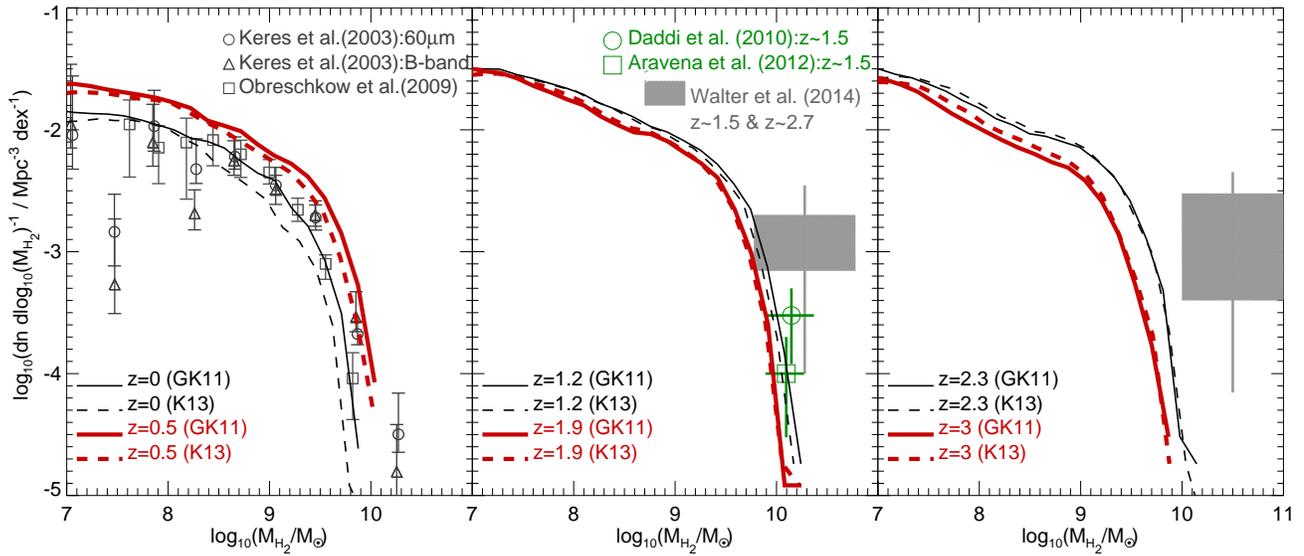}
\caption{The H$_2$ mass functions for the reference simulation
  Ref-L100N1504 using the GK11 (solid lines) and K13 (dashed lines)
  prescriptions applied, at $z=0$ and $z=0.5$ in the left-hand panel; $z=1.2$
  and $z=1.9$ in the middle panel; and $z=2.3$ and $z=3$ in the
  right-hand panel. The part affected by resolution is approximately
  $M_{\rm H_2}\lesssim 5\times 10^7\,\rm M_{\odot}$ (see
  Appendix~\ref{ConvTests}).  Observations shown in the
  left panel are as in Fig.~\ref{H2MFEagleSF}. The solid grey regions
  in the middle and right-hand panels correspond to the blind CO
  measurements presented of \citet{Walter14} and \citet{Decarli14}, at
  $z\approx 1.5$ and $z\approx 2.7$, respectively. The upper side of
  the grey regions show the number density obtained if all the
  high-quality line candidates are considered real, while the bottom
  side of the boxes show the number density if only secure detections
  are considered. The error bars in the boxes indicate the Poisson
  errors on the number densities.  In the middle panel we also show
  the observational inferences from \citet{Daddi10} and
  \citet{Aravena12}, as labelled, which are based on CO emission line
  follow-up of $z\approx 1.5$ galaxies or fields with spectroscopic or
  photometric redshifts.}
\label{H2MFHighz}
\end{center}
\end{figure*}

At $z=0$ the H$_2$ mass function has been measured by \citet{Keres03}
and \citet{Obreschkow09}.  At $z=1.5$ and $z=2.7$ constraints on the
H$_2$ mass function have been provided from a blind molecular emission
line survey by \citet{Decarli14} and \citet{Walter14}. Although the
uncertainties in these measurements are significant (see
Fig.~\ref{H2MFHighz}), these are currently the best available measurements of
the H$_2$ mass function at these redshifts. Other constraints on the H$_2$ mass functions
at $z\approx 1-2$ have been obtained by following-up galaxies at
mm wavelengths that have spectroscopic or photometric
redshifts. This is the case for \citet{Aravena12}, who performed a
JVLA survey towards a candidate cluster at $z\sim 1.5$.  Aravena et
al. obtained two detections and were able to place constraints on the
number density of bright CO($1-0$) galaxies. Another interesting
measurement at $z\approx 1.5$ was provided by \citet{Daddi10}, who
detected CO$(2-1)$ emission lines for six galaxies from a sample of
colour-selected star-forming galaxies.

Fig.~\ref{H2MFHighz} shows the H$_2$ mass function of the simulation
Ref-L100N1504 using both the GK11 and K13 prescriptions for several
redshift bins ranging from $z=0$ to $3$. The H$_2$ mass function shows
an increase in the number density of galaxies of all masses from $z=0$
to $z\approx 1.2-1.5$. At $z>1.5$ the trend reverses and the number
density of galaxies starts to decrease with redshift. 
{ The driver of this reversal is that at $z\gtrsim 1$ the gas metallicity decreases faster than 
at $z\lesssim 1$, which hampers H$_2$ formation. In addition, the median interstellar radiation field in galaxies increases monotonically 
with redshift (at least up to $z=4.5$), hampering H$_2$ formation even further. We come back to this point in $\S$~\ref{OmegaH2Sec}.}
Interestingly,
the differences at $z=0$ between the GK11 and K13 prescriptions
applied to EAGLE disappear at $z\approx 1-2$, and at $z\approx 2-3$
the K13 prescription gives slightly higher abundances of galaxies with
$M_{\rm H_2}<10^9\,\rm M_{\odot}$ than the GK11 prescription.  This is
because the GK11 prescription gives lower H$_2$ fractions than the K13
prescription in the case of intense interstellar radiation fields
($G^{'}_{\rm 0}\gtrsim 100$). The latter values of $G^{'}_{\rm 0}$ are
typical at $z\gtrsim 1.5$ (see bottom panel of Fig.~\ref{nHweighted}
and Appendix~\ref{H2Prescriptions} for an analysis of the differences
between the H$_2$ masses predicted when applying the GK11 and K13
prescriptions).

We find that {EAGLE} predictions at $z\approx 1.2$ (middle panel of
Fig.~\ref{H2MFHighz}), with either the GK11 and K13 prescriptions applied, are
within the uncertainties of the observations of \citet{Daddi10},
\citet{Aravena12} and \citet{Walter14}.  At $z\approx 2.5$ (right-hand
panel of Fig.~\ref{H2MFHighz}), the predicted abundance of galaxies
with $M_{\rm H_2}>10^{10}\,\rm M_{\odot}$ is low compared to the
constraints of \citet{Walter14}, particularly if we compare with the
observational constraints at $z=3$.  However, the redshift range of
that data point in the right panel of Fig.~\ref{H2MFHighz} is large,
$z\approx 2-3.1$, which can partially explain the offset with 
the predictions.  Another important
uncertainty is the CO-H$_2$ conversion factor: $X=2$ was assumed
in \citet{Walter14} and \citet{Decarli14}. If this conversation factor
were on average lower in galaxies with large abundances of H$_2$, as
suggested by the theoretical studies of \citet{Lagos12},
\citet{Narayanan12} and \citet{Popping14}, the H$_2$ masses would be
overestimated by \citet{Walter14}, improving the
consistency with the EAGLE results.

Note that the small number of galaxies with $M_{\rm H_2}\gtrsim
10^{10}\,\rm M_{\odot}$ is not an effect of the limited box size, as
we find from comparing the Ref-L100N1504 and Ref-L050N0752
simulations which produce a very similar number density of galaxies
with $M_{\rm H_2}\gtrsim 10^{10}\,\rm M_{\odot}$ and a very similar
value of the cosmic mean H$_2$ density.  (Fewer than $10$ objects per
dex bin in Fig.~\ref{H2MFHighz} corresponds to a number density of
$10^{-4.3}$ in the units of the $y$-axis.)

The {EAGLE} results in Fig.~\ref{H2MFHighz} are slightly different
from those given by semi-analytic models. The latter predict the peak
of the number density of galaxies with $M_{\rm H_2}\gtrsim
10^{10}\,\rm M_{\odot}$ to be at $z\approx 2-2.5$
(\citealt{Obreschkow09d}; \citealt{Lagos12}).  A possible explanation
for this is that the models that have been preferred in semi-analytic
work to calculate the ratio of H$_2$-to-HI gas mass (for example the
empirical model of \citealt{Blitz06}) do not consider gas metallicity.
Future observations can distinguish between these different
predictions, for example with telescopes such as the Atacama Large
Millimeter Array (ALMA).

\subsection{H$_2$ scaling relations at high redshift}\label{ScalRelationsHighz}

\begin{figure}
\begin{center}
\includegraphics[width=0.5\textwidth]{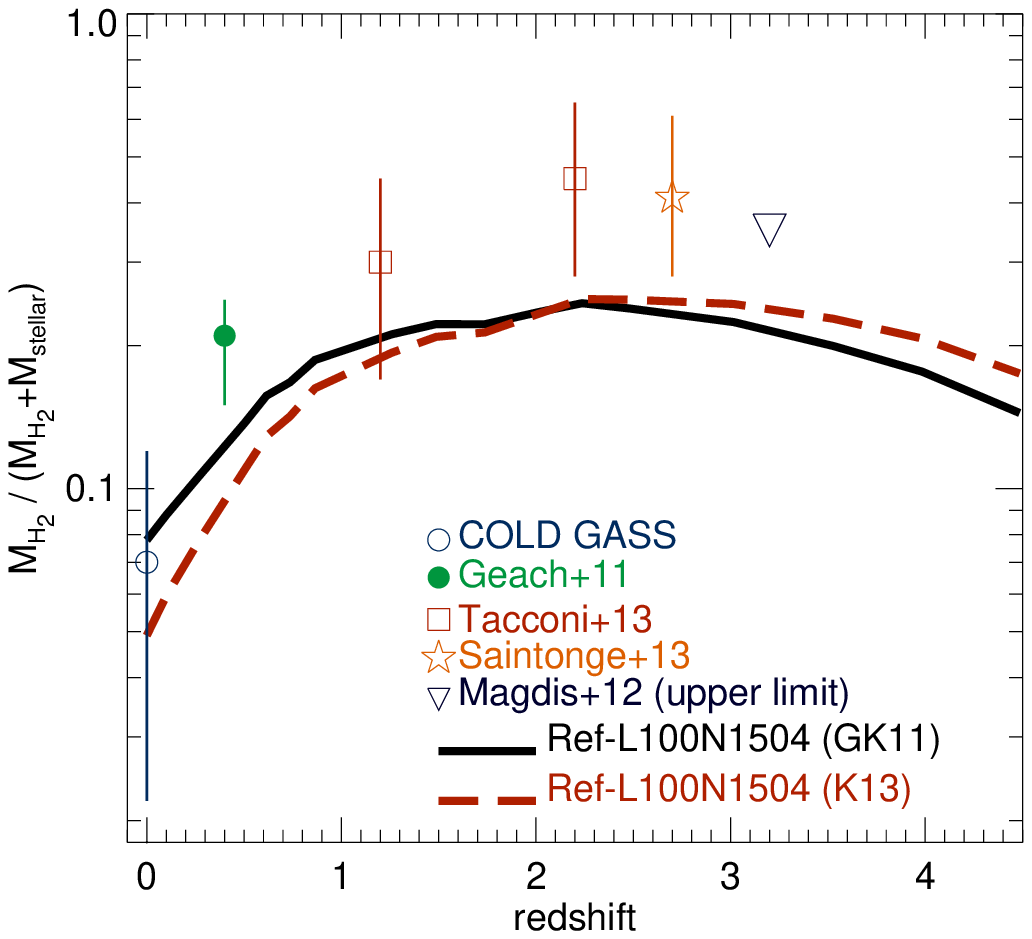}
\includegraphics[width=0.5\textwidth]{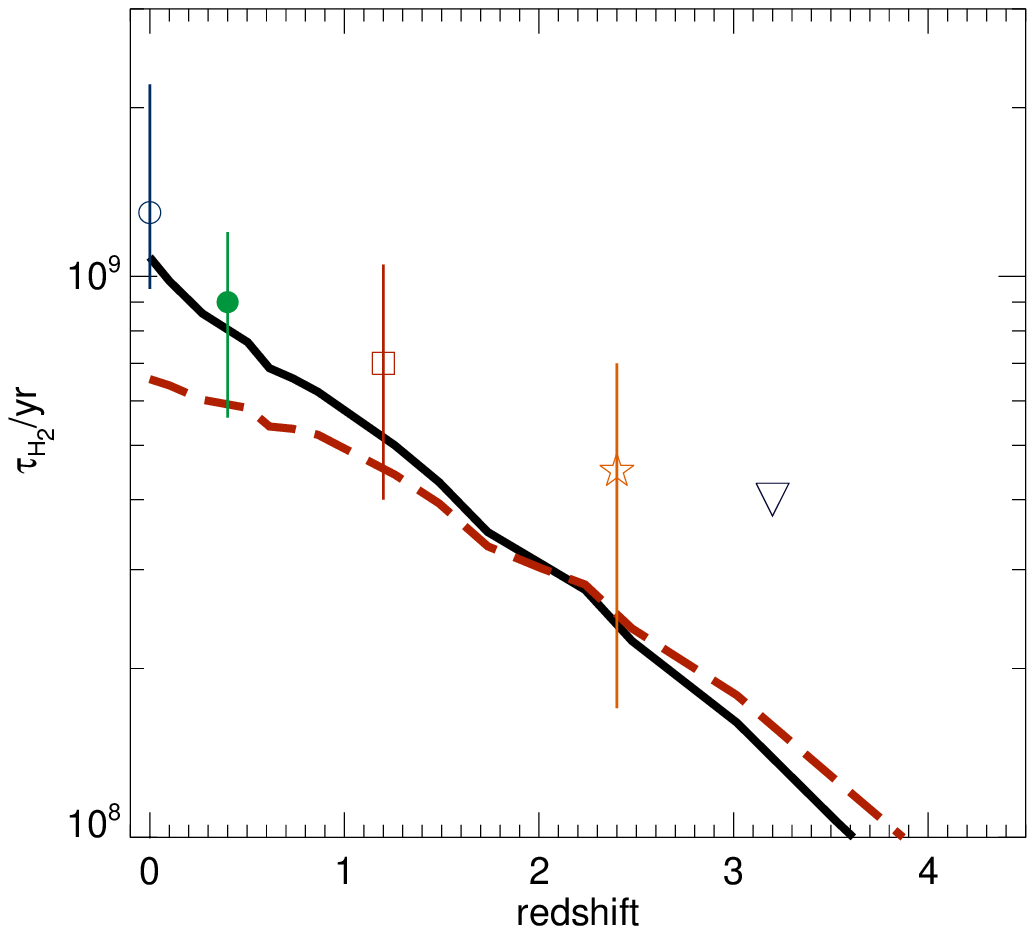}
\caption{{\it Top panel:} The mean H$_2$ gas fraction, $M_{\rm
    H_2}/(M_{\rm H_2}+M_{\rm stellar})$ as a function of redshift for
  the Ref-L100N1504 simulation using the GK11 (solid line) and the K13
  (dashed line) prescriptions.  The mean H$_2$ gas fraction was
  calculated in { EAGLE} using only main sequence galaxies with
  $M_{\rm stellar}>5\times 10^9\,\rm M_{\odot}$.  The observational
  data also correspond to the median H$_2$ gas fraction for main
  sequence galaxies at different redshifts. Data was taken from
  \citet{Saintonge11}, \citet{Geach11}, \citet{Magdis12},
  \citet{Saintonge13} and \citet{Tacconi13}.  Magdis et al. presented
  a detection and a non-detection of the CO(3-2) emission line of
  galaxies at $z\approx 3$, and therefore the combined estimate
  presented here is only an upper limit.  {\it Bottom panel:} The
  H$_2$ depletion timescale, $M_{\rm H_2}/{\rm SFR}$, as a function of
  redshift. The observational data are as in the top panel except for the
  measurement at $z\approx 2.4$ which is a combination of the
  \citet{Tacconi13} and the \citet{Saintonge13} measurements, as was
  presented by Saintonge et al.}
\label{TauH2FracH2}
\end{center}
\end{figure}

Fig.~\ref{TauH2FracH2} shows the ratio $M_{\rm H_2}/(M_{\rm
  H_2}+M_{\rm stellar})$ and the H$_2$ depletion timescale, $\tau_{\rm
  H_2}\equiv M_{\rm H_2}/{\rm SFR}$, as a function of redshift,
calculated for galaxies on the main sequence of star formation and
with stellar masses $M_{\rm stellar}>5\times 10^9\,\rm M_{\odot}$ in
the Ref-L100N1504 simulation using the GK11 and K13 prescriptions. We
define the main sequence of galaxies as $\pm 0.5$~dex around the
median $\rm log_{\rm 10}(SFR)$ at a given stellar mass, calculated
from the simulation outputs.
Observational data are from \citet{Saintonge11}, \citet{Geach11},
\citet{Magdis12}, \citet{Saintonge13} and \citet{Tacconi13}, and also
refer to quantities measured for main sequence galaxies over a range
of stellar masses similar to the one chosen here.  The definition of
the main sequence of galaxies in the observational data is also based
on the median SFR of galaxies of a given stellar mass. However, since
observations are flux limited, the possibility exists that the
observational data are biased towards higher SFRs.  

\begin{figure}
\begin{center}
\includegraphics[width=0.5\textwidth]{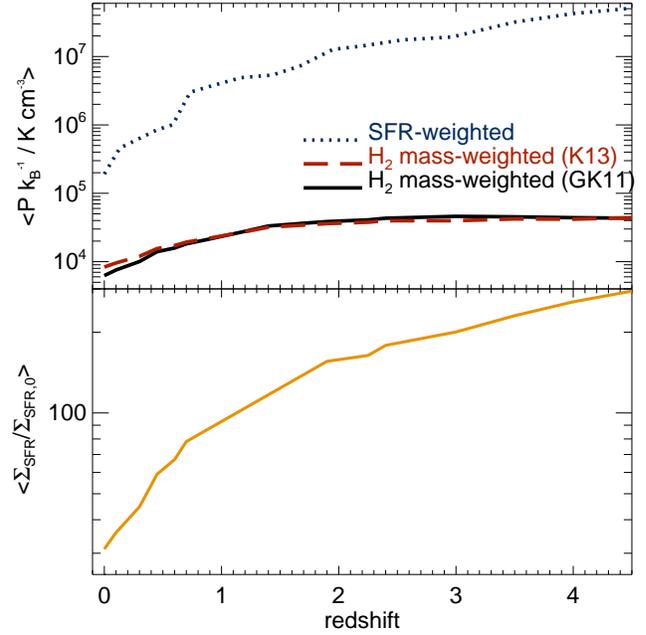}
\caption{{\it Top panel:} The median of the $M_{\rm H_2}$-weighted
  pressure, $P$, when the GK11 (solid line) and the K13 (long dashed
  line) prescriptions are applied to the simulation Ref-L100N1504, as
  a function of redshift.  The SFR-weighted $P$ (dotted line) is also
  shown as a function of redshift in the same simulation. Pressure is
  expressed in terms of $P\,k^{-1}_{\rm B}$, where $k_{\rm B}$ is
  Boltzmann's constant. {\it Bottom panel:} The median $\Sigma_{\rm
    SFR}/\Sigma_{\rm SFR,0}$, where $\Sigma_{\rm SFR,0}=10^{-3}\,\rm
  M_{\odot}\, \rm yr^{-1}\,\rm kpc^{-2}$ is the SFR surface density in
  the solar neighborhood \citep{Bonatto11}, calculated using all the
  star-forming particles at each time, as a function of redshift.}
\label{nHweighted}
\end{center}
\end{figure}

{EAGLE} predicts
the ratio $M_{\rm H_2}/(M_{\rm H_2}+M_{\rm stellar})$ to increase by
$0.5-0.7$~dex from $z=0$ to $1$ and to evolve very little from
$z\approx 1$ to $4$. These trends are also observed. The GK11
prescription applied to the Ref-L100N1504 simulation predicts a
slightly lower median value of $M_{\rm H_2}/(M_{\rm H_2}+M_{\rm
  stellar})$ at $z\gtrsim 2.5$ than is the case for the K13
prescription. The reason is that the GK11 prescription is more
sensitive to intense interstellar radiation fields, $G^{'}_0\gtrsim
100$, than the K13 prescription, and the latter values are typical for
$z\gtrsim 2$ (see bottom panel of Fig.~\ref{nHweighted}).

There is a slight discrepancy of a factor of $\approx 1.5$ between the
predicted ratio $M_{\rm H_2}/(M_{\rm H_2}+M_{\rm stellar})$ and the
observationally inferred ones at $z\approx 2-3$, regardless of whether
the GK11 or the K13 prescription is applied.  It is unclear at this
stage whether this offset is a problem for EAGLE, since the
observations have so far been biased towards gas-rich systems (as they
only quote detections, except for the values in the local Universe
given by  COLD GASS). However, the offset between the
predicted and observed ratio $M_{\rm H_2}/(M_{\rm H_2}+M_{\rm
  stellar})$ is, in any case, within the scatter seen in observations.

The increase of the H$_2$ mass fraction with redshift is driven by the
greater fraction of gas particles that reach the regions of high H$_2$
fractions ($T<10^{4.5}\,\rm K$ and $P/k_{\rm B}\gtrsim 5\times
10^{3}\,\rm cm^{-3}$).  Such a trend can be seen from the evolution of
the $M_{\rm H_2}$-weighted pressure in the top panel of
Fig.~\ref{nHweighted}.  The trend shows that at higher redshift more
H$_2$ is locked up in particles of higher pressure than at $z=0$.

The bottom panel of Fig.~\ref{TauH2FracH2} shows the evolution of the
depletion timescale, $\tau_{\rm H_2}$, measured for the same main
sequence galaxies used to calculate the evolution of $M_{\rm
  H_2}/(M_{\rm H_2}+M_{\rm stellar})$ shown in the top panel.  We find
that $\tau_{\rm H_2}$ decreases with redshift by a factor of $\sim 10$
between $z=0$ and $z=4$, with the GK11 prescription predicting a
slightly larger decrease than the K13 prescription. Such evolution
indicates that EAGLE galaxies at higher redshift are more efficient at
forming stars.
{ The non-monotonic behaviour of the H$_2$ fraction with redshift, seen at the top panel of Fig.~\ref{TauH2FracH2}, 
affects $\tau_{\rm H2}$ by accelerating the decrease with redshift at $z\gtrsim 2.3$. This can be 
seen from fitting a power-law relation before and after $z=2.3$: at $z\le 2.3$, $\tau_{\rm H2}\propto (1+z)^{-1.2}$, while at 
$z>2.3$, $\tau_{\rm H2}\propto (1+z)^{-3}$.}

To help visualize the reasons behind the increase in the efficiency of
star formation with redshift, we show in the top panel of
Fig.~\ref{nHweighted} the SFR-weighted and $M_{\rm H_2}$-weighted
pressures for the Ref-L100N1504 simulation, which we refer to as
$\langle P\rangle_{\rm SFR}$ and $\langle P\rangle_{\rm M_{\rm H_2}}$
respectively.  Applying the GK11 or K13 prescriptions in this
simulation does not make any significant difference to $\langle
P\rangle_{\rm M_{\rm H_2}}$. Another important observation is that
both $\langle P\rangle_{\rm SFR}$ and $\langle P\rangle_{\rm M_{\rm
    H_2}}$ increase with redshift, but $\langle P\rangle_{\rm SFR}$
does it to a greater extent.  Since the star formation law adopted in
EAGLE has a power-law, $n=1.4$ (see Eq~\ref{SFlaw}), forming stars at
higher pressure implies having higher gas surface density and
therefore higher $\Sigma_{\rm SFR}/\Sigma_{\rm gas}$. Although
$\langle P\rangle_{\rm M_{\rm H_2}}$ also increases, its weaker
evolution compared to $\langle P\rangle_{\rm SFR}$ helps $\tau_{\rm
  H_2}$ to decrease faster with redshift.

\citet{Lilly13} have suggested that the observations shown in
Fig.~\ref{TauH2FracH2} are consistent with the ``equilibrium model'',
where the evolution in the specific SFR and the H$_2$ depletion
timescale are mainly driven by the evolution of the gas reservoir
(neutral gas in the galaxy). { We find that besides the gas reservoir, an additional parameter plays an
important role:
the typical pressures at which star
formation takes place and at which H$_2$ forms.} In other words,
galaxies move along the Kennicutt-Schmidt relation towards higher gas
surface densities as redshift increases. { The change in the H$_2$
depletion timescale is therefore a result of 
adopting a superlinear
star formation law (i.e. power-law index $>1$ in Eq.~\ref{SFlaw}) and 
the transition from HI- to H$_2$-dominated gas evolving towards higher mean ISM pressures
as the redshift increases.} 
Note that our conclusion is subject to the assumption that the Kennicutt-Schmidt relation holds 
at redshifts higher than $0$. This is a reasonable assumption since observations so far show 
that galaxies at $z>0$ follow the same Kennicutt-Schmidt relation 
(e.g. \citealt{Genzel10}; \citealt{Genzel13}).

{ The trends obtained in EAGLE for the evolution of H$_2$ gas fractions and H$_2$ depletion timescales
 are qualitatively the same as those inferred from observations. However, observational results have so far presented contradictory conclusions.
For example, \citet{Genzel15} presented CO and dust mass observations for $\approx 500$ galaxies from
$z\approx 0$ to $z\approx 3$, and concluded that the H$_2$ depletion timescale is only a weak function of
redshift, $\tau_{\rm H_2}\propto (1+z)^{-0.3}$, while EAGLE favours a stronger dependence,
$\tau_{\rm H_2}\propto (1+z)^{-1.1}$ for the Ref-L100N1504 with the K13 prescription applied and $\tau_{\rm H_2}\propto (1+z)^{-1.4}$
with the GK11 prescription applied. However, recent ALMA observations by \citet{Scoville15} favour a much stronger
 redshift dependence of $\tau_{\rm H_2}$\footnote{\citet{Scoville15} measure ISM masses rather than H$_2$ masses. However,
from the very high H$_2$ gas fractions measured in high-$z$ galaxies, one can safely assume that the ISM masses of
high-$z$ galaxies are dominated by H$_2$, on average.},
with $z\approx 1$ galaxies having $\tau_{\rm H_2}$ lower by an order of magnitude
than $z=0$ galaxies, in good agreement with the EAGLE results. Moreover, \citet{Scoville15} found no difference in
$\tau_{\rm H_2}$ between galaxies on and off the main sequence, while \citet{Genzel15} found an order of magnitude difference.
This argues for further observations and the need for galaxy samples with simpler selection functions that can then be applied to     
 simulations to make a one-to-one comparison (as we did in $\S$~\ref{SecScalingrelations} for COLD GASS).}

 \subsection{The properties of H$_2$ mass-selected samples}\label{H2selectedSamps}

\begin{figure}
\begin{center}
\includegraphics[width=0.5\textwidth]{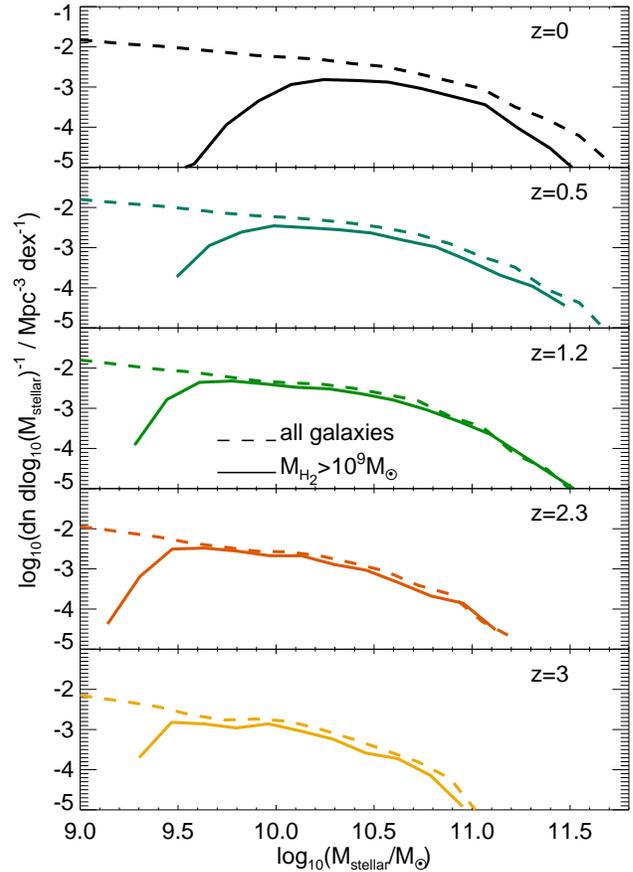}
\caption{Galaxy stellar mass functions of all galaxies (dashed lines)
  and those having $M_{\rm H_2}>10^9\,\rm M_{\odot}$ (solid
  lines) at different redshifts, as labelled in each panel, for the
  simulation Ref-L100N1504 using the GK11 prescription. The H$_2$
  mass-selected sample traces galaxies with stellar masses, $M_{\rm
    stellar}\gtrsim 10^{10}\,\rm M_{\odot}$, but at $z\lesssim 0.5$
  about half of those massive galaxies have $M_{\rm H_2}<10^9\,\rm
  M_{\odot}$.}
\label{SMH2Selected}
\end{center}
\end{figure}

\begin{figure}
\begin{center}
\includegraphics[width=0.5\textwidth]{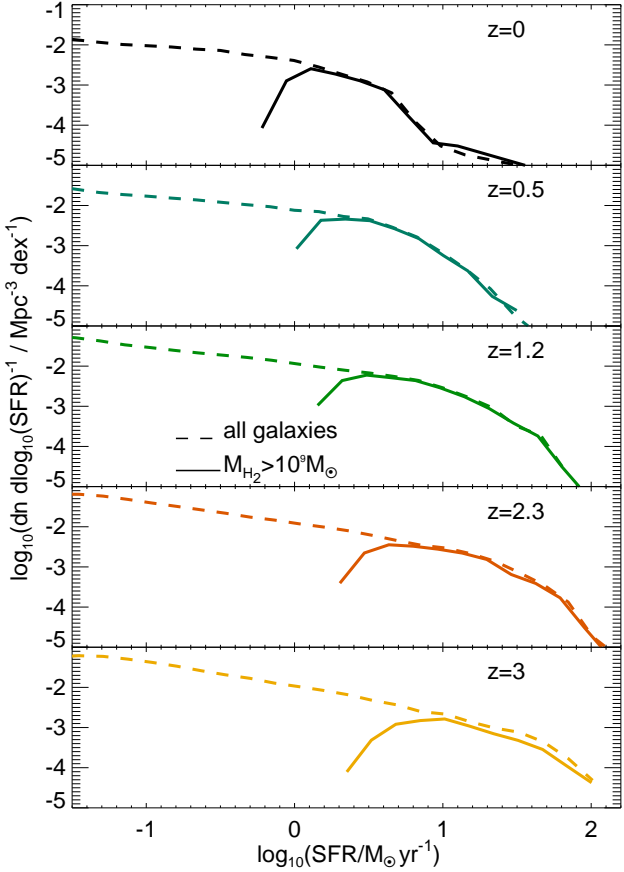}
\caption{SFR functions of all galaxies (dashed lines) and those having
  $M_{\rm H_2}>10^9\,\rm M_{\odot}$ (solid lines) at different
  redshifts, as labelled in each panel, for the simulation
  Ref-L100N1504 using the GK11 prescription. The H$_2$ mass-selected
  sample traces galaxies with SFR$\gtrsim 1\,\rm M_{\odot}\,yr^{-1}$
  at $z=0$, while at $z=3$ the typical galaxies present in H$_2$
  mass-selected samples have SFR$\gtrsim 10\,\rm M_{\odot}\,yr^{-1}$
  at $z=3$.}
\label{SFRH2Selected}
\end{center}
\end{figure}

An interesting question is what are the stellar masses and SFRs of
galaxies selected by their H$_2$ mass? This question is relevant for
planned blind CO surveys as an important aspect of these surveys is
finding optical and near-IR counterparts (see for example
\citealt{Decarli14}).  We show in Fig.~\ref{SMH2Selected} the galaxy
stellar mass functions for all galaxies (dashed lines) and for the
subsample that have $M_{\rm H_2}>10^9\,\rm M_{\odot}$ (solid lines) at
$0<z<3$. Typically, galaxies with large reservoirs of H$_2$ also have
high stellar masses. However, at $z\lesssim 0.5$ we observe an
increasing number of massive galaxies with $M_{\rm H_2}<10^9\,\rm
M_{\odot}$, which is clear from the difference in normalisation
between the galaxy stellar mass function of all galaxies and of those
with $M_{\rm H_2}>10^9\,\rm M_{\odot}$ in the top two panels of
Fig.~\ref{SMH2Selected}. At $z\gtrsim 1$, galaxies with $M_{\rm
  stellar}>10^{10}\,\rm M_{\odot}$ almost always have $M_{\rm
  H_2}>10^9\,\rm M_{\odot}$. For a comparison between the predicted
galaxy stellar mass function in EAGLE and observations see
\cite{Furlong14}.

Fig.~\ref{SFRH2Selected} shows the SFR function of all galaxies
(dashed lines) and of those with $M_{\rm H_2}>10^9\,\rm M_{\odot}$ (solid
lines). The H$_2$ mass-selected sample is a very good tracer of the
most actively star-forming galaxies at any time. For example, at $z=0$ galaxies
with $M_{\rm H_2}>10^9\,\rm M_{\odot}$ have SFR$\gtrsim 1\,\rm
M_{\odot}\,yr^{-1}$. The average SFR of the H$_2$ mass-selected sample
increases with redshift, not only because galaxies have on average
higher SFRs at fixed stellar mass, but also because the ratio between
the H$_2$ mass and the SFR decreases with redshift (see bottom panel
of Fig.~\ref{TauH2FracH2}). This means that the H$_2$ mass at a fixed
SFR decreases with redshift. The outcome of this is that the typical
SFRs of galaxies with $M_{\rm H_2}>10^9\,\rm M_{\odot}$ at $z=3$ are
SFR$\gtrsim 10\,\rm M_{\odot}\,yr^{-1}$.

\subsection{The evolution of $\Omega_{\rm H_2}$}\label{OmegaH2Sec}

\begin{figure}
\begin{center}
\includegraphics[width=0.5\textwidth]{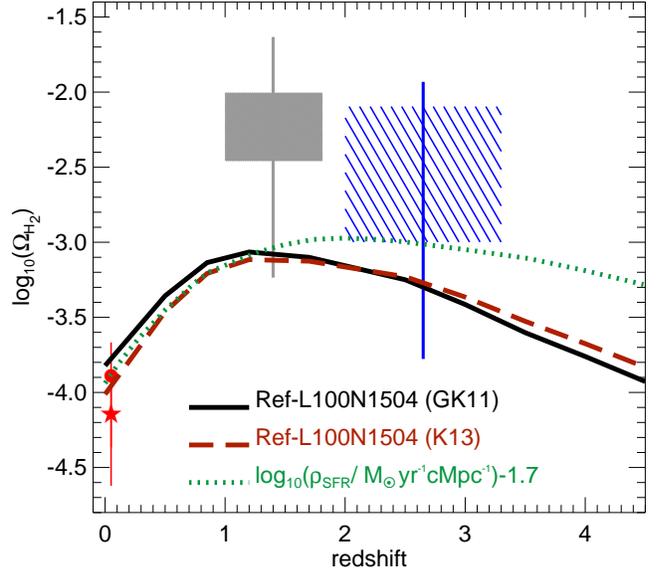}
\caption{$\Omega_{\rm H_2}$, defined as the ratio of the H$_2$ density
  to the critical density of the universe, $\Omega_{\rm H_2}\equiv
  \rho_{\rm H_2}/\rho_{\rm crit}$, as a function of redshift for the
  simulation Ref-L100N1504 using the GK11 (solid line) and the K13
  (dashed line) prescriptions.  For reference, we show the SFR cosmic
  density evolution as dotted line, offset by $1.7$~dex so that it
  lies in a similar range than $\Omega_{\rm H_2}$.  Observations at
  $z=0$ by \citet{Keres03} and \citet{Obreschkow09b} are shown as a
  filled circle and star, respectively, with $1\sigma$ error bars,
  while the observations from the blind CO survey of \citet{Walter14}
  are shown as solid and hatched regions. The boxes show the
  contribution of the CO blind detections without extrapolating to
  account for the undetected sources below the detection limits in the
  millimeter (which would translate into H$_2$ masses $<5\times
  10^{9}\,\rm M_{\odot}$). The error bars in the boxes indicate
  $1\sigma$ Poisson errors.}
\label{OmegaH2}
\end{center}
\end{figure}

Fig.~\ref{OmegaH2} shows the evolution of $\Omega_{\rm H_2}\equiv
\rho_{\rm H_2}/\rho_{\rm crit}$, where $\rho_{\rm crit}$ is the
critical density of the Universe at the given redshift, for
Ref-L100N1504 using both the GK11 and K13 prescriptions.  The global
density of H$_2$ was calculated using the H$_2$ abundance in galaxies
measured in a $30$~pkpc aperture. If we instead use all the gas
particles that are bound to the sub-halo of a galaxy, we find a
value higher by $1.5-3$\%, which shows that most of the H$_2$ in the
universe resides close to galaxy centres. We also show for reference
the evolution of the SFR cosmic density (dotted line) in the
Ref-L100N1504 simulation.

The GK11 prescription applied to EAGLE predicts a peak at a slightly
lower redshift, $z\approx 1.2$, compared with $z\approx 1.4$ for the
K13 prescription. The latter results in an increase of a factor of
$\approx 4.3$ between $z=0$ and $z\approx 1.4$, while GK11 yields an
increase of a factor of $\approx 3.5$ over the same redshift
range. The difference is driven by the GK11 prescription being more
sensitive to intense interstellar radiation fields, as discussed in
$\S$~\ref{ScalRelationsHighz}.  The peak of $\Omega_{\rm H_2}$ is
located at a redshift which in both cases is lower than the peak of the
SFR density in {EAGLE}, at $z\approx 2$ (see dotted line in
Fig.~\ref{OmegaH2} and \citealt{Furlong14} for a detailed analysis on
the SFR density evolution). The reason for this offset is the faster
decrease in gas metallicity at $z>1$ and the monotonic increase of the
median $\Sigma_{\rm SFR}$ with redshift (see the bottom panel of
Fig.~\ref{nHweighted}). At $z<1$, the H$_2$ mass-weighted metallicity
decreases as ${\rm log_{\rm 10}}(Z)\propto (1+z)^{-0.6}$, while at
$z>1$ the decrease accelerates and scales as ${\rm log_{\rm
    10}}(Z)\propto (1+z)^{-1.8}$. Note that we still see a rapid
increase of $\Omega_{\rm H_2}$ between $0$ and $1$ even though some of
the conditions for H$_2$ formation become less favourable (lower gas
metallicities and higher interstellar radiation fields). This is due
to an increase in the pressure of the ISM in galaxies and to an
increase in the total neutral gas content.

We also show in Fig.~\ref{OmegaH2} the observational estimates of
\citet{Keres03}, \citet{Obreschkow09b} and \citet{Walter14}. The error
bars on the Walter et al. inferences are as described in
$\S$~\ref{H2MFevolution}.  Compared to current observational
estimates, {EAGLE} captures the increase in the H$_2$ content of
galaxies, but predicts a lower $\Omega_{\rm H_2}$ in the redshift
range $z=1-3$, although within the error bars of the observations.
Some possibilities to explain this tension are: (i)~EAGLE is truly
predicting a lower value of $\Omega_{\rm H_2}$ than in the real
Universe, and (ii)~the observational bias $\Omega_{\rm H_2}$ towards
higher values. Concerning (i), this could be related to the 
SFR cosmic density in EAGLE being below the observations by a factor of 
$\approx 2$ \citep{Furlong14}.
Related to (ii), \citet{Walter14} adopted a value of
$X$ to convert their CO measurements to H$_2$ similar to the Milky-Way
value. However, theoretical evidence points to a lower value of $X$
being more reasonable for these high-redshift galaxies. For example,
\citet{Narayanan12} show that the value of $X$ decreases with
increasing velocity width of the CO line. Since the CO velocity widths
of these high-redshift galaxies are much larger than for $z=0$ galaxies,
one would expect lower values of $X$. These systematics in the
observations need to be carefully addressed before we can conclude
that the tension between EAGLE and the observational inferences of
$\Omega_{\rm H_2}$ is a true shortcoming of the simulation.

\begin{figure}
\begin{center}
\includegraphics[width=0.5\textwidth]{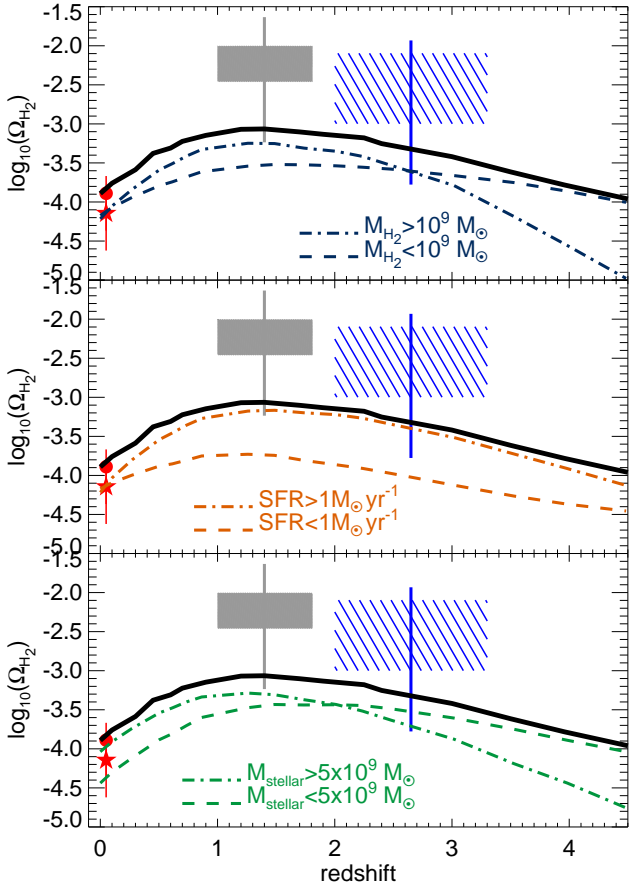}
\caption{$\Omega_{\rm H_2}$ as a function of redshift for the
  simulation Ref-L100N1504 using the GK11 prescription separated into
  the contribution from galaxies with different H$_2$ masses (top
  panel), star formation rates (middle panel) and stellar masses
  (bottom panel), as labelled.  The solid line in each panel
  corresponds to the total H$_2$ density also shown as a solid line in
  Fig.~\ref{OmegaH2}.  Observations are as in Fig.~\ref{OmegaH2} and
  are shown in all panels.}
\label{OmegaH2v2}
\end{center}
\end{figure}

\citet{Walter14} argue that the tension seen between several
predictions of $\Omega_{\rm H_2}$ in the literature and their
observational estimates are, in reality, worse than it seems from
comparisons such as the one presented here.  This is because their
estimate of $\Omega_{\rm H_2}$ accounts only for detections (secure
and candidates) and it does not include the contribution from galaxies
with $M_{\rm H_2}\lesssim 5\times 10^9\,\rm M_{\odot}$.  However, to
assess whether this really worsens the tension with theoretical
predictions, one needs to estimate how much of $\Omega_{\rm H_2}$ is
contributed by galaxies with different values of $M_{\rm H_2}$ or
other properties. We show in Fig.~\ref{OmegaH2v2} the value of
$\Omega_{\rm H_2}$ for subsamples of galaxies in { EAGLE} selected by
their H$_2$ mass, SFR and stellar mass.  Galaxies with $M_{\rm
  H_2}>10^9\,\rm M_{\odot}$ dominate the global $\Omega_{\rm H_2}$ at
$z<2.5$, which means that we expect the H$_2$ mass function integrated
over the massive end only (above the break) to recover most of the
H$_2$ in the universe for these redshifts.

In terms of stellar mass and SFR, $\Omega_{\rm H_2}$ is dominated by
galaxies with:
\begin{itemize}
\item SFR~$>10\,\rm M_{\odot}\,\rm yr^{-1}$ at $z\gtrsim 0.1$, 
\item $M_{\rm stellar}>5\times 10^{9}\,\rm M_{\odot}$ at $z\lesssim 2$ 
\item $M_{\rm stellar}<5\times 10^{9}\,\rm M_{\odot}$
at $z\gtrsim 2$. 
\end{itemize}
\noindent The dominant contribution of galaxies with $M_{\rm stellar}<5\times
10^{9}\,\rm M_{\odot}$ at $z\gtrsim 2$ is unlikely to be a consequence of the
limited simulated volume ($100$~cMpc box size).  In fact, $\Omega_{\rm
  H_2}$ convergences to better than $15$\% in the simulations
Ref-L100N1504 and Ref-L050N0752.

These results are slightly different to those found in semi-analytic
models.  For example \citet{Lagos14b} find that galaxies with $1\,\rm
M_{\odot}\,\rm yr^{-1}<{\rm SFR}<10\,\rm M_{\odot}\,\rm yr^{-1}$ make
a larger contribution to $\Omega_{\rm H_2}$ than those with
SFR~$>10\,\rm M_{\odot}\,\rm yr^{-1}$. These differences are
interesting and will be tested by future observations.

The galaxies that dominate $\Omega_{\rm H_2}$ in { EAGLE} are readily
observable in current optical and near-IR surveys (for example the
CANDELS survey; \citealt{Grogin11}). This suggests that a cheap and
effective observing strategy to unveil $\Omega_{\rm H_2}$ is to 
follow up existing optical and near-infrared surveys to measure  CO,
rather than blind CO surveys that are more expensive and harder to
characterise.  In follow-up strategies, molecular emission line
surveys can be guided by well-known UV, optical, IR and/or radio
surveys, with well identified positions and spectroscopic or
photometric redshifts. Even in the case that only galaxies with
SFR$>10\,\rm M_{\odot}\,yr^{-1}$ are followed up, we expect them to
uncover $50$\% or more of $\Omega_{\rm H_2}$, placing strong
constraints on theoretical models of galaxy formation.

\section{Conclusions}\label{ConcluSec}

We have presented a comprehensive comparison between two hydrodynamic
simulations of the {EAGLE} suite and surveys of molecular gas in
galaxies at low and high redshift.  
To determine H$_2$ masses in simulated galaxies, we first extracted the fraction of
neutral gas (atomic and molecular) in each gas particle using the
fitting functions of \citet{Rahmati13}, which were derived applying
 radiative transfer algorithms to cosmological simulations in order to
calculate local neutral fractions as a function of the collisional
and photoionisation rates, temperature and gas density. After
estimating the fraction of the gas that is neutral, we
proceeded to calculate the fraction of that gas that is molecular.
  We applied two schemes to EAGLE to calculate the H$_2$ fraction
of individual gas particles, those advanced by \citet{Gnedin11} and by
\citet{Krumholz13}. The first one, GK11, provides fitting functions
for the H$_2$ fraction that depend on the dust-to-gas mass ratio and
the local interstellar radiation field. These fitting functions were
calculated from a set of small, high-resolution cosmological
simulations, which included on-the-fly radiative transfer, a simple
model for H$_2$ formation and dissociation, where the main catalyst
for H$_2$ formation are dust grains.  The second prescription, K13, is
a theoretical model where the H$_2$ fraction is calculated assuming
pressure balance between the warm and cold neutral media, and 
depends again on the dust-to-gas mass ratio and the local interstellar
radiation field.  After computing the H$_2$ fraction in each
gas particle, we used apertures of $30$~pkpc to calculate the H$_2$
mass in individual galaxies.

We compared the predicted H$_2$ masses with observations, and
our main conclusions are:
\begin{itemize}
\item The H$_2$ galaxy mass functions at $z=0$ predicted by EAGLE for the two
  prescriptions tested here are very similar to each other and in good
  agreement with the observational estimates of \citet{Keres03} and
  \citet{Obreschkow09b} (see Fig.~\ref{H2MFEagleSF}). We showed that the total star-forming gas mass
  in EAGLE becomes a poor proxy for the H$_2$ mass for $M_{\rm
    H_2}\gtrsim 5\times 10^8\,\rm M_{\odot}$. We find that H$_2$ masses
  become affected by resolution below $M_{\rm H_2}\approx 5\times
  10^{7}\,\rm M_{\odot}$, and for $M_{\rm
    stellar}<10^9\,\rm M_{\odot}$.
\item We made an extensive comparison with the COLD GASS survey \citep{Saintonge11},
  applying the same selection criteria as were applied to the observations:
  $M_{\rm stellar}>10^{10}\,\rm M_{\odot}$ and $M_{\rm H_2}/M_{\rm
    stellar}>0.015$ in galaxies with $M_{\rm stellar}>10^{10.6}\,\rm
  M_{\odot}$ or H$_2$ masses $>10^{8.8}\,\rm M_{\odot}$ in galaxies
  with $10^{10}\,\rm M_{\odot}<M_{\rm stellar}<10^{10.6}\,\rm
  M_{\odot}$. We found generally good agreement in the scaling between the H$_2$ mass and
  other properties such as stellar mass (Fig.~\ref{COLDGASScomparison}), SFR (Fig.~\ref{COLDGASScomparison2}) and stellar surface
  density (Fig.~\ref{COLDGASScomparison3}). An interesting discrepancy was seen in the scaling between
  the SFR and the H$_2$ mass. Observations reveal a population of
  galaxies that have low SFRs, but a considerable amount of H$_2$ (for
  example the early-type galaxies in \citealt{Davis14}; see Fig.~\ref{AllGASScomparison}).  These galaxies
  are offset from the main relation traced by spiral galaxies. In
  EAGLE we find that all galaxies follow the latter relation
  regardless of their morphology, with little scatter, and therefore
  no galaxies lie in the region of low SFR for a fixed H$_2$ mass.
  This is because at small scales the star formation law in EAGLE
  follows the standard \citet{Kennicutt98} star formation relation,
  which has a higher normalisation than the star formation relation
  inferred for early-type galaxies \citep{Davis14}.
\item We presented the SFRs and stellar masses of H$_2$ mass-selected 
samples of galaxies in EAGLE. The most H$_2$-rich galaxies are also the most actively star-forming galaxies 
at any redshift (Fig.~\ref{SFRH2Selected}), while also being the most massive 
at $z\gtrsim 0.5$ (Fig.~\ref{SMH2Selected}). At $z\lesssim 0.5$ a significant fraction of 
galaxies with $M_{\rm stellar}>10^{10}\,M_{\odot}$ are H$_2$-poor ($M_{\rm H_2}<10^{9}\,\rm M_{\odot}$).
\item EAGLE predicts (for both the GK11 or K13 prescriptions
  applied) that $\Omega_{\rm H_2}$ increases by a factor of $\approx
  4$ from $z=0$ to $z \approx 1.3-1.5$, followed by a decrease towards
  higher redshifts (Fig.~\ref{OmegaH2}). In EAGLE the peak redshift of $\Omega_{\rm H_2}$
  is lower than the peak redshift of the SFR density, $z\approx 2$. We find that this is due to the gas metallicity
  decreasing faster at $z>1$ compared to the decrease at $z<1$, and
  the SFR surface density increasing with increasing redshift, on
  average. Both trends hamper the formation of H$_2$ at $z\gtrsim 1$.
  We find that the EAGLE H$_2$ mass functions and $\Omega_{\rm H_2}$
  agree well with the observational inferences in the redshift range
  $0<z<1.5$, but underestimate the H$_2$ abundance at
  $2<z<3$ (see Fig.~\ref{H2MFHighz}). However, it is unclear how much of the latter discrepancy
  is driven by systematic errors in the observations.  We also find
  that $\Omega_{\rm H_2}$ in { EAGLE} is dominated by galaxies with
  $M_{\rm H_2}>10^9\,\rm M_{\odot}$ at $z<2.5$, SFR$>10\,\rm
  M_{\odot}\,\rm yr^{-1}$ at $0.1<z<5$ and $M_{\rm stellar}>5\times
  10^{9}\,\rm M_{\odot}$ at $z\lesssim 2$ (Fig.~\ref{OmegaH2v2}). The latter implies that an
  efficient strategy for unveiling $\Omega_{\rm H_2}$ is to use current
  optical and near-IR surveys with spectroscopic or good photometric
  redshifts and accurate galaxy positions and to follow them up using millimeter
  telescopes targeting  the CO emission.
\end{itemize}

The agreement between the properties of molecular gas in the {EAGLE}
simulations and observations at low and high redshift is remarkable,
particularly when one considers that all the free parameters of the
subgrid physics models in the simulation were fixed by requiring a
match to the observed galaxy stellar mass function and sizes at
the present day. No parameters were adjusted to match any of the H$_2$
observations discussed in this paper. We now plan to study the
distribution of H$_2$ within galaxies in  EAGLE and to compare with
resolved observations, in an effort to further our understanding of how
the physical processes modelled in EAGLE determine the properties
of dense gas in galaxies. 

\section*{Acknowledgements}

We thank Matt Bothwell, Paola Santini and Tim Davis for providing observational
datasets, and Thorsten Naab, Alessandro Romeo and Padelis Papadopoulosfor useful discussions.  
CL is funded by a Discovery Early Career Researcher Award (DE150100618).
RAC is a Royal Society University Research Fellow.
The research leading to these results has received funding
from the European Community's Seventh Framework Programme
($/$FP7$/$2007-2013$/$) under grant agreement no 229517 and from the 
ERC grants: 278594-GasAroundGalaxies. 
This work
used the DiRAC Data Centric system at Durham University, operated by
the Institute for Computational Cosmology on behalf of the STFC DiRAC
HPC Facility ({\tt www.dirac.ac.uk}). This equipment was funded by BIS
National E-infrastructure capital grant ST/K00042X/1, STFC capital
grant ST/H008519/1, and STFC DiRAC Operations grant ST/K003267/1 and
Durham University. DiRAC is part of the National E-Infrastructure.
Support was also provided by the European Research Council (grant
numbers GA 267291 ``Cosmiway'' and GA 238356 ``Cosmocomp''), the
Interuniversity Attraction Poles Programme initiated by the Belgian
Science Policy Office ([AP P7/08 CHARM]), the National Science
Foundation under Grant No. NSF PHY11-25915, and the UK Science and
Technology Facilities Council (grant numbers ST/F001166/1 and
ST/I000976/1) through rolling and consolidating grants awarded to the ICC.

\bibliographystyle{mn2e_trunc8}
\bibliography{EAGLE}

\label{lastpage}
\appendix
\section[]{The transition from neutral to molecular hydrogen}\label{H2Prescriptions}

In this Appendix we provide the equations used to apply the GK11 and K13 prescriptions 
to EAGLE.

\subsection{The prescription of \citet{Gnedin11} applied to { EAGLE}}\label{GnedinSec}

In this section we describe the analytic formulae that
\citet{Gnedin11} obtained for computing the H$_2$ fraction and the way
we apply them to {EAGLE}. The H$_2$ fraction is calculated as

\begin{equation}
f_{\rm H_2}\equiv \frac{\Sigma_{\rm H_2}}{\Sigma_{\rm n}} \approx \left(1+\frac{\Sigma_{\rm c}}{\Sigma_{\rm n}}\right)^2,
\label{fH2n}
\end{equation}

\noindent where, $\Sigma_{\rm H_2}$ is the surface density of H$_2$ and $\Sigma_{\rm n}$ is the surface density of neutral hydrogen.
The parameter $\Sigma_{\rm c}$ is defined as 

\begin{equation}
\Sigma_{\rm c}=20\,{\rm M}_{\odot}\,{\rm pc}^{-2} \frac{\Lambda^{4/7}}{D_{\rm MW}}\frac{1}{\sqrt{1+G^{'}_0\,D^2_{\rm MW}}}, 
\end{equation}

\noindent where $D_{\rm MW}$ is the dust-to-gas mass ratio in units of
the Milky Way dust-to-gas mass ratio, which we take to be $D_{\rm
  MW}\equiv Z/Z_{\odot}$ under the assumption that the dust-to-gas
ratio scales linearly with gas metallicity, $Z$, and that this
relation is universal, $Z_{\odot}=0.0127$ is the solar metallicity
\citep{Allende-Prieto01}, and $G^{'}_0$ is the interstellar radiation
field (defined in the wavelength range $912\AA$ to $2400 \AA$) in
units of the Habing radiation field, $1.6\times 10^{-3}\, \rm erg\,
cm^{-2}\, s^{-1}$ \citep{Habing68}.  Radiation at these frequencies is
produced mainly by OB stars.  In EAGLE we adopt a universal IMF and
therefore the rate at which OB stars are formed is proportional to the
SFR, such that we can express $G^{'}_0$ as $G^{'}_0=\Sigma_{\rm
  SFR}/\Sigma_{\rm SFR,0}$, where $\Sigma_{\rm SFR,0}=10^{-3}\,\rm
M_{\odot}\, \rm yr^{-1}\,\rm kpc^{-2}$ is the SFR surface density in
the solar neighborhood \citep{Bonatto11}.  Finally, $\Lambda$ is
defined as

\begin{equation}
\Lambda \equiv {\rm ln}\left[1+g\,D^{3/7}_{\rm MW}\,\left(\frac{G^{'}_0}{15}\right)^{4/7}\right].
\label{LambdaEq}
\end{equation}

\noindent Here, the function $g$ is 

\begin{equation}
g=\frac{1+\alpha s+s^2}{1+s},
\end{equation}

\noindent where $\alpha$ and $s$ are defined as 

\begin{eqnarray}
s&\equiv& \frac{0.04}{D_{\ast}+D_{\rm MW}},\\
\alpha &\equiv& 5\,\frac{G^{'}_0/2}{1+(G^{'}_0/2)^2}.
\end{eqnarray}

\noindent Here $D_{\ast}$ is defined as

\begin{equation}
D_{\ast}= 1.5\times 10^{-3}\,{\rm ln}\left(1+[3\,G^{'}_0]^{1.7}\right).
\label{Dast}
\end{equation}

Following \citet{Schaye01}, we estimate the local neutral hydrogen surface density, $\Sigma_{\rm n}$,          
by multiplying the volume density of hydrogen by the neutral fraction and the Jeans length, $\lambda_{\rm J}$:

\begin{equation}
\Sigma_{\rm n}=\eta \rho_{\rm H}\,\lambda_{\rm J}=\eta \rho_{\rm H}\,\frac{c_{\rm s}}{\sqrt{G\,\rho_{\rm H}}}.
\label{SigmaH}
\end{equation}

\noindent Here, $c_{\rm s}$ is the effective sound speed which is a function of the 
pressure, $P$, and the gas density, $\rho$ (see \citealt{Schaye08} for details).   
From Eqs.~\ref{fH2n}~and~\ref{SigmaH}, we can estimate $\Sigma_{\rm H_2}$.

We apply the above prescriptions to individual gas particles, where
$\Sigma_{\rm SFR}=\rho_{\rm SFR}\,\lambda_{\rm J}$, $\rho_{\rm SFR}$
is the SFR density (computed from the pressure using Eq.~$12$ of
\citealt{Schaye08}), $\lambda_{\rm J}$ is calculated as in
Eq~\ref{SigmaH}, $n_{\rm n}$ is the particle density of neutral
hydrogen and $Z$ the particle smoothed metallicity.

For non star-forming particles (those that have densities below the 
density threshold for star formation, see Eq.~\ref{critn}), $\rho_{\rm SFR}=0$, so we adopt the UV photoionisation background of 
\citet{Haardt01}, which is a function of redshift, to determine $G^{'}_0$.
{ \citet{Rahmati13b} showed that star-forming galaxies produce a galactic scale photoionisation 
rate of $\approx 1.3\times 10^{-13}\rm \, s^{-1}$, which is of a similar magnitude than the UV background at $z=0$, and smaller 
than it at $z>0$. This favours our approximation, which uses the UV photoionisation background as the 
relevant photo-ionisation rate for non star-forming particles.}

\begin{figure}
\begin{center}
\includegraphics[width=0.49\textwidth]{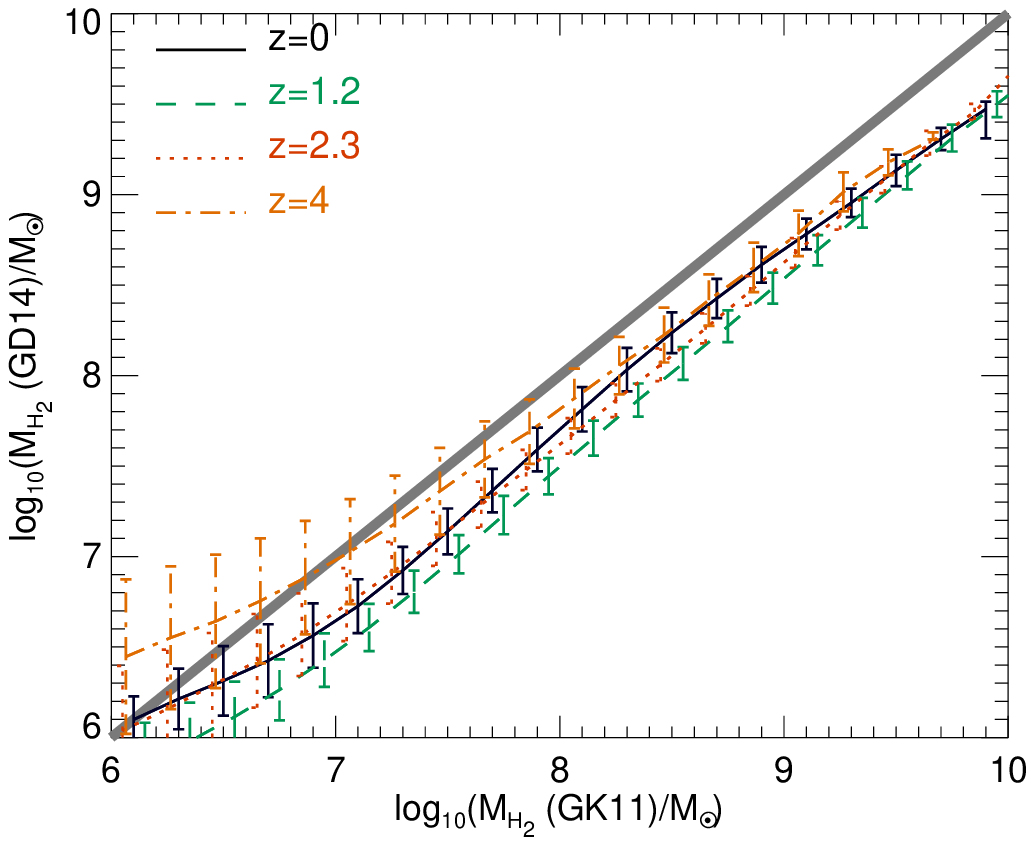}
\includegraphics[width=0.49\textwidth]{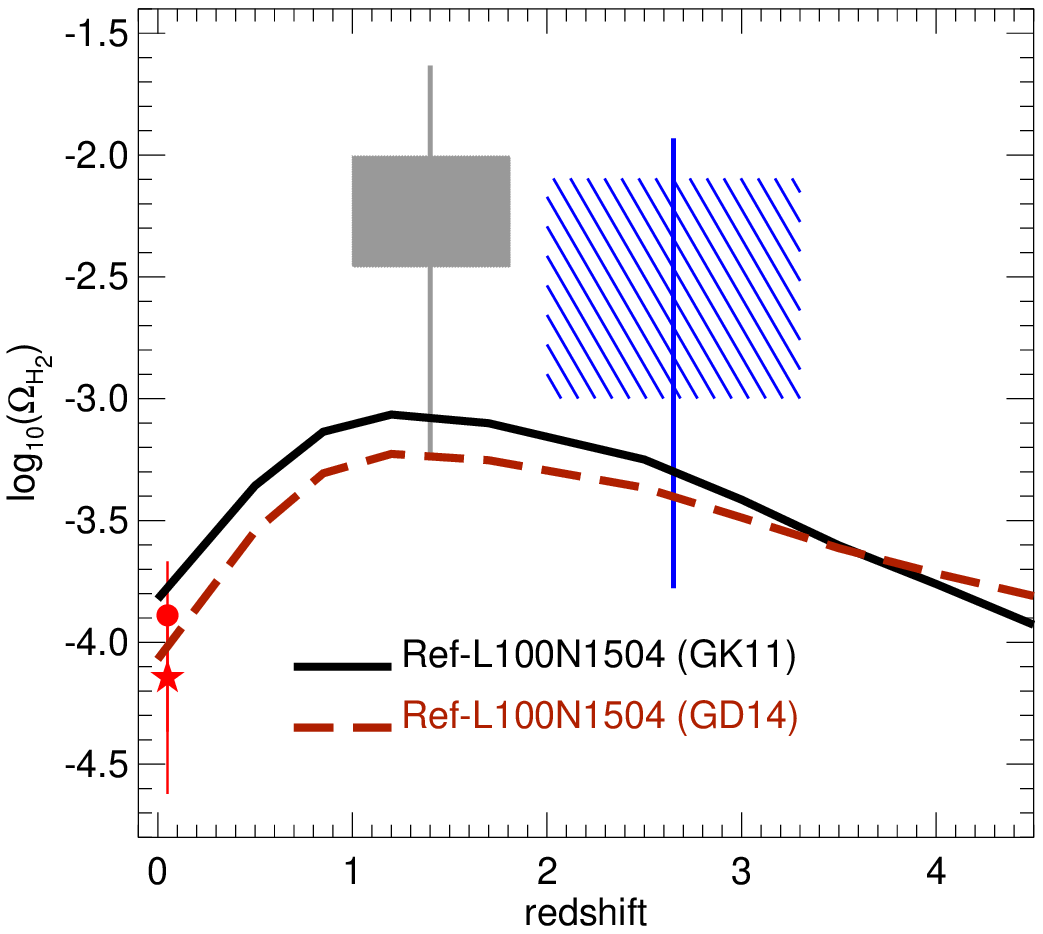}
\caption{{\it Top panel:} The H$_2$ masses in the Ref-L100N1504 simulation when the GD14 prescription is applied compared 
to when the GK11 is applied, at different redshifts, as labelled. 
The diagonal solid line shows the relation $M_{\rm H_2}(\rm GD14)\equiv M_{\rm H_2}(\rm GK11)$. 
Lines with errorbars show the median and $16^{\rm th}$ and $84^{th}$ percentiles. {\it Bottom panel:} The evolution of 
$\Omega_{\rm H_2}$ in the Ref-L100N1504 simulation with the GK11 and GD14 prescriptions applied, as labelled. Symbols 
are as in Fig.~\ref{OmegaH2}.}
\label{H2GD14}
\end{center}
\end{figure}

Note that, since Eqs.~\ref{fH2n} to \ref{Dast} are fits to the simulation 
set of \citet{Gnedin11}, there are no free parameters we can adjust. Instead, we apply the prescription 
as presented in Gnedin \& Kravtsov directly to EAGLE. { We also implemented the updated version of the GK11 
prescription, described in \citet{Gnedin14} and found that H$_2$ masses were typically lower when the updated prescription 
is applied (which 
we refer to as GD14) than with the GK11 prescription. The top panel of Fig.~\ref{H2GD14} shows the H$_2$ masses predicted by 
the GD14 prescription as a function of the H$_2$ masses resulting from applying the GK11 prescription, at different redshifts. 
The differences seen in the H$_2$ masses with the GK11 and GD14 prescriptions depends on redshift, because the correction 
introduced by GD14 to the GK11 prescriptions changes the dependency of $f_{\rm H_2}$ on the ISRF. In the bottom panel of 
Fig.~\ref{H2GD14} we show the effect the changed on $M_{\rm H_2}$ have on $\Omega_{\rm H_2}$. Although there is a change in 
overall normalisation, the behaviour of the two models is quite similar: there is a peak in $\Omega_{\rm H_2}$ at 
$z\approx 1.2$ followed by a decline at higher redshift.}

\subsection{The prescription of \citet{Krumholz13} applied to EAGLE}

\citet{Krumholz13} considered an ISM composed of a warm neutral medium (WNM) and 
a cold neutral medium (CNM) which are in pressure equilibrium. This implies a minimum density 
of the CNM, first described by \citet{Wolfire03}, which is

\begin{equation}
n_{\rm CNM,min} \approx 31\,G^{'}_0\,\left(\frac{Z_{\rm d}/Z}{1+3.1(G^{'}_0\,Z_{\rm d}/\zeta_{\rm t})^{0.365}}\right)\, {\rm cm}^{-3},
\label{ncnmW03}
\end{equation}

\noindent where $G^{'}_0$ is the interstellar radiation field in units of the Habing radiation field,
$Z_{\rm d}$ and $Z$ are the dust and gas phase metallicities, respectively, and 
$\zeta_{\rm t}$ is the ionisation rate due to cosmic rays and X-rays. Assuming that the cosmic rays density and the UV intensity scale 
with the local SFR, \citet{Krumholz13} approximates $G^{'}_0/\zeta_{\rm t}= 1$. Similarly, 
the metallicity of the dust and gas phase are assumed to be equal, $Z_{\rm d}/Z\equiv 1$.
In $2$-phase equilibrium the CNM can have densities between $n_{\rm CNM,min}$ and $\approx 5\,n_{\rm CNM,min}$. 
With this in mind \citet{Krumholz13} writes a fiducial CNM density for 
$2$-phase equilibrium 

\begin{equation}
n_{\rm CNM,2p} \approx 23\,G^{'}_0\,\left(\frac{1+3.1\,D^{0.365}_{\rm MW}}{4.1}\right)^{-1}\, {\rm cm}^{-3}.
\label{ncnm}
\end{equation}

\noindent Note that we slightly changed the notation of \citet{Krumholz13} to make it consistent with the notation used in $\S$~\ref{GnedinSec}.

In the regime where $G^{'}_0 \rightarrow 0$, $n_{\rm CNM,2p}$ in Eq.~\ref{ncnm} also tends to $0$, which would imply 
that the pressure of the CNM tends to $0$ too. This regime is unphysical, and therefore \citet{Krumholz13} argues that 
the thermal pressure of the gas becomes relevant in this regime and sets a minimum CNM density to maintain hydrostatic balance, 
$n_{\rm CNM,hydro}$ (see also \citealt{Ostriker10}). 
The latter depends on the maximum temperature at which the CNM can exist ($\approx 243$~K; \citealt{Wolfire03}), 
$T_{\rm CNM,max}$, the matter density (of stars and dark matter), $\rho_{\rm sd}$, and the neutral gas surface density, $\Sigma_{\rm n}$ 
(which we compute using Eq.~\ref{SigmaH}). \citet{Krumholz13} writes:

\begin{equation}
n_{\rm CNM,hydro}=\frac{P_{\rm th}}{1.1\,k_{\rm B}\,T_{\rm CNM,max}},
\end{equation}

\noindent where

\begin{equation}
P_{\rm th}\approx \frac{\pi\,G\,\Sigma^2_{\rm n}}{4\,\alpha}\left[1+\sqrt{1+\frac{32\,\zeta_{\rm d}\,\alpha\,f_{\rm w}\,c^2_{\rm w}\, \rho_{\rm sd}}{\pi\,G\,\Sigma^2_{\rm n}}}\right].
\end{equation}

\noindent Here $\alpha\approx 5$ represents how much of the midplane pressure support comes from turbulence, magnetic fields and cosmic rays, 
compared to the thermal pressure \citep{Ostriker10}, $\zeta_{\rm d}\approx 0.33$ is a numerical factor that depends on 
the shape of the gas surface isodensity contour, 
$f_{\rm w}=0.5$ is the ratio between the mass-weighted mean square thermal velocity dispersion and the square of the sound speed of 
the warm gas (the value adopted here originally comes from \citealt{Ostriker10}) and $c_{\rm w}=8\,\rm km\,s^{-1}$ 
is the sounds speed in the warm neutral medium (e.g. \citealt{Leroy08}). The value of the gas density in the CNM 
is then taken to be $n_{\rm CNM}={\rm max}(n_{\rm CNM,2p},n_{\rm CNM,hydro})$.

\citet{Krumholz13} defines a dimensionless radiation field parameter:

\begin{equation}
\chi=7.2\,G^{'}_0 \left(\frac{n_{\rm CNM}}{10\,{\rm cm}^{-3}}\right)^{-1},
\end{equation}

\noindent and writes the ratio between the H$_2$ density and the total neutral hydrogen density as

\begin{eqnarray}
f_{\rm H_2}= \left\{ \begin{array}{rl}
 1-{0.75\,s}/({1+0.25\,s}), &\mbox{ $s<2$} \\
  0, &\mbox{ $s \ge 2$}
       \end{array} \right.
\label{H2FracK13}
\end{eqnarray}

\noindent where 

\begin{eqnarray}
s&\approx& \frac{{\rm ln}(1+0.6\,\chi+0.01\,\chi^2)}{0.6\,\tau_{\rm c}},\\
\tau_{\rm c}&=& 0.066\,f_{\rm c}\,D_{\rm MW}\,\left(\frac{\Sigma_{\rm n}}{\rm M_{\odot} {\rm pc}^{-2}}\right).\\
\label{sdefinition}
\end{eqnarray}
 
\noindent Here $f_{\rm c}$ is a clumping factor representing the ratio between the surface density of a 
gas complex and the surface density of the diffuse medium. 
This value is suggested to be $f_{\rm c}\approx 5$ on scales of $\approx 1\,{\rm kpc}$ by \citet{Krumholz05} based 
on observations of nearby galaxies. Since the spatial resolution 
of EAGLE is of the order of $1\,{\rm kpc}$ at low redshift (see Table~\ref{TableSimus}), we adopt $f_{\rm c}= 5$.

As in $\S$~\ref{GnedinSec}, we apply Eqs.~\ref{ncnm} to \ref{sdefinition} to individual 
gas particles to obtain the H$_2$ mass per particle, and then calculate the total H$_2$ mass within $30$~pkpc of the 
centre of individual subhalos.  

\subsection{Differences between the GK11 and K13 prescriptions}
\begin{figure}
\begin{center}
\includegraphics[width=0.49\textwidth]{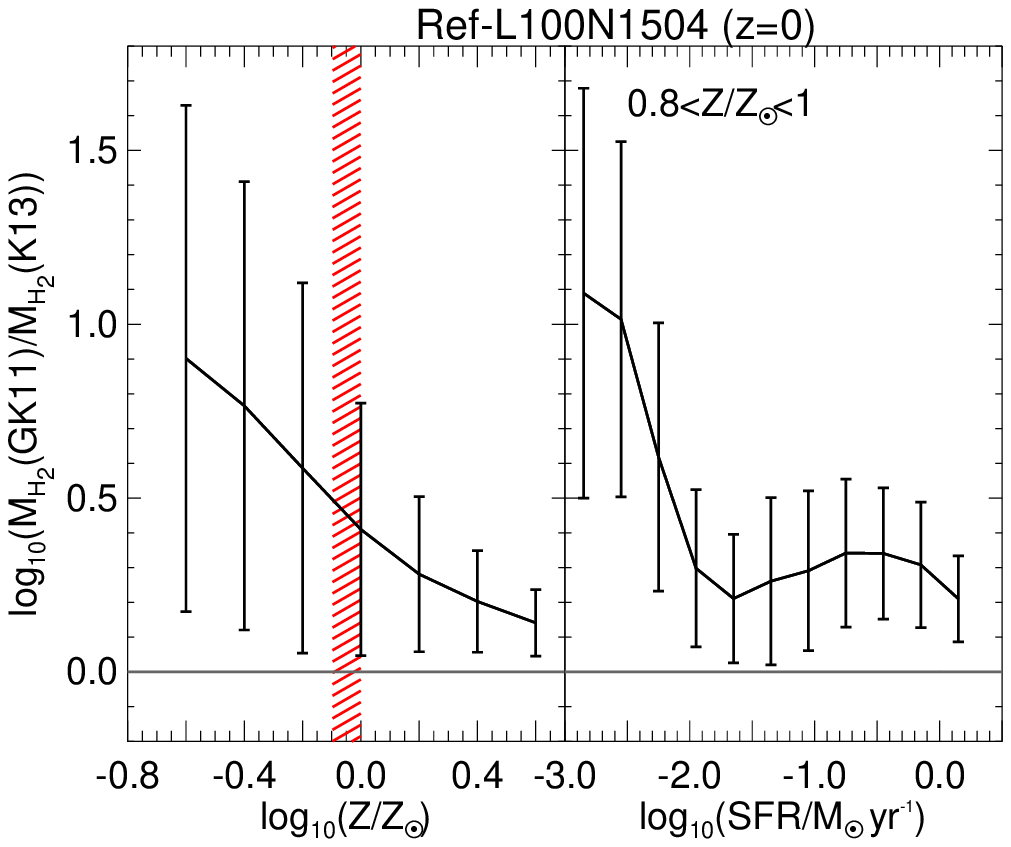}
\caption{The ratio between the H$_2$ mass calculated using the GK11 prescription 
and the mass calculated using the K13 prescription, $M_{\rm H_2}{\rm (GK11)}/M_{\rm H_2}\rm (K13)$, 
as a function of the neutral gas metallicity at $z=0$ in the 
 simulation Ref-L100N1504 (left-hand panel). There is a trend of increasing $M_{\rm H_2}{\rm (GK11)}/M_{\rm H_2}\rm (K13)$ 
with decreasing metallicity. Lines with error bars show the median and the $16^{\rm th}$ and $84^{\rm th}$ percentiles. 
We select galaxies in a narrow range in metallicity, shown by the hatched strip in the left-hand panel, 
and show in the right-hand panel the ratio $M_{\rm H_2}{\rm (GK11)}/M_{\rm H_2}\rm (K13)$ as a function of the SFR. 
For reference, we show in both panels the ratio $M_{\rm H_2}{\rm (GK11)}/M_{\rm H_2}\rm (K13)=1$ as horizontal line. 
The major driver of the dispersion at a fixed metallicity is the SFR.} 
\label{RhoT}
\end{center}
\end{figure}
\begin{figure}
\begin{center}
\includegraphics[width=0.49\textwidth]{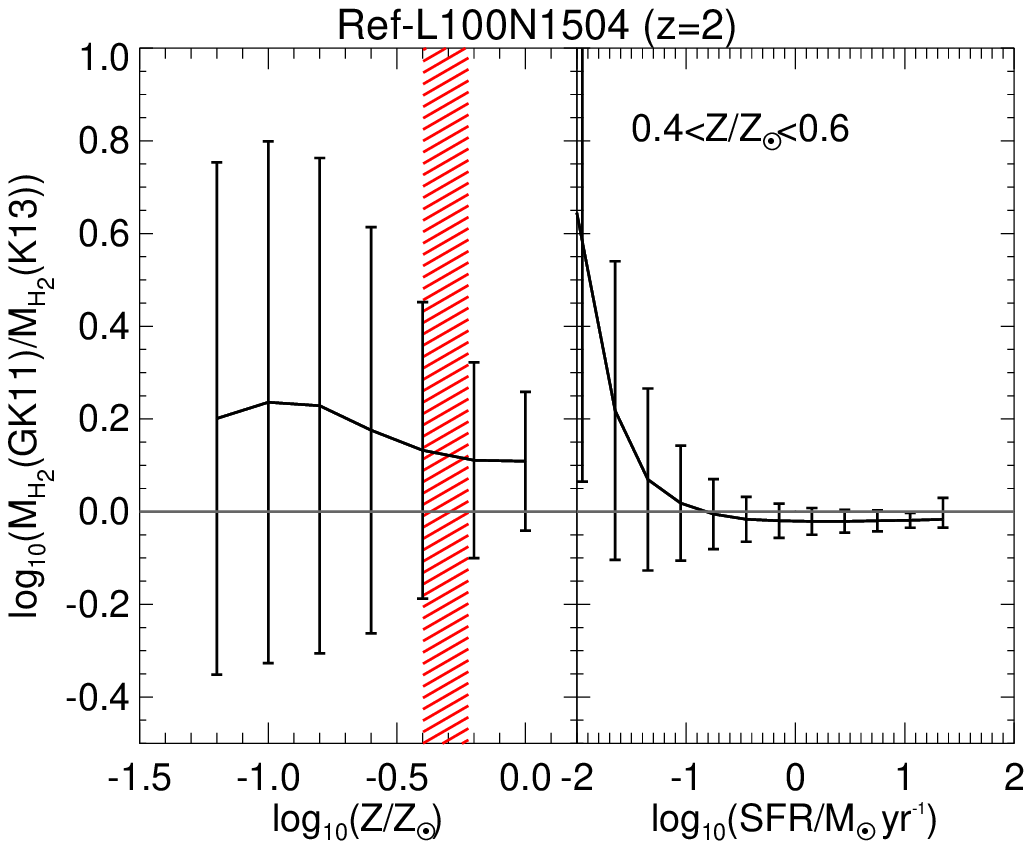}
\caption{As Fig.~\ref{RhoT}, but for galaxies at $z=2$. 
Note that to produce the right-hand panel we select galaxies with lower metallicities than in the right-hand 
panel of Fig.~\ref{RhoT} to increase the number of galaxies selected.}
\label{RhoT2}
\end{center}
\end{figure}

Fig.~\ref{RhoT} shows the ratio between 
the H$_2$ mass calculated using the GK11 prescription
and the mass calculated using the K13 prescription, $M_{\rm H_2}{\rm (GK11)}/M_{\rm H_2}\rm (K13)$,
as a function of the neutral gas metallicity at $z=0$ in the
 simulation Ref-L100N1504. The largest differences between $M_{\rm H_2}{\rm (GK11)}$ and $M_{\rm H_2}\rm (K13)$ are
obtained in the low-metallicity 
regime, $Z\lesssim0.5\,Z_{\odot}$, where $M_{\rm H_2}{\rm (GK11)}/M_{\rm H_2}\rm (K13)>5$. However, the variation around 
the median value of $M_{\rm H_2}{\rm (GK11)}/M_{\rm H_2}\rm (K13)$ seen at a fixed gas metallicity is large. Fow example, 
at $Z\approx 0.25\,\rm Z_{\odot}$, the median value of $M_{\rm H_2}{\rm (GK11)}/M_{\rm H_2}\rm (K13)$ is $\approx 8$, but 
the 16$^{\rm th}$ and 84$^{\rm th}$ percentiles are $\approx 1.5$ and $\approx 38$, respectively, revealing that there is another important parameter driving 
the differences seen in the H$_2$ mass resulting from applying the GK11 and K13 prescriptions. 

To gain insight into what is driving the large dispersion in $M_{\rm H_2}{\rm (GK11)}/M_{\rm H_2}\rm (K13)$ at a fixed gas metallicity, 
we select galaxies in a narrow range of metallicities (shown by the hatched strip in the left-panel of Fig.~\ref{RhoT}) and plot for those 
galaxies 
 the ratio $M_{\rm H_2}{\rm (GK11)}/M_{\rm H_2}\rm (K13)$ as a function of SFR in the right-hand panel of Fig.~\ref{RhoT}. It can be seen 
that galaxies with SFR$<0.01\,\rm M_{\odot}\,\rm yr^{-1}$ exhibit the 
largest differences between $M_{\rm H_2}{\rm (GK11)}$ and $M_{\rm H_2}\rm (K13)$. We conclude that large deviations between 
H$_2$ mass calculated using the GK11 or the K13 prescriptions appear in regimes of low metallicity and weak interstellar 
radiation fields. Note there is still a large scatter in the distribution shown at the right panel of Fig.~\ref{RhoT}, which is due to 
the SFR entering in the H$_2$ fraction equations as SFR surface density.

At $z\gtrsim 1$ an interesting regime appears, which is the one of intense interstellar
radiation fields, $G^{'}_0\gtrsim 100$ in units of the Habing field (see Fig.~\ref{nHweighted}). In this regime we observe 
that the K13 prescription gives similar or slightly higher H$_2$ fractions than the GK11.
Fig.~\ref{RhoT2} is analogous to Fig.~\ref{RhoT} but for $z\approx 2$. As can be seen from the right-hand panel of Fig.~\ref{RhoT2}, 
galaxies with SFR$>1\,\rm M_{\odot}\,\rm yr^{-1}$, have a ratio 
$M_{\rm H_2}{\rm (GK11)}/M_{\rm H_2}\rm (K13)\sim 1$. This is the regime of intense interstellar radiation fields. 
 Note that for the same metallicity, galaxies at $z=0$ have a higher ratio $M_{\rm H_2}{\rm (GK11)}/M_{\rm H_2}\rm (K13)$ 
than galaxies at $z=2$, on average. 

\subsubsection{Comparison with previous work}

In previous works, the empirical prescription of \citet{Blitz06} has been applied to gas particles 
in hydrodynamic simulations (e.g. \citealt{Duffy12}), but restricting
the calculation to gas particles that are star-forming. 
\citet{Duffy12} adopted the following prescription (following the observations presented in \citealt{Leroy08}),

\begin{equation}
f_{\rm H_2} = \left(\frac{P/k_{\rm b}}{10^{4.23}\rm K\,cm^{-3}}\right)^{0.8},
\label{BR06}
\end{equation}

\noindent where $k_{\rm b}$ is Boltzmann's constant. To get a better insight into the main differences between the theoretical prescriptions 
adopted here and the empirical prescription adopted in Duffy et al., we first estimate how much of the H$_2$ in the Ref-L100N1504 simulation 
with the prescriptions GK11 and K13 applied is locked up in particles
that are star-forming. We find that on average 
$82$\% of the H$_2$ is located in gas particles that are star-forming when the GK11 prescription is applied. This percentage rises to 
$99$\% when the K13 prescription is applied. However, we find that variations around that median are large, and galaxies can have as little at 
$20$\% of their H$_2$ mass in gas particles that are star-forming.

\begin{figure}
\begin{center}
\includegraphics[width=0.49\textwidth]{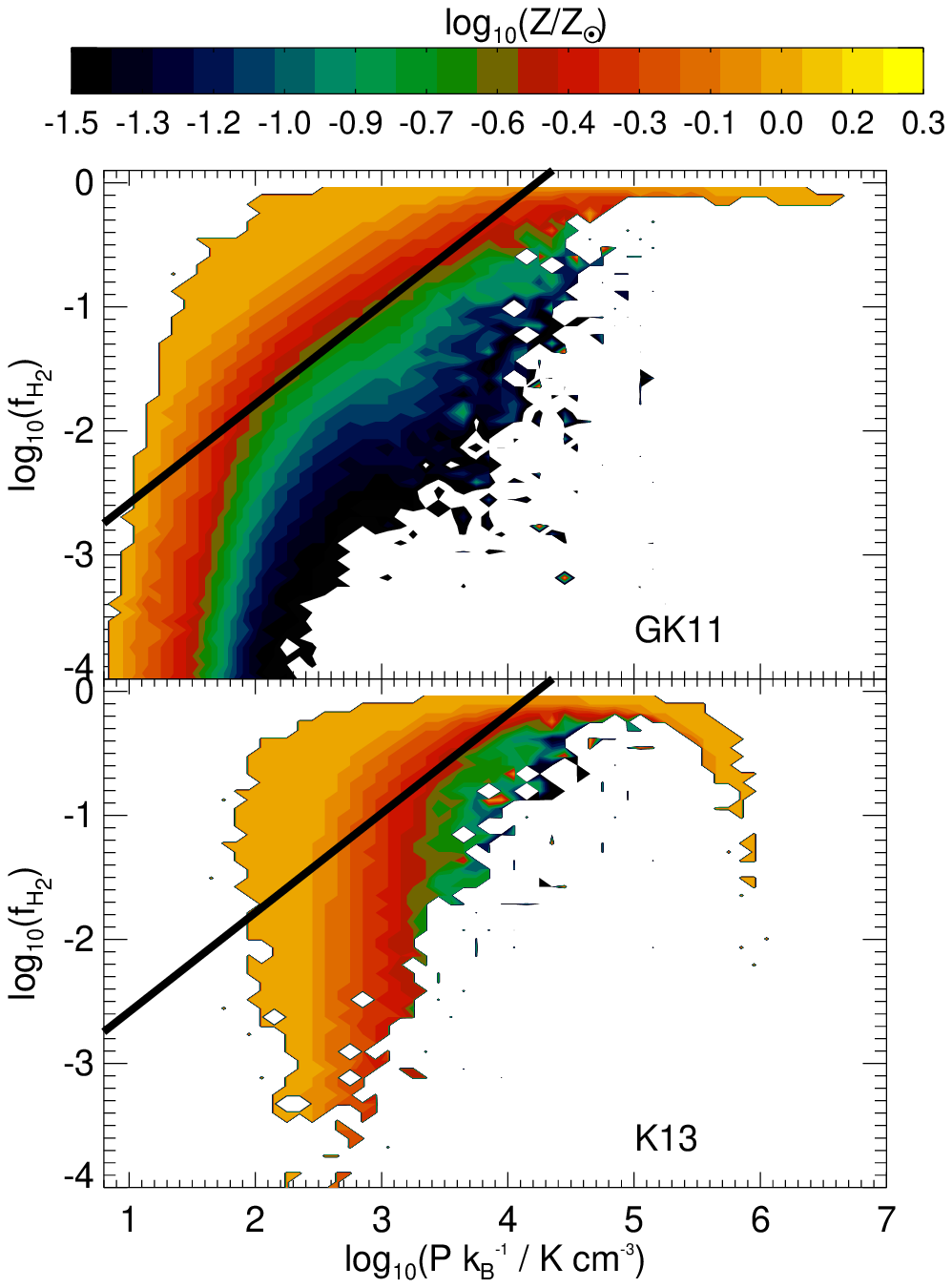}
\caption{The fraction $f_{\rm H_2}$ as a function of pressure for the 
simulation Ref-L100N1504 when the GK11 (top panel) and the K13 (bottom panel) 
prescriptions are applied. Pixels are coloured by the median gas metallicity, as labelled at the top.
 The solid line shows the relation of Eq.~\ref{BR06}.}
\label{CompBR06}
\end{center}
\end{figure}

Second, we study the dependency of $f_{\rm H_2}$, calculated using the GK11 and K13 prescriptions, on pressure in the 
simulation Ref-L100N1504 and compare it 
with Eq.~\ref{BR06}. This is shown in Fig.~\ref{CompBR06}. The relations resulting from applying the GK11 and K13 prescriptions
deviate significantly from Eq.~\ref{BR06} particularly in regions of low metallicity.
This is a good example of the differences seen between applying the empirical relation of Eq.~\ref{BR06} 
and more physically-motivated prescriptions that include a dependence on gas metallicity. 

\section[]{Strong and weak convergence tests}\label{ConvTests}

\begin{figure}
\begin{center}
\includegraphics[width=0.49\textwidth]{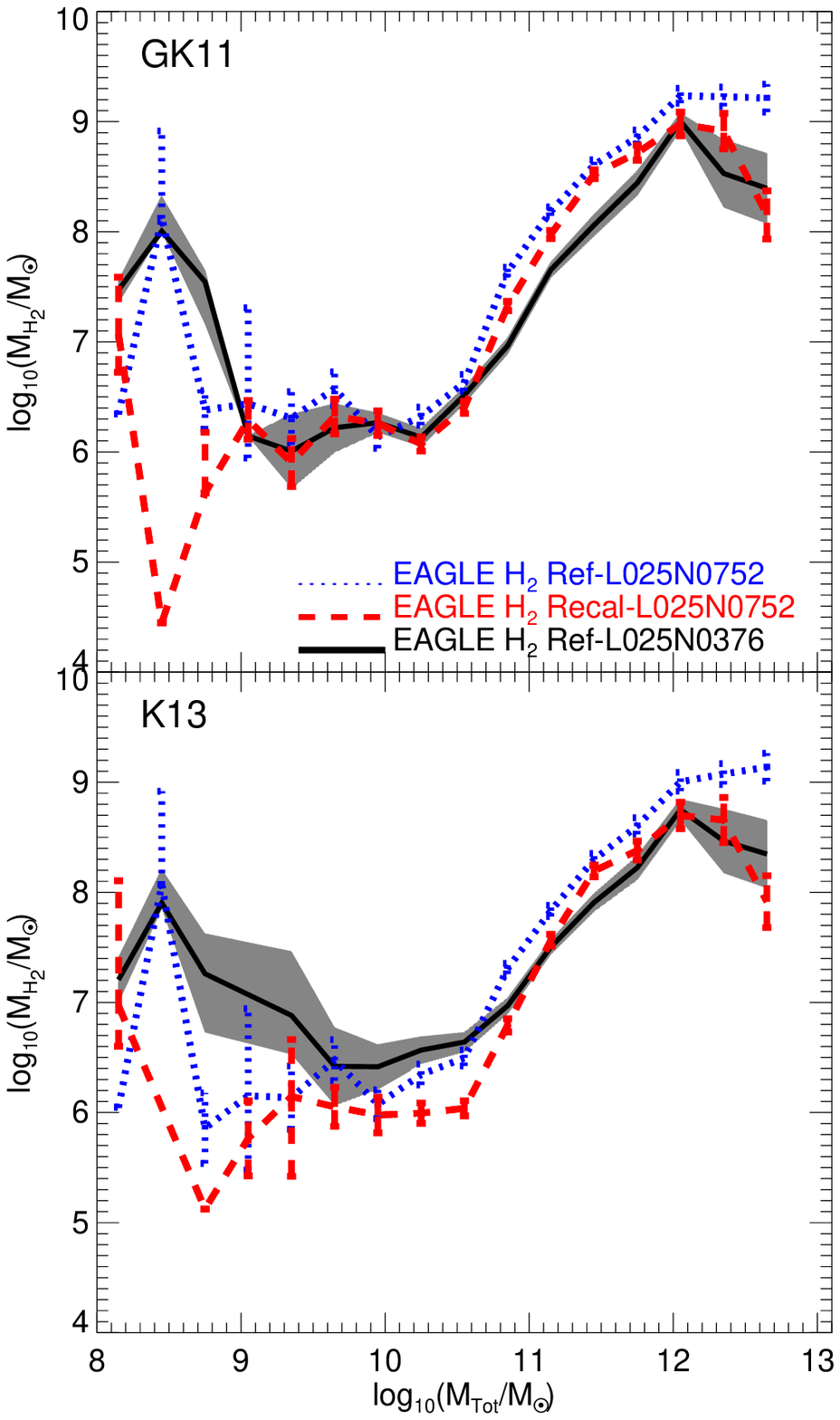}
\caption{The H$_2$ mass as a function of the total matter, $M_{\rm Tot}$ (dark plus baryonic matter), at 
$z=0$ for the simulations Ref-L025N0376, Ref-L025N0752 and Recal-L025N0752, as labelled, using the GK11 (top panel) 
and K13 (bottom panel) prescriptions. 
The H$_2$ mass here corresponds to the mass measured in an aperture of $30$pkpc, while 
$M_{\rm Tot}$ is the total subhalo mass.
Lines show the medians, while the shaded region and errorbars show the uncertainty on the median (e.g. the 
$16^{\rm th}$ and $84^{\rm th}$ percentiles of the distributions divided by $\sqrt(N)$, where $N$ is the number of objects at 
each bin). The simulations 
Ref-L025N0376 and Recal-L025N0752 agree very well at $M_{\rm Tot}>10^{10}\,\rm M_{\odot}$ 
regardless of the prescription used to calculate H$_2$. At 
$M_{\rm Tot}<10^{10}\,\rm M_{\odot}$ the simulation Ref-L025N0376 starts to differ significantly from the 
higher resolution simulations which implies that resolution problems become important.}
\label{H2MFEagle}
\end{center}
\end{figure}

Here we compare the simulations Ref-L025N0376, Ref-L025N0752 and
Recal-L025N0752 to address the effect of resolution on the H$_2$
masses of galaxies in a fixed cosmological volume.  The comparison at
a fixed volume is needed as simulations with different cosmological
volumes sample different halos, which has the problem that differences
can come from either resolution or halo sampling.  We remind the
reader that the simulation Recal-L025N0752 has four parameters that
are modified slightly compared to Ref-L025N0376 and Ref-L025N0752 (see
$\S$~\ref{EagleSec} for details).

Fig.~\ref{H2MFEagle} shows the H$_2$ mass as a function of total
subhalo mass (dark plus baryonic) at $z=0$ for the simulations
Ref-L025N0376 (solid lines), Ref-L025N0752 (dotted lines) and
Recal-L025N0752 (dashed line) using either the GK11 (top panel) or K13
(bottom panel) prescriptions.  The H$_2$ mass is measured in an
aperture of $30$~pkpc, while $M_{\rm Tot}$ is the total subhalo
mass. We decide to use $M_{\rm Tot}$ to characterise when the
resolution becomes important in the H$_2$ mass of galaxies in the
intermediate-resolution simulations  because the subhalo mass is a well
converged quantity.  For the strong convergence test we need to
compare Ref-L025N0376 with Ref-L025N0752 (i.e. fixed subgrid
parameters).  We see that the simulation Ref-L025N0752 shows H$_2$
masses for a fixed $M_{\rm Tot}$ that are a factor of $2-3$ larger
than the H$_2$ masses in Ref-L025N0376 for $M_{\rm Tot}>5\times
10^{10}\,\rm M_{\odot}$. This difference is similar to the difference
seen by S15 and \citet{Furlong14} in the $z=0.1$ galactic stellar mass
function in the same set of simulations.

For the weak convergence test we compare Ref-L025N0376 with
Recal-L025N0752.  We find that the medians are closer to each other
than in the strong convergence test.  The difference seen between the
median relations in the simulations Ref-L025N0376 and Recal-L025N0752
is a factor of $\lesssim 1.5-2$ for $M_{\rm Tot}>5\times 10^{10}\,\rm M_{\odot}$,
which again is consistent with the differences seen in the galactic
stellar mass function in S15 and \citet{Furlong14}. Note that the
exact prescription we use to calculate H$_2$ makes little difference
to the offset obtained between the simulations we are comparing here.

The higher resolution of Ref-L025N0752 and Recal-L025N0752 enables us
to establish that, at the resolutions of the simulations  Ref-L025N0376 and
Ref-L100N1504, we can trust the H$_2$ content of subhalos with $M_{\rm
  Tot}>10^{10}\,\rm M_{\odot}$. In the three simulations analysed here
there are $\approx 1500$ subhalos with $M_{\rm Tot}>10^{10}\,\rm
M_{\odot}$.

Note that it is difficult to establish a H$_2$ mass below which
resolution effects become significant. This is because each gas
particle has a weight, the H$_2$ fraction, and therefore a low H$_2$
mass can be obtained by either having a small H$_2$ fraction but a
large number of gas particles, or by having a larger H$_2$ fraction
but few gas particles. Thus, we find it more appropriate to express
the resolution limit in terms of $M_{\rm Tot}$.  We adopt $M_{\rm
  Tot}=10^{10} \,\rm M_{\odot}$ as our resolution limit, and show
unresolved results in Fig.~\ref{H2MFEagleSF} as a dotted line.  In
terms of H$_2$ mass, this resolution limit corresponds to $M_{\rm
  H_2}\approx 5\times 10^7\,\rm M_{\odot}$ on average for the
intermediate-resolution simulations, both for the GK11 and K13
prescriptions.

\end{document}